\RequirePackage{fix-cm}
\RequirePackage{graphicx}

\documentclass[numbook]{svjour3}  

\usepackage{amsmath,amsfonts,amssymb}
\usepackage[utf8]{inputenc}
\usepackage{bm}
\usepackage{braket}
\usepackage{xcolor}

\newcommand{\MSbar}{\ensuremath{\overline{\rm MS}}}
\newcommand{\bea}{\begin{eqnarray}} 
\newcommand{\eea}{\end{eqnarray}}

\newcommand{\str}[1]{{}}

\newcommand{\inst}[1]{\textsuperscript{#1}}

\allowdisplaybreaks
\begin{document}

\title{Challenges in Semileptonic $\bm{B}$ Decays\quad\hfill{\normalsize MITP/20-035}}
\titlerunning{Challenges in Semileptonic $B$ Decays} 

\author{P.~Gambino\inst{1}
\and A.S.~Kronfeld\inst{2}
\and M.~Rotondo\inst{3}
\and C.~Schwanda\inst{4}
\and F.~Bernlochner\inst{5}
\and A.~Bharucha\inst{6}
\and C.~Bozzi\inst{7}
\and M.~Calvi\inst{8}
\and L.~Cao\inst{5}
\and G.~Ciezarek\inst{9}
\and C.T.H.~Davies\inst{10}
\and A.X.~El-Khadra\inst{11}
\and S.~Hashimoto\inst{12}
\and M.~Jung\inst{1}
\and A.~Khodjamirian\inst{13}
\and Z.~Ligeti\inst{14}
\and E.~Lunghi\inst{15}
\and V.~Lüth\inst{16}
\and T.~Mannel\inst{13}
\and S.~Meinel\inst{17}
\and G.~Paz\inst{18}
\and S.~Schacht\inst{19,1}
\and S.~Simula\inst{20}
\and W.~Sutcliffe\inst{5}
\and A.~Vaquero~Avilés-Casco\inst{21}
}

\institute{
\inst{1}Physics Department, University of Torino \& INFN Torino, 10125 Torino, Italy \and
\inst{2}Theoretical Physics Department, Fermilab, Batavia, IL 60510-5011, USA \and 
\inst{3}Laboratori Nazionali di Frascati, INFN, 00044 Frascati RM, Italy \and 
\inst{4}Institut für Hochenergiephysik,  Österreichische Akademie der Wissenschaften, 1050 Wien, Austria \and
\inst{5} Physikalisches Institut der Rheinischen Friedrich-Wilhelms-Universit\"at Bonn, 53115 Bonn, Germany \and
\inst{6}Aix Marseille Univ, Universit\'e de Toulon, CNRS, CPT, Marseille, France \and
\inst{7} INFN, Sezione di Ferrara, Ferrara, Italy \and
\inst{8} University of Milano Bicocca and INFN, Milano Bicocca, Milano, Italy \and
\inst{9}European Organization for Nuclear Research (CERN), Geneva, Switzerland \and
\inst{10}SUPA, School of Physics and Astronomy, University of Glasgow, Glasgow, G12 8QQ, UK\and
\inst{11}Department of Physics, University of Illinois, Urbana, IL 61801, USA\and
\inst{12}Theory Center, High Energy Accelerator Research Organization (KEK) and School of High Energy Accelerator Science (SOKENDAI),
    Tsukuba 305-0801, Japan\and
\inst{13 }Theoretische Physik 1,
Naturwissenschaftlich-Technische Fakult\"at, 
Universit\"at Siegen, 57068 Siegen, Germany\and
\inst{14}Lawrence Berkeley National Laboratory, University of California, Berkeley, CA 94720, USA\and
\inst{15} Physics Department, Indiana University, Bloomington, IN 47405, USA \and
\inst{16} SLAC National Accelerator Laboratory, Stanford, California 94309, USA \and
\inst{17}Department of Physics, University of Arizona, Tucson, AZ 85721, USA\and
\inst{18}Department of Physics and Astronomy, Wayne State University, Detroit, MI 48201, USA\and
\inst{19}Department of Physics, LEPP, Cornell University, Ithaca, NY 14853, USA \and
\inst{20}INFN, Sezione di Roma Tre, Roma, Italy\and
\inst{21}Department of Physics and Astronomy, University of Utah, Salt Lake City, UT 84112 USA
}


\maketitle

\begin{abstract}
Two of the elements of the Cabibbo-Koba\-yashi-Maskawa quark mixing matrix, $|V_{ub}|$ and $|V_{cb}|$,
are extracted from semileptonic $B$ decays. The results of the $B$ factories, analysed in the light of the most recent theoretical 
calculations, remain puzzling, because for both $|V_{ub}|$ and $|V_{cb}|$ the exclusive and inclusive determinations are in clear tension.
Further, measurements in the $\tau$ channels at Belle, Babar, and LHCb show discrepancies with the Standard Model predictions,
pointing to a possible violation of lepton flavor universality. LHCb and  Belle~II  have the potential to
resolve these issues in the next few years.
This article summarizes the discussions and results obtained at the MITP workshop held on April 9--13, 2018, in Mainz, Germany, with
the goal to develop a medium-term strategy of analyses and calculations aimed at solving the puzzles. Lattice and continuum theorists
working together with experimentalists have discussed how to reshape the semileptonic analyses in view of the much higher luminosity 
expected at Belle~II, searching for ways to systematically validate the theoretical predictions in both exclusive and inclusive $B$ 
decays, and to exploit the rich possibilities at LHCb.
\end{abstract}

\tableofcontents

\section{Executive Summary}
\label{intro}


The magnitudes of two of the elements of the Cabibbo-Koba\-yashi-Maskawa (CKM) quark mixing
matrix~\cite{Cabibbo:1963yz,Kobayashi:1973fv}, $|V_{ub}|$ and $|V_{cb}|$, are extracted from semileptonic
$B$-meson decays.
The results of the $B$ factories, analysed in the light of the most recent theoretical calculations, remain puzzling, because -- for
both $|V_{ub}|$ and $|V_{cb}|$ -- the determinations from exclusive and inclusive decays are in tension by about 3$\sigma$.
Recent experimental and theoretical results reduce the tension, but the situation remains unclear.
Meanwhile, measurements in the semitauonic channels at Belle, Babar, and LHCb show discrepancies with the Standard Model (SM)
predictions, pointing to a possible violation of lepton-flavor universality.
LHCb and the upcoming experiment Belle~II have the potential to resolve these issues in the next few years.

Thirty-five participants met at the Mainz Institute for Theoretical Physics to develop a medium-term strategy of analyses and 
calculations aimed at the resolution of these issues.
Lattice and continuum theorists discussed with experimentalists how to reshape the semileptonic analyses in view of the much larger
luminosity expected at Belle~II and how to best exploit the new possibilities at LHCb, searching for ways to systematically validate
the theoretical predictions, to confirm new physics indications in semitauonic decays, and to identify the kind of new physics
responsible for the deviations.


\subsubsection*{Format of the workshop}
The program took place during a period of five days, allowing for ample discussion time among the participants.
Each of the five workshop days was devoted to  specific topics: the inclusive and exclusive determinations of $|V_{cb}|$ and
$|V_{ub}|$, semitauonic $B$ decays and how they can be affected by new physics, as well as related subjects such as purely leptonic
$B$ decays and heavy quark masses.
In the mornings, we had overview talks from the experimental and theoretical sides, reviewing the main aspects and summarizing the
state of the art.
In the late afternoon, we organized discussion sessions led by experts of the various topics, addressing questions that have been
brought up before or during the morning talks.

\subsubsection*{Exclusive heavy-to-heavy decays}
The  $B\to D^{(*)}\ell \nu$ decays have received significant attention in the last few years.
New Belle results for the $q^2$ and angular distributions have allowed studies of the role 
played by the parametrization of the form factors in the extraction of $|V_{cb}|$. It turns out that 
the extrapolation to zero-recoil is very sensitive to the parametrization employed, a problem 
that can be solved only by precise calculations of the form factors at non-zero recoil. Until 
these are completed, the situation remains unclear, with repercussions on the calculation of 
$R(D^*)$ as well, with diverging views on the theoretical uncertainty of present estimates
based on Heavy Quark Effective Theory (HQET) expressions. 

Beside a critical reexamination of these recent developments, we discussed several incremental and qualitative improvements in
lattice QCD, also in baryonic decays.
Though unlikely to carry much weight in determining $|V_{cb}|$, the latter offer great opportunities to test lepton-flavor
universality violation (LFUV) and lattice QCD.
The discussions also addressed the fact that QCD errors are now almost as small as effects from QED.
Thus, further improvement must be theoretically made by properly studying the effect of QED radiation, especially the treatment of
soft photons and photons that are neither soft nor hard and their sensitivity to the meson wave functions.

Concerning studies of LFUV, we discussed the role played by higher excited charmed states in establishing new physics and
the challenges that the present $R(D^{(*)})$ measurements represent for model building.

\subsubsection*{Exclusive heavy-to-light decays}

This determination of $|V_{ub}|$ relies on nonperturbative calculations of the form factor of $B\to \pi\ell\nu$, which is the most
precise channel. 
We discussed the status of the light-cone sum rule (LCSR) calculations and several recent improvements in lattice QCD,
in particular the most recent results from the Fermilab Lattice \& MILC Collaborations and from the RBC \& UKQCD Collaborations,
as well as future prospects.
The Fermilab/MILC calculation alone leads to a remarkably small total error on $|V_{ub}|$, about $4\%$.
While at present the most precise extraction of $|V_{ub}|$ comes from $B\to \pi\ell\nu_\ell$, it is worth considering the channel
$B_s \to K\ell\nu$ as well, because here the lattice-QCD calculations are affected by somewhat smaller uncertainties.
$B_s \to K\ell\nu$ can be accessible at Belle~II in a run at the $\Upsilon(5S)$ and a precision of about 5--10\% could be achieved
with $1~\text{fb}^{-1}$.
On the other hand, LHCb has an ongoing analysis of the ratio $B(B_s \to K\ell\nu)/B(B_s\to D_s\ell\nu)$, which will provide a new
determination of $|V_{ub}/V_{cb}|$.
This approach follows the success that LHCb demonstrated for semileptonic baryon decays via the precise measurement of the ratio
$B(\Lambda_b\to p\mu\nu)/(\Lambda_b\to \Lambda_c \mu\nu)$ in the high-$q^2$ region. 
This measurement, combined with precise lattice-QCD calculations of the form factors, allowed the extraction of ratio
$|V_{ub}/V_{cb}|$ with an uncertainty of $7\%$.
We discussed also other channels, in particular how to study $B\to\pi\pi \ell\nu$ including the resonant structures. 
Careful studies of other heavy-to-light channels will also be crucial to improve the signal model for the inclusive $|V_{ub}|$
measurements.

\subsubsection*{Inclusive heavy-to-heavy decays}
The theoretical predictions in this case are based on an operator product expansion.
Theoretical uncertainties already dominate current determinations, and better control of all higher-order corrections
is needed to reduce them.
In this respect, it would be important to have the perturbative-QCD corrections to the complete coefficient of the Darwin operator and to
check the treatment of QED radiation in the experimental analyses.
A full $O(\alpha_s^3)$ calculation of the total width may be within reach with recently developed techniques.
From the experimental point of view, new and more accurate measurements will be most welcome, in particular to better understand
the correlations between different moments and moments with different cuts.
A better determination of the higher hadronic mass moments and a first measurement of the forward-backward asymmetry would benefit 
the global fit, as would a better understanding of higher power corrections.
The importance of having global fits to the moments in different schemes and by different groups has also been stressed.
This calls for an update of the $1S$ scheme fit and could lead to a cross-check of the present theoretical uncertainties.
Lattice QCD already provides inputs to the fit with the calculation of the heavy quark masses, which have been reviewed.
New developments discussed at the workshop may soon be able to provide additional information that can be fed into the fits,
such as constraints on the heavy-quark quantities $\mu_\pi^2$ and $\mu_G^2$.
The two main approaches are \emph{i)}~computing inclusive rates  directly with lattice QCD and \emph{ii)}~using the heavy quark 
expansion for meson masses, precisely computed at different quark mass values.
The state of theoretical calculations for inclusive semitauonic decays has also been discussed, as they represent an important 
cross-check of the LFUV signals.  


\subsubsection*{Inclusive heavy-to-light decays}
This determination is based on various well-founded theoretical methods, most of which agree well.
The 2017 endpoint analysis by BaBar seems to challenge this consolidated picture, suggesting discrepancies between some of the
methods and a lower value of~$|V_{ub}|$.
For the future, the complete NNLO corrections in the full phase space should be implemented and the various methods should be
upgraded in order to make the best use of the Belle~II differential data based on much higher statistics.
These data will make it possible to test the various methods and to calibrate them, as they will contain information on the shape
functions.
The SIMBA and NNVub methods seem to have the potential to fully exploit the $B\to X_u \ell \nu$ (and possibly radiative) measurements
through combined fits to the shape function(s) and $|V_{ub}|$.
The separation of $B^\pm$ and $B^0$ in the experimental analyses will certainly help to constrain weak annihilation, but the real
added value of Belle~II could be precise measurements of kinematic distributions in $M_X$, $q^2$, $E_l$, etc.
A detailed measurement of the high $q^2$ tail might be very useful, also in view of attempts to check quark-hadron duality.
Experimentally, better hybrid (inclusive+exclusive) Monte Carlos are badly needed; $s$-$\bar s$ popping should be investigated to 
develop a better understanding of kaon vetos.
The $b\to c$ background will be measured better, which will benefit these analyses.

\subsubsection*{Leptonic decays}

The measurement of $B\to\tau\nu$ is not yet competitive with semileptonic decays for measuring $|V_{ub}|$, 
because of a 20\% error on the rate. Belle~II will improve on this. 
The corresponding lattice-QCD calculation is however very precise, with an error below~1\%, according to the 2019 report from 
FLAG~\cite{Aoki:2019cca} and based mainly on a result from Fermilab/MILC that was presented at the workshop.
That said, the mode is useful today to model builders trying to understand new physics explanations of the 
tension between inclusive and exclusive determinations of $|V_{ub}|$.
Belle~II will also access $B\to\mu\nu(\gamma)$ with the possibility to reach
an uncertainty on the branching fraction of about $5\%$ with $50~ab^{-1}$, allowing for  a new determination 
of $|V_{ub}|$ in the long term.
We discussed also the LHCb contribution to leptonic decays with the process $B\to\mu\mu\mu\nu_\mu$ where two of the muons
come from virtual $\gamma$ or light vector meson decays.
A study of this channel has been published in 
~\cite{Aaij:2018pKa}
and a very stringent upper limit obtained, inconsistent with the existing branching fraction predictions, calls for new reliable theoretical calculations.


\section{Heavy-to-heavy exclusive decays}
\label{h2h_excl}


%
The aim of this section is to present an overview of $b\to c$ exclusive decays.
After an introduction to the parametrization of the relevant form factors between hadronic states we describe the status of current
lattice QCD calculations with particular focus on $B\to D^*$ and $\Lambda_b\to \Lambda_c$.
Next, we discuss experimental measurements of $B\to D^{(*)}$ semileptonic decays with special focus on the ratios $R(D^{(*)})$, and
several phenomenological aspects of these decays: the extraction of $V_{cb}$, theoretical predictions for
$R(D^{(*)})$, the role of $B\to D^{**}$ transitions and constraints on new physics.
We also briefly discuss the information that is required to reproduce results presented in experimental analyses and to incorporate
older measurements into approaches based on modern form factor parametrizations.
We conclude with the description of HAMMER, a tool designed to more easily calculate the change in signal acceptancies,
efficiencies and signal yields in the presence of new physics.

\subsection{Parametrization of the form factors}
\label{sec:param}

In this section, we introduce the form factors for the hadronic matrix elements that arise in semileptonic decays.
Several different notations appear in the literature, often using different conventions depending on whether the final-state meson
is heavy (e.g., $D$) or light (e.g., $\pi$).
A general decomposition relies, however, only on Lorentz covariance and other symmetry properties of the matrix elements.
As discussed below, it is advantageous to choose the Lorentz structure so that the form factors have definite parity and spin.

In this spirit, let us consider the matrix elements for a meson decay $B_{(l)}\to X^{(\ast)}\ell\nu$, where the quark content of
the $\bar{B}$ is $bl$ with $l$ a light quark ($u$, $d$, or $s$), and the quark content of the $X$ is $\bar{q}l$ where $q$ can be
either a light quark or the~$c$ quark.
The desired decomposition can be written as
\begin{align}
\left\langle X(p')             |S         |B_{(l)}(p) \right\rangle &= \frac{M^2-m^2}{m_b - m_q} f_0(q^2),
    \label{eq:ff-scalar} \\
\left\langle X(p')             |V^{\mu}   |B_{(l)}(p) \right\rangle &= \left[(p+p')^\mu - \frac{M^2-m^2}{q^2}q^\mu\right]f_+(q^2) + 
    \frac{M^2-m^2}{q^2}q^\mu\,f_0(q^2),
    \label{eq:ff-vector} \\
\left\langle X(p')             |T^{\mu\nu}|B_{(l)}(p) \right\rangle &= 2\frac{p^\mu p^{\prime\nu} - p^\nu p^{\prime\mu}}{M+m}
    f_T(q^2), \label{eq:ff-pseudo-tensor} \\
\left\langle X^\ast(p')        |P         |B_{(l)}(p) \right\rangle &= \frac{2m}{m_b + m_q}(\epsilon^\ast \cdot q) A_0(q^2),
    \label{eq:ff-pseudo} \\
\left\langle X^\ast(p')        |V^{\mu}   |B_{(l)}(p) \right\rangle &= \frac{2i}{M+m}\varepsilon^{\mu\nu\alpha\beta}
    \epsilon^\ast_\nu p_\alpha p'_\beta\,V(q^2),
    \label{eq:ff-vector-vector} \\
\left\langle X^\ast(p')        |A^{\mu}   |B_{(l)}(p)  \right\rangle &= 2m\frac{\epsilon^\ast\cdot q}{q^2}q^\mu\,A_0(q^2) +
        (M+m) \left(\epsilon^{\ast\mu} - \frac{\epsilon^\ast\cdot q}{q^2}q^\mu\right) A_1(q^2) \nonumber \\
        &-  \frac{\epsilon^\ast\cdot q}{M+m}\left[(p+p')^\mu - \frac{M^2 - m^2}{q^2}q^\mu\right]A_2(q^2),
    \label{eq:ff-axial} \\
\left\langle X^\ast(p')        |T^{\mu\nu}|B_{(l)}(p)  \right\rangle &= i\epsilon^{\mu\nu\sigma\rho}
    \left\{\epsilon^\ast_\sigma\left[(p+p')_\rho T_1(q^2) - \vphantom{\frac{M^2-m^2}{q^2}} \right.\right. \nonumber \\
        & \left.\left. \hspace{2em} q_\rho\frac{M^2-m^2}{q^2}\left(T_1(q^2) - T_2(q^2)\right)\right]\right.
    \label{eq:ff-tensor} \\
        +& \left.(\epsilon^\ast\cdot p)\frac{(p+p')_\sigma q_\rho}{q^2}\left[T_1(q^2) - T_2(q^2) -
            \frac{q^2}{M^2 - m^2}T_3(q^2)\right]\right\}, \nonumber
\end{align}
where $q^\mu= (p - p')^\mu$ is the momentum transfer, 
$S=\bar{b}q$ is the scalar current,
$P=\bar{b}\gamma^5q$ is the pseudoscalar current,
$V^\mu = \bar{b}\gamma^\mu q$ is the vector current,
$A^\mu = \bar{b}\gamma^\mu\gamma_5 c$ is the axial current,
$T^{\mu\nu} = \bar{b}\sigma^{\mu\nu} c$ is the tensor current,
$m_q$ is the mass of the quark~$q$,
$M$ is the mass of the parent meson ($B$ in this case), 
$m$ (without subscript) is the mass of the daughter meson,
and $r=m/M$.
Contracting Eqs.~(\ref{eq:ff-vector}) and (\ref{eq:ff-axial}) with $q_\mu$ and using the appropriate Ward identities shows that the 
scalar form factor, $f_0$, and pseudoscalar form factor, $A_0$, appear in the vector and axial vector transitions.
The $J^P$ quantum numbers of the form factors are given in Table~\ref{tab:qn}.
\begin{table} \centering
    \caption{Quantum numbers of various meson form factors.}
    \label{tab:qn}
    \begin{tabular}{lccccc}
    \hline\hline
                                            & $0^+$ &      $0^-$      & $1^-$ &   $1^+$    & $2^+$ \\
    \hline
        $B_{(l)}\to X\ell\bar{\nu}$         & $f_0$ &        --       & $f_+$ &     --     & $f_T$ \\
        $B_{(l)}\to X^\ast\ell\bar{\nu}$    &  --   &       $A_0$     & $V_0$ & $A_1, A_2$ & $T_1, T_2, T_3$ \\
    \hline\hline
    \end{tabular}
\end{table}
The tensor form factors in Eqs.~(\ref{eq:ff-pseudo-tensor}) and~(\ref{eq:ff-tensor}) appear in extensions of the Standard Model.

One can impose bounds on the shape of these form factors by using QCD dispersion relations for a generic decay
$H_b\to H_q\ell\bar{\nu}$.
Since the amplitude for production of $H_b H_q$ from a virtual $W$ boson is determined by the analytic continuation of the form 
factors from the semileptonic region of momentum transfer $m_\ell^2 < q^2 < M^2- m^2$ to the pair production region
$q^2\geq M^2 + m^2$, one can find constraints in the pair-production region, amenable to perturbative QCD calculations, and then
propagate the constraint to the semileptonic region by using analyticity.
The result of this process applied to the form factors is the model-independent Boyd-Grinstein-Lebed (BGL) 
parametrization~\cite{Boyd:1995cf,Boyd:1997kz}, which expands a form factor $F(z)$ in the dimensionless variable $z$ as
\begin{align}
    F(z) &= \frac{1}{B_F(z)\phi_F(z)}\sum_{j=0}^\infty a^F_j z^j,
    \label{BGLseries} \\
    z(q^2;t_0) &= \frac{\sqrt{t_+ - q^2} - \sqrt{t_- - q^2}}{\sqrt{t_+ - q^2} + \sqrt{t_+ - t_0}},
    \label{eq:z-def}
\end{align}
where $t_\pm = (M\pm m)^2$, $B_F(z)$ are known as the \emph{Blaschke factors}, which incorporate the below- or 
near-threshold~\cite{Caprini:2017ins} poles in the $s$-channel process $\ell\nu\to \bar{B}X$, and $\phi_F(z)$ is called the
\emph{outer function}.
The poles, and hence the Blaschke factor, depend on the spin and the parity of the intermediate state, which is why it is useful to
use fixed $J^P$ for the form factors.
See Sec.~\ref{sec:z-remarks} for more details.%
\footnote{In particular, there are cases when one should \emph{not} use the naive choice $t_+=(M+m)^2$ in Eq.~(\ref{eq:z-def}).
The correct choice is the branch point of a cut in the complex-$q^2$ plane, which sometimes is at~$t_\text{cut}<(M+m)^2$.}
Of course, in practical applications the series (\ref{BGLseries}) is truncated at some power $z^{n_F}$. 

By taking certain linear combinations of form factors with the same spin and parity one obtains the BGL notation for the helicity 
amplitudes,
\begin{align}
    f^\text{BGL}_+ &= f_+,                                          \label{fpBGL} \\
    f^\text{BGL}_0 &= (M^2 - m^2) f_0,                              \label{f0BGL} \\
    g              &= \frac{2}{M+m} V,                              \label{gBGL}  \\
    f              &= (M+m)A_1,                                     \label{fBGL}  \\
    \mathcal{F}_1  &= M(M+m)(w-r)A_1 - \frac{2Mm(w^2-1)}{1+r}A_2,   \label{F1BGL} \\
    \mathcal{F}_2  &= 2A_0,                                         \label{F2BGL}
\end{align}
leaving aside the (BSM) tensor form factors.
Here the velocity transfer
\begin{equation}
  w = v_M\cdot v_m = \frac{M^2 + m^2 - q^2}{2Mm},
  \label{eq:wdef}
\end{equation}
with $v_M=p/M$ and $v_m=p'/m$, is often used in heavy-to-heavy decays.
For heavy-to-light decays it can be helpful to work with the energy of the daughter meson in the rest frame of the parent, i.e.,
\begin{equation}
    E = p'\cdot v_M = \frac{M^2 + m^2 - q^2}{2M}.
\end{equation}
These form factors are subject to three kinematic constraints, namely
\begin{align}
(M^2 - m^2)f^\text{BGL}_+(q^2=0)                    &= f^\text{BGL}_0(q^2=0),             \label{ffKin1}\\
(M-m)f(q^2=q_\text{max}^2)                          &= \mathcal{F}_1(q^2=q_\text{max}^2), \label{ffKin2}\\
\frac{2}{M^2-m^2} \mathcal{F}_1(q^2=0) &= \mathcal{F}_2(q^2=0),              \label{ffKin3}
\end{align}
where $q^2_\text{max}=(M-m)^2$, corresponding to $w=1$ and $E=m$.

The variable $z$ can also be expressed via $w$,
\begin{equation}
    z = \frac{\sqrt{w+1} - \sqrt{2N}}{\sqrt{w+1} + \sqrt{2N}},
\end{equation}
where $N=(t_+ - t_0)/(t_+ - t_-)$, is real for $q^2 \leq (M + m)^2$, and it becomes a pure phase beyond that limit.
The constant $t_0$ defines the point at which $z=0$.
Often $t_0 = t_-$, one end of the kinematic range, so $z$ ranges from 0 at maximum $q^2$ to
$z_\text{max}=(1-\sqrt{r})^2/(1+\sqrt{r})^2$ when $m_\ell\approx 0$.
Alternatively, the choice $t_0=(M+m)(\sqrt{M}-\sqrt{m})^2$ sets $z=0$ exactly in the middle of the kinematic range.
Even for $B\to\pi\ell\nu$, $z$ is always a small quantity, which ensures a fast convergence of the
power series defined in \eqref{BGLseries}.

Unitarity constraints from the QCD dispersion relations are translated into constraints for the coefficients of the BGL expansion.
In general,
\begin{equation}
    \sum^\infty_{j=0} \left(a^F_j\right)^2 \leq 1,
\end{equation}
for each form factor $F$,  but in the particular case of $\bar{B}\to D^\ast\ell\bar{\nu}$ the bound becomes
\begin{equation}
  \sum^\infty_{j=0} \left[\left(a^f_j\right)^2 + \left(a^{\mathcal{F}_1}_j\right)^2\right] \leq 1,
\end{equation}
for the $f$ and $\mathcal{F}_1$ form factors, because they have the same quantum numbers.
These bounds are known as the \emph{weak} unitarity constraints.

A modification of the BGL parametrization by Bourrely, Lellouch and Caprini (BCL)~\cite{Bourrely:2008za} is often
chosen in analyses of heavy-to-light decays.
The BCL parametrization improves BGL by fixing two artifacts of the truncated BGL series.
In particular, it removes an unphysical singularity at the pair production threshold and corrects  the large $q^2$
behavior (see~\cite{Lepage:1980fj,Akhoury:1993uw}) in the functional form.
These two modifications improve the convergence of the expansion.
However, the kinematic range is much more constrained in the heavy-to-heavy case, and lies farther from both the production
threshold and the large $q^2$ region.
Therefore, the presence of far singularities or an incorrect asymptotic behavior are not expected to spoil the $z$-expansion in
that case.

In the heavy-to-heavy case, one can sharpen the \emph{weak} unitarity constraints on the BGL coefficients using 
heavy quark symmetry (HQS) which relates the different $B^{(*)}\to D^{(*)}\ell \bar\nu$ channels and their form factors: each form
factor is either proportional to the Isgur-Wise function $\xi(w)$ or zero. 
Using heavy quark effective theory (HQET) one can improve the precision by introducing radiative and power (i.e.\ in inverse powers
of the heavy masses) corrections.
Then we can define any form factor in such a way that it admits the expansion in both $\alpha_s$ and the heavy quark masses 
\begin{equation}
    F(w) = \xi(w)\left(1+c_{\alpha_s}\frac{\alpha_s}{\pi} + c_b\frac{\Lambda_{\textrm{QCD}}}{m_b} +
        c_c\frac{\Lambda_{\textrm{QCD}}}{m_c} + \cdots\right).\label{eq:ffexp}
\end{equation}
These expansions can be used to link the $z$~expansion coefficients of different form factors, leading to the so-called
\emph{strong} unitarity constraints \cite{Caprini:1997mu,Bigi:2017jbd}.
The power corrections depend on subleading Isgur-Wise functions that have been estimated with QCD sum
rules~\cite{Neubert:1992wq,Neubert:1992pn,Ligeti:1993hw}.


Previous analyses of $B\to D^\ast\ell\nu$ have used the Caprini-Lellouch-Neubert (CLN) parametrization~\cite{Caprini:1997mu}.
CLN  employ a notation for the form factors that satisfies (\ref{eq:ffexp}),\footnote{See~\cite{Bigi:2017jbd} for a comprehensive table including other decays.}
\begin{align}
    S^\text{CLN}_1 &= \frac{f^\text{BGL}_0}{M^2(1-r)\sqrt{r}(1+w)}, & \quad &
    P^\text{CLN}_1  = \frac{\sqrt{r}}{1+r}\mathcal{F}_2, \\
    V^\text{CLN}_1 &= \frac{2\sqrt{r}}{1+r}f^\text{BGL}_+,          & \quad &
    V^\text{CLN}_4  = M\sqrt{r}g, \\
    A^\text{CLN}_1 &= \frac{f}{M\sqrt{r}(1+w)},                     & \quad &
    A^\text{CLN}_5  = \frac{\mathcal{F}_1}{M^2(1-r)\sqrt{r}(1+w)}, \\
    R^\text{CLN}_1 &= \frac{V^\text{CLN}_4}{A^\text{CLN}_1},        & \quad &
    R^\text{CLN}_2  = \frac{w - r}{w-1} - \frac{(1-r)}{w-1}\frac{A^\text{CLN}_5}{A^\text{CLN}_1},
\end{align}
where the letter naming the form factor ($S$, $P$, $V$ and~$A$) encodes its quantum numbers (scalar, pseudoscalar, vector and
axial vector), and $R^\text{CLN}_{1,2}$ are two convenient ratios of form factors.
Sometimes the ratio $R^\text{CLN}_0 = P^\text{CLN}_1/A^\text{CLN}_1$ is considered.

In the CLN parametrization the
\emph{strong} unitarity constraints obtained with HQET at NLO are used to remove some of the coefficients of the $z$ expansion.
Further, specific numerical coefficients are introduced in a polynomial in~$w$ for $R_{1,2}^\text{CLN}$.
The numerical values were determined using information available in 1997, which has been partly superseded but not updated.
The numerical values also omit error estimates  (which were discussed in the original CLN paper~\cite{Caprini:1997mu}, although in an optimistic manner) because at the time the experimental statistical errors dominated,
which is no longer the case.
A consensus of the workshop recommends that CLN no longer be used, certainly not unless the numerical coefficients have been updated and 
the ensuing theoretical uncertainties are accounted for.
It is better to use a general form of the $z$~expansion.

\str{
The strong unitarity constraints are used to remove some of the coefficients of the expansion, and in the end a simplified polynomial
is presented per each form factor:
\begin{align}
    S^\text{CLN}_1(w) &= V^\text{CLN}_1(w)\left(\frac{2\sqrt{r}}{1+r}\right)^2\frac{1+w}{2} 1.0036 \times \nonumber \\
        & \left(1 - 0.0068(w-1) + 0.0017(w-1)^2 - 0.0013(w-1)^3\right), \\
    V^\text{CLN}_1(w) &= V^\text{CLN}_1(1)\left(1 - 8\rho_1^2 z + (r1\rho^2_1 - 10)z^2 - (252\rho^2_1 - 84)z^3\right), \\
    A^\text{CLN}_1(w) &= A^\text{CLN}_1(1)\left(1 - 8\rho_{A_1}^2z + (53\rho_{A_1}^2 - 15)z^2 - (231\rho_{A_1}^2 - 91)z^3\right), \\
    R^\text{CLN}_1(w) &= R^\text{CLN}_1(1) - 0.12(w-1) + 0.05(w-1)^2, \\
    R^\text{CLN}_2(w) &= R^\text{CLN}_2(1) + 0.11(w-1) - 0.06(w-1)^2, \\
    P^\text{CLN}_1(w) &= A^\text{CLN}_1(w) 1.218  \times \nonumber \\
        & \left(1 - 0.2367(w-1) - 0.0508(w-1)^2 + 0.0988(w-1)^3\right).
\end{align}
where the fit parameters are the slopes $\rho^2_1$ and $\rho^2_{A_1}$, the normalization factors $V^\text{CLN}_1(1)$ and 
$A^\text{CLN}_1(1)$, and the constants $R^\text{CLN}_1(1)$ and $R^\text{CLN}_2(1)$.
Actually, CLN declares $R^\text{CLN}_1(1) \approx 1.27$ and $R^\text{CLN}_2(1) \approx 0.80$, but the late approach in experimental 
fits is to fit them to the available data, relaxing the HQS constraints.
This procedure is known as the \emph{practical CLN parametrization}.
Since this expansion was developed for the $B^{(\ast)}\to D^{(\ast)}$ decays, the hadronic form factors are notably missing.
An alternative and less restrictive way of imposing the strong unitarity constraints has been developed in~\cite{Boyd:1997kz}
and expanded in~\cite{Bigi:2016mdz,Bigi:2017jbd}.
}

HQET naturally presents another basis for the form factors of the $\bar{B}\to D^{(\ast)}\ell\bar{\nu}$ processes.
Using velocities instead of momenta and otherwise mimicking the Lorentz structure of Eqs.(\ref{eq:ff-vector}), 
(\ref{eq:ff-vector-vector}), and~(\ref{eq:ff-axial}), the notation is $h_+$ and $h_-$ for $\bar{B}\to D\ell\bar{\nu}$,
and $h_V$ and $h_{A_{1,2,3}}$ for $\bar{B}\to D^{\ast}\ell\bar{\nu}$.
In the heavy quark limit, these form factors tend to
\begin{equation}
    h_X(w) = \eta(\alpha_s)\xi(w) + O\big(\frac{\Lambda_{\rm QCD}}{m_{b,c}}\big),  \label{hqetFF1}
\end{equation}
for $X=+$, $A_1$, $A_3$, $V$, and
\begin{equation}
    h_Y(w) = \beta(\alpha_s)\xi(w) + O\big(\frac{\Lambda_{\rm QCD}}{m_{b,c}}\big), \label{hqetFF2}
\end{equation}
with $Y=-$,~$A_2$.
Here $\eta(\alpha_s) = 1 + O(\alpha_s)$, while $\beta(\alpha_s) = O(\alpha_s)$.
In this representation, the identities expressed in Eqs.~(\ref{ffKin1})--(\ref{ffKin3}) become evident.

Finally, for the case of a baryonic decay $\Lambda_b\to Y_{(q)}\ell\nu$, with $Y=p,\Lambda_c$, we define
\begin{align}
\left\langle Y(p') |     S           |\Lambda_b(p)\right\rangle &= \bar{u}_q(p')        \frac{M-m}{m_b - m_q} f_0(q^2) u_b(p), \\
\left\langle Y(p') |     P           |\Lambda_b(p)\right\rangle &= \bar{u}_q(p')\gamma_5\frac{M+m}{m_b + m_q} g_0(q^2) u_b(p), \\
\left\langle Y(p') |     V^{\mu}     |\Lambda_b(p)\right\rangle &= \bar{u}_q(p')        \left[(M-m)\frac{q^\mu}{q^2} f_0(q^2) + \right. \nonumber \\
                                                                &  \left. \frac{M+m}{s_+}\left((p + p')^\mu -
                                                                \frac{q^\mu}{q^2}(M^2- m^2)\right) f_+(q^2) +\right.\nonumber \\
                                                                &  \left.\left(\gamma^\mu - 2\frac{mp^\mu + Mp^{\prime\mu}}{s_+}\right)
                                                                f_\bot(q^2)\right] u_b(p), \\
\left\langle Y(p') |     A^{\mu}     |\Lambda_b(p)\right\rangle &=-\bar{u}_q(p')\gamma_5\left[(M+m)\frac{q^\mu}{q^2} g_0(q^2) + \right. \nonumber \\
                                                                &  \left.\frac{M-m}{s_-}\left((p + p')^\mu -
                                                                \frac{q^\mu}{q^2}(M^2- m^2)\right) g_+(q^2) +\right.\nonumber \\
                                                                &  \left.\left(\gamma^\mu + 2\frac{mp^\mu - Mp^{\prime\mu}}{s_-}\right) g_\bot(q^2)\right] u_b(p), \\
\left\langle Y(p') |q_\nu T^{\mu\nu} |\Lambda_b(p)\right\rangle &=-\bar{u}_q(p')\left[\left((p + p')^\mu - \frac{q^\mu}{q^2}(M^2-m^2)\right)\frac{q^2}{s_+} h_+(q^2)\right. \nonumber \\
                                                                &  \left.+ (M + m) \left(\gamma^\mu - 2\frac{mp^\mu + Mp^{\prime\mu}}{s_+}\right) h_\bot(q^2)\right] u_b(p),
\end{align}
where $M$ is the mass of the $\Lambda_b$, $m$ is the mass of the daughter baryon and $s_\pm = (M\pm m)^2 - q^2$.
The $z$ expansions for the baryonic form factors employed in Ref.~\cite{Detmold:2015aaa}
use trivial outer functions and do not impose unitarity bounds on the coefficients of the expansion.
As a result, the coefficients are unconstrained and reach values as high as $\sim10$. See also Sec.~\ref{sec:z-remarks}.

\subsection{Heavy-to-heavy form factors from lattice QCD}
\label{sec:hth_lat}

The lattice QCD calculation of the form factors for the semileptonic 
decay of a hadron uses 
two- and three-point correlation functions, which are constructed 
from valence quark propagators obtained by solving the Dirac equation 
on a set of gluon field configurations. Averaging the correlation 
functions over the gluon field configurations then yields the appropriate Feynman 
path integral. 
The two-point correlation functions give the amplitude for a hadron to be created 
at the time origin and then destroyed at a time $T$. The three-point 
correlation functions include the insertion of a current $J$ at 
time $t$ on the active quark line, changing the active quark 
from one flavor to another. 
Usually calculations are performed with the initial hadron at 
rest. Momentum is inserted at the current so that a range of 
momentum transfer, $q$, from initial to final hadron can be mapped 
out. 

The three-point correlation functions (for multiple $q$ values) and 
the two-point correlation functions (with multiple momenta in the 
case of the final-state hadron) are fit 
as functions of $t$ and $T$ to determine the matrix elements of 
the currents between initial and final hadrons that yield the 
required 
form factors. An important point here is that the initial and final 
hadrons that we focus on are the ground-state particles in their respective 
channels. However, terms corresponding to excited 
states must be included in the fits in order to make sure that systematic effects 
from excited-state contamination are taken into account in the 
fit parameters that yield the ground-state
to ground-state matrix element of $J$ and hence the form factors. 

Statistical uncertainties in the form factors obtained obviously 
depend on the numbers of samples of gluon-field configurations 
on which correlation functions are calculated. To improve statistical 
accuracy further, calculations usually include 
multiple positions of the time origin
for the correlation functions on each configuration. 
The numerical cost of the calculation of quark propagators falls 
as the quark mass increases and so heavy ($b$ and $c$) quark 
propagators are typically numerically inexpensive. 
The accompanying light quark propagators for heavy-light hadrons 
are much more expensive, especially if $u/d$ quarks with physically 
light masses are required. It is this cost that limits the 
statistical 
accuracy that can be obtained, especially since the statistical 
uncertainty for a heavy-light hadron correlation function 
(on a given number of gluon field configurations) 
also grows as the separation in mass between the heavy and light  
quarks increases.

A key issue for heavy-to-heavy ($b$ to $c$) form factor calculations 
is how to handle heavy quarks on the lattice. Discretization of 
the Dirac equation on a space-time lattice gives systematic discretization 
effects that depend on powers of the quark mass in lattice units. 
The size of these effects depends on the value of the lattice spacing 
and the power with which the effects appear (i.e., the level of 
improvement used in the lattice Lagrangian). 

Since the $b$ quark is so heavy, its mass in lattice units will be larger 
than 1 on all but the finest lattices ($a < $ 0.05~fm) currently in 
use. Highly-improved discretizations of the Dirac equation are 
needed to control the discretization effects. A good example of such 
a lattice quark formalism is the highly improved 
staggered quark (HISQ) action developed   
by HPQCD~\cite{Follana:2006rc} for both light and heavy quarks with 
discretization errors appearing at $O(\alpha_s(am)^2)$ and $O((am)^4)$.   
An alternative approach is to make use of the fact 
that $b$ quarks are nonrelativistic inside their bound states. 
This means that a discretization of a nonrelativistic action 
(NRQCD) can be used, expanding the action to some specified order in the 
$b$ quark velocity. Discretization effects then depend on the 
scales associated with the internal dynamics and 
these scales are all much smaller 
than the $b$ quark mass. Relativistic effects can be included  
and discretization effects corrected at the cost of complicating the action 
with additional operators. 
A third possibility is to start from the
Wilson quark action and improved versions of it but to tune the 
parameters (such as the quark mass) using a nonrelativistic 
dispersion relation for the meson, which is known as the Fermilab method~\cite{ElKhadra:1996mp}. 
This removes the leading source 
of mass-dependent discretization effects, whilst retaining a 
discretization that connects smoothly to the continuum limit.    
Again, improved versions of this approach (such as the Oktay-Kronfeld action \cite{Oktay:2008ex}) 
include additional operators. 

The $c$ quark has a mass larger than $\Lambda_{\mathrm{QCD}}$
but within lattice QCD it can be treated successfully as a light quark 
because its mass in lattice units is less than 1 on lattices in 
current use (with $a <  0.15$~fm). This means that, although 
discretization effects are visible in lattice QCD calculations 
with $c$ quarks, they are not large and can easily be extrapolated 
away accurately for a continuum result. For example, discretization 
effects are less than 10\% at $a=0.15$~fm in calculations of 
the decay constant of the $D_s$ using the HISQ action~\cite{Davies:2010ip}. 
Purely nonrelativistic approaches to the $c$ quark are therefore not 
useful on the lattice. 
There can be some advantage for $b$-to-$c$ form factor 
calculations in using the same action for $b$ and $c$, however, as we 
discuss below. 

Because lattice and continuum QCD regularize the theory in a 
different 
way, the lattice current $J$ needs a finite renormalization factor to match its 
continuum counterpart so that matrix elements of $J$, and form 
factors 
derived from them, can be used 
in continuum phenomenology. 
For NRQCD and Wilson/Fermilab quarks the current 
$J$ must be normalized using lattice QCD perturbation theory. 
Since this is technically rather challenging it has only been 
done through $O(\alpha_s)$ and this 
leaves a sizeable (possibly several percent) systematic error from 
missing higher-order terms in the perturbation theory. 
If Wilson/Fermilab quarks are used for both $b$ and $c$ quarks, 
then arguments 
can be made about the approach to the heavy-quark limit 
that can reduce, but not eliminate, this uncertainty~\cite{Harada:2001fj}. 

Relativistic treatments of the $b$ and $c$ quarks 
have a big advantage here, because $J$ can generally 
be normalized in a fully nonperturbative way 
within the lattice QCD calculation and without additional 
systematic errors.  
The advantages of this approach were first demonstrated by the 
HPQCD collaboration using the HISQ action to determine 
the decay constant of the $B_s$ \cite{McNeile:2012qf}. The HISQ PCAC relation 
normalizes the axial-vector current in this case. 
Calculations for multiple 
quark masses on lattices with multiple values of the lattice spacing 
allow both the physical dependence of the decay constant on quark 
mass and the dependence of the discretization effects to be mapped 
out 
so that the physical result at the $b$ quark mass can be determined. 
This calculation has now been updated and extended to the $B$ meson 
by the Fermilab Lattice and MILC collaborations \cite{Bazavov:2017lyh}, achieving 
better than 1\% uncertainty. HPQCD is now carrying out a similar approach 
to $b$-to-$c$ form factor calculations \cite{McLean:2019sds}, and the JLQCD 
collaboration is also working in that direction 
\cite{Kaneko:2019vkx} with M\"{o}bius domain-wall quarks. 

An equivalent approach, using ratios of hadronic quantities at different 
quark masses where normalization factors cancel, has been developed by the 
European Twisted Mass collaboration using the twisted-mass action \cite{Blossier:2009hg,Bussone:2016iua}
for Wilson fermions.

\subsubsection{$B \rightarrow D^{(*)}$ form factors from lattice QCD} 

Early lattice QCD calculations of $B\to D$ form factors  were limited to the determination of
$\mathcal{G}^{B\to D} (w) = 4 r f_+ (q^2)/(1+r)$ (with notation defined near \eqref{eq:wdef})
at the zero-recoil point $w=1$. Results include the $N_f=2+1$ calculation of Fermilab/MILC~\cite{Okamoto:2004xg,Qiu:2013ofa} and the
$N_f=2$ calculation of Atoui et al.~\cite{Atoui:2013zza}.
More recently Fermilab/MILC~\cite{Lattice:2015rga} and HPQCD~\cite{Na:2015kha,Monahan:2017uby} have presented $N_f=2+1$ calculations of
the $B\to D$ form factor at non-zero recoil based on partially overlapping subsets of the same MILC asqtad ($a^2$ tadpole improved) ensembles.

The Fermilab/MILC calculation~\cite{Lattice:2015rga} uses configurations with four different lattice spacings and with pion masses in
the range $[260,670]$~MeV. The bottom and charm quarks are implemented in the Fermilab approach. The form factors $f_{+,0}^{B\to D}(w)$
are extracted from double ratios of three point functions up to a matching factor which is calculated at 1-loop in lattice perturbation
theory.
The results are presented in terms of three synthetic data points which can be subsequently fitted using any form factor
parametrization. 
The systematic uncertainty due to the joint continuum-chiral extrapolation is about 1.2\% and dominates the error budget.

The HPQCD calculations~\cite{Na:2015kha,Monahan:2017uby} rely on ensembles with two different lattice spacings and two/three light-quark 
masses values, respectively.
The treatment of heavy quarks is different from that used in the Fermilab/MILC papers: the bottom  quark is described in NRQCD and the
charm quark using HISQ.
The form factors are extracted from appropriate three-point functions and the results are presented in terms of the parameters of a
modified BCL $z$~expansion that incorporates dependence on lattice spacing and light-quark masses into the expansion coefficients. 

In order to combine the Fermilab/MILC and HPQCD results~\cite{Aoki:2019cca}, it is necessary to generate a set of synthetic data which
is (almost exactly) equivalent to the HPQCD calculation.
The two sets of synthetic data can then be combined while taking into account the correlation due to the fact the Fermilab/MILC and HPQCD
share MILC asqtad configurations.
As mentioned above, dominant uncertainties are of systematic nature, implying that this correlation (whose estimate is rather uncertain)
is a subdominant effect.
A simultaneous fit of Fermilab/MILC and HPQCD synthetic data together with the available Belle and Babar data
yields a determination of $|V_{cb}|$ with an overall 2.5\% 
uncertainty (dominated by the experimental error which contributes about 2\% to the total error).

Finally, both collaborations present values for both the $f_+$ and $f_0$ form factors, which allow for a lattice only calculation of the 
SM prediction for $R(D)$.
The uncertainty on the Fermilab/MILC and HPQCD combined determination of $R(D)$, without experimental input, is about 2.5\% and is 
 negligible compared to current experimental errors.

The advantage of an approach in which currents can be nonperturbatively normalized has 
been demonstrated by HPQCD for $B_s \rightarrow D_s$ form factors in~\cite{McLean:2019qcx}. They use the HISQ action for all quarks, extending the method developed for decay constants. The range of heavy quark masses can be increased on successively finer lattices (keeping the value in lattice units below 1) until the full range from $c$ to $b$ is reached. The full $q^2$ range of the decay can also be covered by this method since the spatial momentum
of the final state meson (which should also be less than 1 in lattice units) grows in step with 
the heavy meson/quark mass. Results from~\cite{McLean:2019qcx} improve on the 
uncertainties obtained in~\cite{Monahan:2017uby} with NRQCD $b$ quarks and this 
promising all-HISQ approach is now being extended to other processes. It is interesting to 
observe that the $B_s\to D_s$ form factors are very close to the $B\to D$ form factors over 
the entire kinematic range, see also \cite{Kobach:2019kfb,Bordone:2019guc}. 

Calculations of $B\to D^*$ form factors at non-zero recoil are considerably more involved due to difficulties in describing the resonant
$D^*\to D \pi$ decay. Up to now, lattice QCD simulations have focused on the single 
$B \rightarrow D^*$ form factor that contributes to the rate at zero 
recoil, $A_1(q^2_{max})$. The quantity generally quoted is 
$h_{A_1}(1)$ where 
\begin{equation}
h_{A_1}(1) = \frac{M_B+M_{D^*}}{2\sqrt{M_BM_{D^*}}} A_1(q^2_\text{max})
\end{equation}
The combination of the lattice QCD result and the experimental rate, extrapolated to zero recoil, yields a value for~$V_{cb}$.

The Fermilab Lattice/MILC Collaborations have achieved the highest precision for this result so far \cite{Bailey:2014tva}.
They use improved Wilson quarks within 
the Fermilab approach for both $b$ and $c$ quarks and work on gluon field 
configurations that include $u/d$ (with equal mass) and $s$ quarks in 
the sea ($n_f=2+1$) using the asqtad action.  By taking a ratio of three-point correlation 
functions they are able simultaneously able to improve their statistical 
accuracy and reduce part of the systematic uncertainty from the normalization 
of their current operator.   
Their result is $h_{A_1}(1)=0.906(4)(12)$ where the uncertainties are 
statistical and systematic respectively. Their systematic error is dominated 
by discretization effects. They take the systematic uncertainty from 
missing higher-order terms in the perturbative current matching~\cite{Monahan:2012dq} to 
be $0.1\alpha_s^2$. 

The HPQCD collaboration have calculated $h_{A_1}(1)$ on gluon field 
configurations that include $n_f=2+1+1$ HISQ sea quarks using NRQCD 
$b$ quarks and HISQ $c$ quarks \cite{Harrison:2017fmw}. Their result, $h_{A_1}(1) = 0.895(10)(24)$ 
has a larger uncertainty, dominated by the systematic uncertainty of  
$0.5\alpha_s^2$ allowed for in the current matching. 
They were also able to calculate the equivalent result for 
$B_s \rightarrow D_s^*$, obtaining $h^s_{A_1}(1) = 0.879(12)(26)$ and 
demonstrating that the dependence on light quark mass is small. 
The $B_s \rightarrow D_s^*$ provides a better lattice QCD comparison 
point than $B \rightarrow D^*$ because it has less sensitivity to 
light quark masses (in particular the $D^*D\pi$ ``cusp") and to the volume.  
\begin{figure}
\centerline{\includegraphics[width=0.8\textwidth]{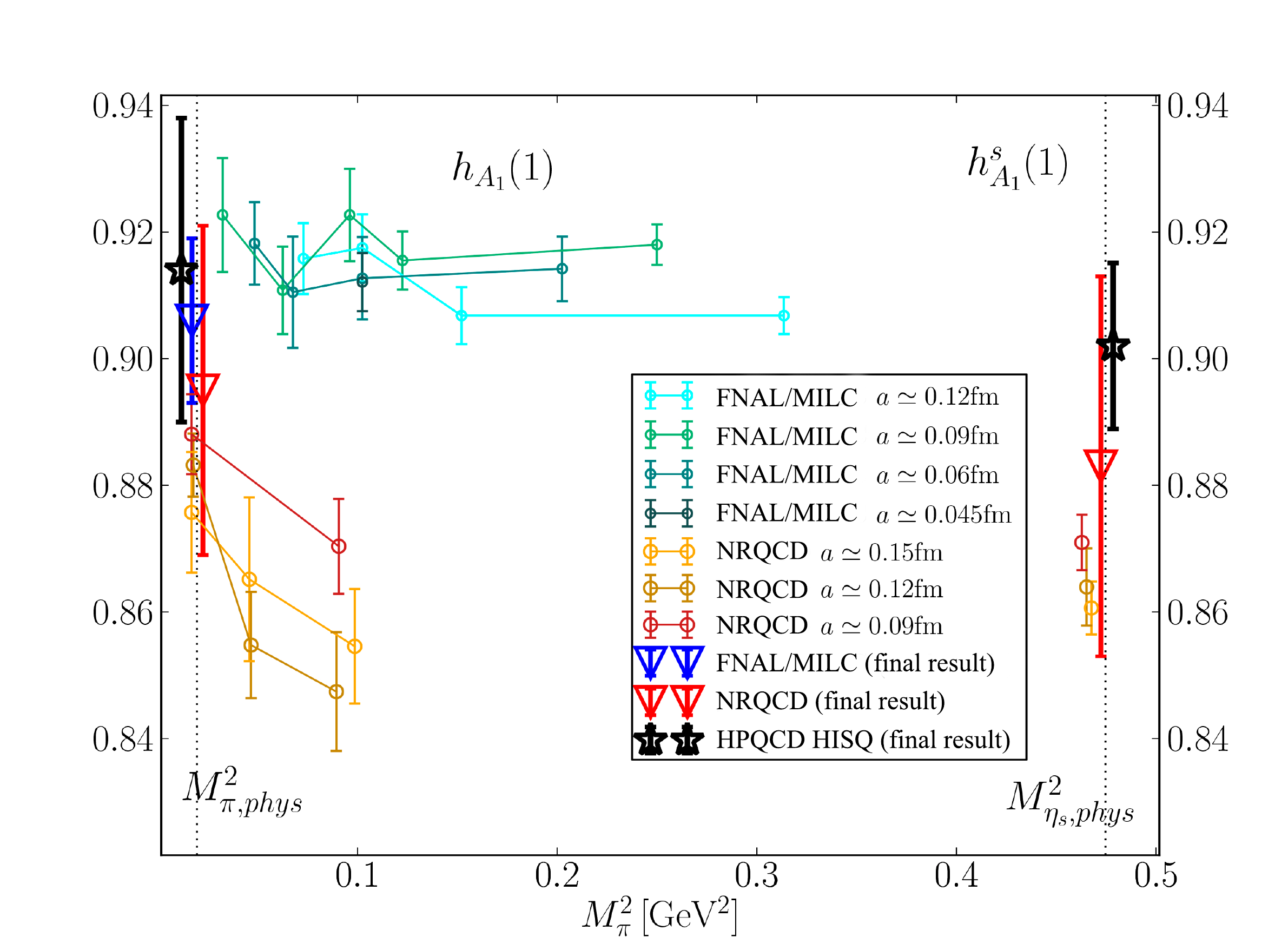}}
\caption{ Plot taken from Ref.~\cite{McLean:2019sds} showing
the comparison of lattice QCD results for 
$h_{A_1}(1)$ (left side) and $h^s_{A_1}(1)$ (right side). 
Raw results for $h_{A_1}(1)$ are
from~\cite{Harrison:2017fmw} and~\cite{Bailey:2014tva} 
and are plotted as a function of valence 
(=sea) light quark mass, given by the square of $M_{\pi}$. 
On the right are points for $h^s_{A_1}(1)$ from~\cite{Harrison:2017fmw} 
plotted at the appropriate valence mass for the $s$ quark, but 
obtained at physical sea light quark masses.  The final result for 
$h_{A_1}(1)$ from~\cite{Bailey:2014tva}, with its full 
error bar, 
is given by the inverted blue triangle. The inverted red triangles 
give the final results for $h_{A_1}(1)$ and $h^s_{A_1}(1)$ 
from~\cite{Harrison:2017fmw}. The HPQCD results of \cite{McLean:2019sds} are given by the black stars.}
\label{fig:hA1-comparison}
\end{figure}
More recently the HPQCD collaboration have used the HISQ action 
for all quarks, with a fully nonperturbative current normalization, 
 to determine $h^s_{A_1}(1)$ \cite{McLean:2019sds}. Their result, 
$h^s_{A_1}(1) = 0.9020(96)(90)$ agrees well 
with the earlier results and has smaller systematic uncertainties.  
Figure~\ref{fig:hA1-comparison} compares the three results. 

The importance of being able to compare lattice QCD and experiment away 
from the zero recoil point is now clear and several lattice QCD calculations 
are underway, attempting to cover the full $q^2$ range of the decay and 
all 4 form factors.
This includes calculations for $B \rightarrow D^*$ from JLQCD~\cite{Kaneko:2019vkx} with M\"{o}bius domain-wall quarks, Fermilab/MILC~\cite{Vaquero:2019ary}
(see also talk at Lattice 2019) with improved Wilson/Fermilab quarks and LANL/SWME with an improved version of this formalism known as the Oktay-Kronfeld action~\cite{Bhattacharya:2020xyb}. Calculations for other $b\to c$ pseudoscalar-to-vector form factors, $B_s \rightarrow D^*_s$~\cite{McLean:2019jll} and $B_c \rightarrow (J/\psi,\eta_c)$ are also underway from
HPQCD~\cite{Lytle:2016ixw,Colquhoun:2016osw} using the all-HISQ approach. At the same time further $B \rightarrow D$ and $B_s \rightarrow D_s$ form factor calculations are in progress, including those using a variant of the Fermilab approach known as Relativistic Heavy Quarks on RBC/UKQCD configurations~\cite{Flynn:2019jbg}. 
In future we should be able to compare results from multiple actions with experiment 
for improved accuracy in determining~$|V_{cb}|$.

\subsubsection{$\Lambda_b \to \Lambda_c^{(*)}$ form factors from lattice QCD}
\label{sec:LbLcLattice}

The $\Lambda_b \to \Lambda_c$ form factors have been calculated with $2+1$ dynamical quark flavors; the vector
and axial vector form factors can be found in Ref.~\cite{Detmold:2015aaa}, while the tensor form factors (which
contribute to the decay rates in many new-physics scenarios) where added in Ref.~\cite{Datta:2017aue}. This calculation used two different lattice spacings of approximately 0.11~fm and 0.08~fm,
sea quark masses corresponding to pion masses in the range from 360 down to 300~MeV, and valence quark masses corresponding to pion masses in the range
from 360 down to 230~MeV. The lattice data for the form factors, which
cover the kinematic range from near $q^2_{\rm max}\approx 11\:{\rm GeV}^2$ down to $q^2\approx 7\:{\rm GeV}^2$,
were fitted with a modified version of the BCL $z$ expansion~\cite{Bourrely:2008za} discussed in Sec.~\ref{sec:param}, where simultaneously to the expansion in $z$, an expansion in powers of the lattice spacing and quark masses is performed.
No dispersive bounds were used in the $z$ expansion here (this is something that can perhaps be improved in the future, see also Sec.~\ref{sec:z-remarks}).
The form factors extrapolated to the continuum limit and physical pion mass yield the following Standard Model predictions:
\begin{equation}
 \frac{1}{|{V_{cb}}|^2}\Gamma (\Lambda_b \to \Lambda_c\: \mu^- \bar{\nu}_\mu)
= (21.5 \:\pm\: 0.8_{\,\rm stat} \:\pm\: 1.1_{\,\rm syst})\:\:{\rm ps}^{-1}
\end{equation}
for the fully integrated decay rate, which has a total uncertainty of 6.3\% (corresponding to a 3.2\% theory uncertainty in a possible $|V_{cb}|$ determination from this decay rate),
\begin{equation}
\frac{1}{|{V_{cb}}|^2}\int_{7\:{\rm GeV}^2}^{q^2_{\rm max}}
\frac{\mathrm{d}\Gamma (\Lambda_b \to \Lambda_c\: \mu^- \bar{\nu}_\mu)}{\mathrm{d}q^2} \mathrm{d} q^2
= (8.37 \:\pm\: 0.16_{\,\rm stat} \:\pm\: 0.34_{\,\rm syst})\:\:{\rm ps}^{-1} \label{eq:LbLcPartialRate}
\end{equation}
for the partially integrated decay rate, which has a total uncertainty of 4.5\% (corresponding to 2.3\% for $|V_{cb}|$), and
\begin{equation}
R(\Lambda_c)=\frac{\Gamma (\Lambda_b \to \Lambda_c\: {\tau^- \bar{\nu}_\tau})}{\Gamma (\Lambda_b \to \Lambda_c\: {\mu^- \bar{\nu}_\mu})} \:=\: 0.3328 \:\pm \:0.0074_{\,\rm stat} \:\pm\:  0.0070_{\,\rm syst}
\end{equation}
for the lepton-flavor-universality ratio, which has a total uncertainty of 3.1\%. The systematic uncertainties of the vector and axial vector form factors are dominated
by finite-volume effects and the chiral extrapolation. Both of these can be reduced substantially in the future by adding a new lattice gauge field ensemble
with physical light-quark masses and a large volume, and dropping the ``partially quenched'' data sets that have $m_\pi^{(\mathrm{val})}<m_\pi^{(\mathrm{sea})}$.
Adding another ensemble at a third, finer lattice spacing will also be beneficial to better control the continuum extrapolation. 

At this workshop, there was some discussion about the validity of the modified $z$ expansion; it has been argued that it would be safer to first
perform chiral/continuum extrapolations and then perform a secondary $z$ expansion fit. This is expected to make a difference mainly if nonanalytic
quark-mass dependence from chiral perturbation theory is included. However, the fits used in Ref.~\cite{Detmold:2015aaa} for the $\Lambda_b$ form factors were analytic in the lattice spacing
and light-quark mass. Note that the shape of the $\Lambda_b \to \Lambda_c\: \mu^- \bar{\nu}_\mu$ differential decay rate was later measured by LHCb, and found to
be in good agreement with the lattice QCD prediction all the way down to $q^2=0$~\cite{Aaij:2017svr}.

Motivated by the prospect of an LHCb measurement of $R(\Lambda_c^*)$, work is now also underway to compute the $\Lambda_b \to \Lambda_c^{*}$ form factors in lattice QCD, for the $\Lambda_c^*(2595)$ and $\Lambda_c^*(2625)$, which have
$J^P=\frac12^-$ and $J^P=\frac32^-$, respectively. Preliminary results were shown at the workshop. For these form factors, the challenge is that, to project the $\Lambda_c^*$ interpolating field
exactly to negative parity and avoid contamination from the lower-mass positive parity states, one needs to perform the lattice calculation in the $\Lambda_c^*$ rest frame. With the $b$-quark action currently in use,
discretization errors growing with the $\Lambda_b$ momentum then limit the accessible kinematic range to a small region near $q^2_{\rm max}$. To predict $R(\Lambda_c^*)$,
it will be necessary to combine the lattice QCD results for the form factors in the high-$q^2$ region with heavy-quark effective theory and LHCb data for the shapes of
the $\Lambda_b \to \Lambda_c^*\, \mu^-\bar{\nu}_\mu$ differential decay rates~\cite{Boer:2018vpx}.

\subsection{Measurements of $B \to D^{(*)} \ell \nu$ and related processes \label{ssec::exp}}

\subsubsection{Measurements with light leptons}

The decays $B\to D^*\ell\nu$ and $B\to D\ell\nu$ have been measured at Belle and BaBar as well as at older experiments (CLEO, LEP). Unfortunately, most of these measurements assume the Caprini-Lellouch-Neubert parametrization of the form factors (see Sec.~\ref{sec:param}) and report results in terms of $|V_{cb}|$ times the only form factors relevant at the zero-recoil point $w=1$, namely $\mathcal{F}(1)\equiv h_{A_1}(1)$ for $B\to D^*\ell\nu$ and $\mathcal{G}(1)\equiv 2\sqrt{M_D M_B}/(M_D + M_B))f_+(1)$ for $B\to D\ell\nu$, 
 and of the other CLN parameters, instead of a general form of the $z$~expansion or the raw spectra. The Heavy Flavor Averaging Group (HFLAV) has performed an average of these CLN measurements~\cite{Amhis:2019ckw} and reports
\begin{eqnarray}
  \eta_\mathrm{EW}\mathcal{F}(1)|V_{cb}| & = & (35.27\pm 0.11(\rm stat)\pm 0.36(\rm syst))\times 10^{-3}~, 
  \label{eq:b2dstar}\\
  \eta_\mathrm{EW}\mathcal{G}(1)|V_{cb}| & = & (42.00\pm 0.45(\rm stat)\pm 0.89(\rm syst))\times 10^{-3}~. \label{eq:b2d}
\end{eqnarray}
Notice that Eq.~(\ref{eq:b2dstar}) together with $h_{A_1}(1)=0.904(12)$ \cite{Aoki:2019cca}
leads to the low  value $|V_{cb}|=38.76(69) 10^{-3}$.
Eq.~(\ref{eq:b2d}) together with $\mathcal{G}(1)=1.0541(83)$ \cite{Lattice:2015rga} leads to a consistent result  $|V_{cb}|=39.58(99) 10^{-3}$.
In the case of $B\to D\ell\nu$ one can also use 
the existing lattice calculations at non-zero recoil \cite{Lattice:2015rga,Na:2015kha} to guide the extrapolation
to zero recoil, together with the $w$ spectrum measured by Belle \cite{Glattauer:2015teq}. In the BGL parametrization, this leads to
a higher value, $|V_{cb}|=40.83(1.13) 10^{-3}$, a more reliable determination than (\ref{eq:b2d}).
In the following we will have a closer look at the most recent measurements by the various experiments.

{\bf Belle} has recently updated the untagged measurement of the $B^0\to D^{*-}\ell^+\nu$ mode~\cite{Waheed:2018djm}. While the new analysis is based on the same 711~fb$^{-1}$ Belle data set, the re-analysis takes advantage of a major improvement of the track reconstruction software, which was implemented in 2011, leading to a substantially higher slow pion tracking efficiency and hence to much larger signal yields than in the previous publication~\cite{Dungel:2010uk}. Again $D^{*+}$~mesons are reconstructed in the cleanest mode, $D^{*+}\to D^0\pi^+$ followed by $D^0\to K^-\pi^+$, combined with a charged, light lepton (electron or muon) and yields are extracted in 10 bins for each of the 4 kinematic variables describing the $B^0\to D^{*-}\ell^+\nu$~decay. These yields are published along with their full error matrix. The updated publication also contains an analysis of these yields using both the CLN and the BGL form factors (where BGL has only 5 free parameters).
 The CLN analysis results in $\eta_\mathrm{EW}\mathcal{F}(1)|V_{cb}|=(35.06\pm 0.15(\rm stat)\pm 0.56(\rm syst))\times 10^{-3}$, while the BGL fit gives $\eta_\mathrm{EW}\mathcal{F}(1)|V_{cb}|=(34.93\pm 0.23(\rm stat)\pm 0.59(\rm syst))\times 10^{-3}$. Both results are thus well consistent. This contrasts with a tagged measurement of $B^0\to D^{*-}\ell^+\nu$ first shown by Belle in November 2016~\cite{Abdesselam:2017kjf}. Analyzing the raw data of this measurement in terms of the CLN and BGL form-factors gives a difference of almost two standard deviations in $|V_{cb}|$~\cite{Bigi:2017njr,Grinstein:2017nlq}. However, this result has remained preliminary and will not be published. A new tagged analysis, using an improved version of the hadronic tag is now underway and should clarify the experimental situation.
 
{\bf Babar}
 has presented a full four-dimensional angular analysis  of $B^0\to D^{*0}\ell^-\nu_\ell$ decays, using both CLN and BGL parametrizations \cite{Dey:2019bgc}. 
This analysis is based on the full data set of 450~fb$^{-1}$, and exploits the hadronic $B$-tagging approach.
The full decay chain $e^+e^-\to \Upsilon(4S)\to B_{\rm tag} B_{\rm sig}(\to D^*\ell\nu_\ell)$ is considered in a kinematic fit that includes constraints on the beam properties, the secondary vertices, the masses of $B_{\rm tag}$, $B_{\rm sig}$, $D^*$ and the missing neutrino. After applying requirements on the probability of the $\chi^2$ of this constrained fit, which is the main discriminating variable, the remaining background is only about $2\%$ of the sample. The resolution on the kinematic variables is about a factor five better than the one possible with untagged measurements. The shape of the form factors is extracted using an unbinned maximum likelihood fit where the signal events are described by the four dimensional differential decay rate. 
The extraction of $|V_{cb}|$ is performed indirectly by adding to the likelihood the constraint that the integrated rate $\Gamma=\mathcal{B}/\tau_B$, where $\mathcal{B}$ is the $B\to D^*\ell\nu$ branching fraction and $\tau_B$ is the $B$-meson lifetime. The values of these external inputs are taken from HFLAV \cite{Amhis:2019ckw}. 
The final result, using $h_{A_1}(1)$ from \cite{Bailey:2014tva}, 
is $|V_{cb}|=(38.36\pm 0.90)\times 10^{-3}$ with a 5-parameter BGL version and 
$|V_{cb}|=(38.40\pm0.84)\times 10^{-3}$ in the CLN case, both compatible with the above 
HFLAV average. Nevertheless, the individual form factors show significant deviations 
from the world average CLN determination by HFLAV.   

{\bf LHCb} has extracted $V_{cb}$ from semileptonic $B_s^0$ decays for the first 
time~\cite{Aaij:2020hsi}. The measurement uses both $B_s^0\rightarrow 
D_s^{-}\mu^+\nu_{\mu}$ and $B_s^0\rightarrow D_s^{*-}\mu^+\nu_{\mu}$ decays using 
$3$~fb$^{-1}$ collected in 2011 and 2012. 
The value of $|V_{cb}|$ is determined from the observed yields of $B_s^0$ decays 
normalized to those of $B^0$ decays after correcting for the relative reconstruction 
and selection efficiencies. The normalization channels are $B^0\to D^-\mu^+\nu_{\mu}$
and $B^0\to D^{*-}\mu^+\nu_{\mu}$ with the $D^-$ reconstructed with the same decay 
mode as the $D_s$, ($D_{(s)}^-\to [K^+K^-]_{\phi}\pi^-$), to minimize the systematic 
uncertainties. 
The shapes of the form factors are extracted as well, exploiting the kinematic 
variable $p_{\perp}(D_s)$ which is the component of the $D_s^-$ momentum 
perpendicular to the $B_s^0$ flight direction. This variable is correlated with  $q^2$. In this analysis both the CLN parametrization and a 5-parameter version of BGL  have been 
used. The results for $V_{cb}$ are
\begin{eqnarray}
|V_{cb}|_{CLN}&=&(41.4\pm0.6(\rm stat)\pm 0.9(\rm syst)\pm 1.2(\rm ext))\times 10^{-3}\nonumber\\ \nonumber
|V_{cb}|_{BGL}&=&(42.3\pm0.8(\rm stat)\pm 0.9(\rm syst)\pm 1.2(\rm ext))\times 10^{-3},
\end{eqnarray}
where the first uncertainty are statistical, the second systematic and the third due to the limited knowledge of the external input, in particular the $B_s^0$ to $B^0$ production ratio $f_s/f_d$ which is known with an uncertainty of about $5\%$. The results are compatible with both the inclusive and exclusive decays. Although not competitive with the results obtained at the $B$~factories, the novel approach used can be extended to the semileptonic $B^0$ decays. 

\subsubsection{Past measurements of $R(D)$ and $R(D^*)$}
$R_{D}$ and $R_{D^*}$ are defined as  the ratios of the semileptonic  decay width of $B_d$ and $B_u$ meson to a $\tau$ lepton and its associated neutrino $\nu_\tau$ over the  $B$ decay width to a light lepton. 
A summary of the currently available measurements of $R_{D}$ and $R_{D^*}$  is presented in Table~\ref{tab_H2H_RDexp}, showing the yield of $B$ signal and $B$ normalization decays and the stated uncertainties.  The data were collected by the BaBar and Belle experiments at $e^+e^-$ colliders operating at the $\Upsilon(4S)$ resonance, which decays exclusively to pairs of  $B^+B^-$ or $B^0\bar B^0$  mesons.  The LHCb experiment operates at the high energy $pp$ collider at CERN at total energies of 7 and 8 TeV, where pairs of $b$-hadrons (mesons or baryons) along with a large number of other charged and neutral particles are produced. While the maximum production rate of the $\Upsilon(4S)\to B\bar B$ events has been 20 Hz, the rates observed at LHCb exceed 100kHz.
\begin{table}[t]
\centering
\begin{tabular}{|lll|cc|c|}\hline
Experiment  & tag & $\tau$ decay & N(D$\tau\nu_{\tau}$) & N$_{norm}$ & R(D) \\ \hline 
Babar \cite{Lees:2012xj,Lees:2013uzd}      & Had.& $\ell\nu_\ell\nu_\tau$ & $489 \pm 63$ & $2891 \pm 65$ & $0.440 \pm 0.058 \pm 0.042$  \\
Belle \cite{Huschle:2015rga}      & Had.& $\ell\nu_\ell\nu_\tau$ & $320 \pm 55$ & $3147 \pm 72$ & $0.375 \pm 0.064 \pm 0.026$  \\ 
Belle \cite{Abdesselam:2019dgh}      & SL  & $\ell\nu_\ell\nu_\tau$ & $1778 \pm 204$& $22896 \pm 471$ & $0.307 \pm 0.037 \pm 0.016$ \\ \hline
\multicolumn{5}{|c|}{HFLAV} & $0.340 \pm 0.027 \pm 0.013$  \\
\multicolumn{5}{|c|}{Theory} & $0.299 \pm 0.003$  \\ \hline
Experiment  & tag & $\tau$ decay & N(D$^* \tau\nu_{\tau}$) & N$_{norm}$ & R(D$^*$) \\ \hline 
Babar \cite{Lees:2012xj,Lees:2013uzd}      & Had.& $\ell\nu_\ell\nu_\tau$ & $888 \pm 63$ & $11953 \pm 122$ & $0.332 \pm 0.024 \pm 0.018$  \\
Belle \cite{Huschle:2015rga}      & Had.& $\ell\nu_\ell\nu_\tau$ & $503 \pm 65$ & $3797 \pm 74$ & $0.293 \pm 0.038 \pm 0.015$  \\ 
Belle \cite{Hirose:2016wfn,Hirose:2017dxl}      & Had.& $\pi\nu_\tau, \rho\nu_\tau$ & $298 \pm 29$ & $7213 \pm 96$ & $0.270 \pm 0.035 \pm 0.028$  \\ 
LHCb \cite{Aaij:2015yra}       & -   & $\mu\nu_\mu\nu_\tau$ & 16480 & 363000 & $0.336 \pm 0.027 \pm 0.030$ \\
LHCb \cite{Aaij:2017uff,Aaij:2017deq}       & -   & $\pi\pi\pi\nu_\tau$ & 1273 & 17660  & $0.280 \pm 0.018 \pm 0.029$ \\
Belle \cite{Abdesselam:2019dgh}      & SL  & $\ell\nu_\ell\nu_\tau$ & $651 \pm 46$& $16942 \pm 148$& $0.283 \pm 0.018 \pm 0.014$ \\ \hline
\multicolumn{5}{|c|}{HFLAV} & $0.295 \pm 0.011 \pm 0.008$  \\
\multicolumn{5}{|c|}{Theory} & $0.253 \pm 0.005$  \\ \hline
\end{tabular}
\caption{Summary of $R_D$ and $R_{D^*}$ measurements and theoretical predictions. The number of observed signal and normalization events is also reported. The normalization channel is B$\to$D$^{(*)}\ell\nu_{\ell}$ for all measurements but the LHCb one with three-prong $\tau$ decays, where the normalization channel is B$\to$ D$^*\pi\pi\pi$. The latter LHCb measurement has been updated using
the latest HFLAV average for ${\cal{B}}(B\to D^*\ell\nu_\ell)$. The quoted theory predictions are arithmetic averages of the values reported in Table \ref{tab:RD-RDstar} below; they are given for illustration only and do not imply consent from the authors of the calculations. }
\label{tab_H2H_RDexp}
\end{table}

Currently we have only two measurements~\cite{Lees:2012xj,Lees:2013uzd,Huschle:2015rga} of the ratios $R_{D}$ and $R_{D^*}$ based on two distinct samples of hadronic tagged $B\bar B$ events with signal $B\to D\tau\nu_\tau$ and $B\to D^*\tau\nu_\tau$ decays and purely leptonic tau decays, $\tau^- \to e^-\bar\nu_e\nu_\tau$ or $\tau^- \to \mu^-\bar{\nu}_\mu\nu_\tau$. In addition, there is a measurement from Belle~\cite{Hirose:2016wfn,Hirose:2017dxl} of $R_{D^*}$ with hadronic tags and a semileptonic one-prong $\tau$ decay ($\tau^-\to\pi^-\nu_\tau$ or $\tau^-\to\rho^-\nu_\tau$).  
A Belle measurement~\cite{Abdesselam:2019dgh} of $R_{D}$ and $R_{D^*}$ with semi-leptonic tags and purely leptonic $\tau$ decays appeared recently, superceding a 
previous measurement~\cite{Sato:2016svk} of $R_{D^*}$ obtained with the same technique. 

At LHCb only decays of neutral $B$ mesons producing a charged $D^*$ meson and a muon of opposite charge are selected, with a single decay chain $D^{*+}\to D^0 (\to K^-\pi^+) \pi^+$.
LHCb published two measurements~\cite{Aaij:2015yra,Aaij:2017uff,Aaij:2017deq} of $R_{D^*}$, the first relying on purely leptonic $\tau$ decays and normalized to the $B^0 \to D^{*+}\mu^-\bar \nu_\mu$ decay rate, and the more recent one using 3-prong semileptonic $\tau$ decays, $\tau^-\to\pi^+\pi^-\pi^- \nu_\tau$ and normalization to the decay $B^0 \to D^{*+}\pi^+\pi^-\pi^-$.
This LHCb measurement extracts directly the ratio of branching fractions ${\cal K}(D^*)={\cal B}(B^0 \to D^{*+}\tau^-\bar \nu_\tau)/{\cal B}(B^0 \to D^{*+}\pi^+\pi^-\pi^-)$. The ratio ${\cal K}(D^*)$ is then converted to $R(D^*)$ by using the known branching fractions of $B^0 \to D^{*+}\pi^+\pi^-\pi^-$ and $B^0 \to D^{*+}\mu^-\bar \nu_\mu$. 

BaBar and Belle analyses rely on the large detector acceptance to detect and reconstruct all final state particles from the decays of the two B mesons, except for the neutrinos. They exploit the kinematics of the two-body $\Upsilon(4S)$ decay and known quantum numbers to suppress non-$B\bar B$ and combinatorial
 backgrounds.  They differentiate the signal decays involving two or three missing neutrinos from decays involving a low mass charged lepton, an electron or muon, plus an associated neutrino.

LHCb isolates the signal decays from very large backgrounds by exploiting the relatively long $B$ decay lengths which allows for a separation of the charged particles from the $B$ and charm decay vertex from many others originating from the pp collision point.  There are insufficient kinematic constraints and  therefore the total $B$ meson momentum is estimated from its transverse momentum, degrading the resolution of kinematic quantities like the missing mass and the momentum transfer squared $q^2$.   Also, the production of $D^{*+} D_s^-$ pairs with the decay $D_s^-\to \tau^-\bar \nu_\tau$ leads to sizable background in the signal sample.

The summary in Table~\ref{tab_H2H_RDexp} indicates that the results are not inconsistent. For BaBar and Belle the systematic uncertainties are comparable for  $R_{D^*}$, 
while Belle systematic uncertainties are smaller for $R_{D}$.
However the differences in the signal yield and the background suppression lead to smaller statistical errors for BaBar.  The Belle measurements based on semileptonic tagged samples result in a 50\% smaller signal yield than for the hadronic tag samples. 
For the two LHCb measurements, the event yields exceed the BaBar yields by close to a factor of 20, but the relative statistical errors on $R_{D^*}$ are comparable to BaBar, and the systematic uncertainties are larger by a factor of 2. 

\subsubsection{Lessons learned} 
All currently available measurements are limited by the difficulty of separating the signal from large backgrounds from many sources, leading to sizable statistical and systematic uncertainties.  The measurement of ratios of two $B$ decay rates with the very similar - if not identical – final state particles,  significantly reduces the systematic uncertainties due to detector effects, tagging efficiencies,  and also from uncertainties in the kinematics due to form factors and branching fractions.  For all three experiments the largest systematic uncertainties are attributed to the limited size of the MC samples, the fraction and shapes of various backgrounds, especially from decays involving higher mass charm states, and uncertainties in the relative efficiency of signal and normalization, the efficiency of other backgrounds, as well as lepton mis-identification.  Though the total number of $B\bar B$ events of the full Belle data set exceeds the one for BaBar by 65\%, the signal BaBar signal yield for $B \to D^{(*)} \tau\nu_\tau$ exceeds Belle by 67\% due to differences in event selection and fit procedures.

While the use by Belle of semileptonic B decays as tags for $B\bar B$ events benefits from the fewer decay modes with higher BFs, the presence of a neutrino in the tag decays results in the loss of stringent kinematic constraints.   The resulting signal yields are lower by 50\% compared to hadronic tags, and the backgrounds are much larger. The use of  the ECL, namely the sum of the energies of the excess photons in a tagged event, in the fit to extract the signal yield is somewhat problematic, since it includes not only the photons left over from incorrectly reconstructed $B \bar B$ events, but also photons emitted from the high intensity beams. As a result the signal contributions are difficult to separate from the very sizable backgrounds.

\subsubsection{ Outlook for $R(D)$ and $R(D^*)$}
Belle II and the upgraded LHCb are expected to collect large data samples with considerably improved detector performances.  This should lead to much reduced detector related uncertainties, higher signal fractions, and opportunities to measure many  related processes.  The goal is to push the sensitivity of many measurements of critical variables and distributions beyond theory uncertainties and thereby increase the  sensitivity to non-Standard Model processes.

Currently there are only two measurements of the ratio $R_{D}$, one each by BaBar and Belle, based on two distinct samples of hadronic tagged $B \bar B$ events for the signal $B\to D\tau\nu_\tau$ and $B\to D^* \tau\nu_\tau$ decays.  The decay $B\to D\tau\nu_\tau$ is dominated by a P-wave, whereas in the $B\to D^* \tau\nu_\tau$ S, P, and D waves contribute and the impact for contributions from new physics  processes is expected to be smaller. A contribution of a hypothetical charged Higgs  would result in an S-wave for $B\to D\tau\nu_\tau$, and a P-wave for $B\to D^*\tau\nu_\tau$, thus measurements of the angular distributions and the polarization of the $\tau$ lepton or $D$ and $D^*$ mesons will be important.  Such measurements would of course also serve as tests of other hypotheses, for instance contributions from leptoquarks.
The studies for many decay modes, the detailed kinematics of the signal events, the four-momentum transfer $q^2$, the lepton momentum, the angles and momenta of $D$ and $D^*$ and the $\tau$ spin should be extended to perform tests for potential new physics contributions.

Belle II will benefit from major upgrades to all detector components, except for the barrel  sections of the calorimeter and the muon detector.  In addition, a new data acquisition and analysis software are being developed to benefit from the very high data rates and improved detector performance.
Upgrades to the precision tracking and lepton identification, especially at lower momenta, are expected to significantly improve the mass resolution and purity of the signal samples.  This should also improve the detector modeling of efficiencies for signal and backgrounds and fake rates that are the major contributions to the current systematic uncertainties. The much larger data rates should allow choice of cleaner and more efficient $B \bar B$ tagging algorithms. 

Major improvements to the MC simulation signal and backgrounds will be needed. They require much better understanding of all semileptonic $B$ decays, contributing to signal and backgrounds, i.e., updated measurements of branching fractions and form factors and theoretical predictions, especially for backgrounds involving higher mass charm mesons, either resonances or states resulting from charm quark fragmentation. The fit to extract the signal yields could be improved by reducing the backgrounds and making use of fully 2D or 3D distributions of kinematic variables, and by avoiding simplistic  parametrizations.  
The suppression of fake photons and $\pi^0$s needs to be scrutinized to avoid unnecessary signal loss and very large backgrounds for $D^{*0}$ decays.  Shapes of distributions entering multi-variable methods to reduce the backgrounds should be scrutinized by comparisons with data or MC control samples, and any significant differences should be addressed. The use of ECL, the sum of the energies of all unassigned photon in an event, may be questionable, given the expected high rate of beam generated background.

The first study by Belle of the $\tau$ spin in $B\to D^*\tau\nu_\tau$ decays with $\tau^-\to\rho^-\nu_\tau$ or $\tau^-\to\pi^-\nu_\tau$  is very promising, it indicates that much larger and cleaner data samples will be needed. The systematic uncertainty on the $R_{D^*}$ measurement of 11\% is dominated by the hadronic $B$ decay composition of 7\%  and the size of the MC sample \cite{Hirose:2017dxl}.  The measured transverse $\tau$ polarization of $P_\tau = -0.38\pm 0.51^{+0.21}_{-0.16}$ is totally statistics dominated, and implies   $P_\tau < 0.5$ at 90\% C.L.

Among the many other measurements Belle II is planning, ratios $R$ for both inclusive and inclusive  semileptonic $B$ decays are of interest, for instance
in addition to  $R_D$, $R_{D^{*}}$, and $R_{D^{**}}$ also $R_{X_c}$, as well as $R_\pi$ and $R_{X_u}$,  which rely on unique capabilities of Belle II.

The LHCb detector is currently 
undergoing a major upgrade  with the goal to switch to an all software trigger and to be able to select and record data up to rates of 100kHz.  Replacements of all tracking devices are planned, ranging from radiation hard pixel detector near interaction region to scintillation fibers downstream. Improvements to electron and muon detection and reduction in pion misidentification will be critical for the suppression of backgrounds, and should also allow rate comparison for decays involving electron or muons.   LHCb relies on large data samples rather than MC simulation to assess signal efficiencies and most importantly the many sources of backgrounds and their suppression. 

Several analyses are underway based on Run 1 and Run 2 data samples, and are benefiting from improved trigger capabilities.  The first analysis based on 3-prong $\tau$ decays showed a clear separation of the $\tau$ decay vertex from both the $D$ and the proton interaction point, improving the signal purity to about 11\%, compared to 4.4\% for the purely leptonic 1-prong $\tau$ decay.  This may therefore be the favored $\tau$ decay mode, and should also be tried for $B^+\to D^0\tau^+\nu_\tau$. Improved measurements of the branching fractions for normalization and the $\tau$ decays will be essential.

As a follow-up on the first LHCb measurement of $R_{D^*}$, a simultaneous fit to two disjoint $D^0 \mu^-$ and $D^{*+}\mu^-$  samples is in preparation, taking into account the large feed-down from $D^*$ decay present in the $D^0 \mu^-$ sample.  As pointed out above, the decay $B^+\to D^0\tau^+\nu_\tau$ is more sensitive to new physics processes than $B^0\to D^{*-}\tau^+\nu_\tau$ and thus this analysis is expected to be very important to establish the excess in these decay modes and its interpretation. This analysis will benefit from the addition of dedicated triggers sensitive to $D^0 \mu^-$, $D^{*+}\mu^-$, $\Lambda_c^+ \mu$ and $D_s^{+}\mu$ final states.

LHCb is considering a series of other ratios measurements, among several $b \to c$ transitions ($\bar B_s^0 \to D_s^- \tau^+\nu_\tau$, $B\to D^{**} \tau^+\nu_\tau$ and $\Lambda_b^+\to \Lambda_c^{(*)} \tau^+\nu_\tau$) and certain $b\to u$ transitions ($B^+\to\rho^0\tau^+\nu_\tau$, $B^+\to p\bar p \tau^+\nu_\tau$ and $\Lambda_b^0\to p\tau^-\nu_\tau$), most of which will be challenging to observe and not trivial to normalize. The decay $\Lambda_b^+\to \Lambda_c^* \tau^+\nu_\tau$ probes a different spin structure, and a precise measurement of $R_{\Lambda_c}$ would be of great interest  for the interpretation of the excess of events in $R_{D}$ .
The observation of the decay $B_c^-\to J/\psi (\to\mu^+\mu^-) \tau^- (\to \mu^-\bar\nu_\mu\nu_\tau) \bar\nu_\tau$ has recently been reported.  It is a very rare process which is only observable at LHCb. The final state of 3 muons is a unique signature, though impacted by sizable backgrounds from hadron misidentification. The measured ratio $R_{J/\psi} = 0.71\pm 0.17 \pm 0.18$  has large uncertainties, dominated systematically by the signal simulation since the form factors are unknown.

\subsection{Extraction of $V_{cb}$ and predictions for $R_{D^{(*)}}$}
\label{sec:Vcb-RD-RDs}

The values of $V_{cb}$ extracted from inclusive and exclusive decays have been in tension for a long time~\cite{Amhis:2016xyh}.
In order to extract $V_{cb}$ from $B\rightarrow D^{(*)}l\nu$ data we need information on the form factors, 
which is mostly provided by lattice QCD.
For the $B\rightarrow D$ form factors $f_{+,0}$ there are lattice results at $w\geq 1$~\cite{Lattice:2015rga,Na:2015kha,Aoki:2019cca}.
A fit to all the available experimental and lattice data of $B\rightarrow Dl\nu$ leads to  \cite{Bigi:2016mdz}
\begin{align}
V_{cb} \cdot 10^3   &= 40.49(97)\,,
\end{align}
with $\chi^2/\mathrm{dof} = 19.0/22$. Similar results have been obtained in
\cite{Aoki:2019cca}.
For $B\rightarrow D^*$ at the moment there is only information on one of the four form factors at zero-recoil, 
$A_1(w=1)$~\cite{Bailey:2014tva,Harrison:2017fmw}, 
however further developments look promising~\cite{Aviles-Casco:2017nge,Kaneko:2018mcr,Aviles-Casco:2019vin}.
At the other end of the $w$ or $q^2$ spectrum there are results available from 
LCSR~\cite{Faller:2008tr,Gubernari:2018wyi}. 
In view of the advanced experimental precision, a key question for the precise extraction of $V_{cb}$ and a 
robust prediction of $R(D^{(*)})\equiv \mathcal{B}(B\rightarrow D^{(*)} \tau\nu)/\mathcal{B}(B\rightarrow D^{(*)}l\nu)$ 
is how large the theoretical uncertainties are. 
For example, whenever relations such as (\ref{eq:ffexp}) are used, how large are HQET corrections beyond NLO, i.e.\ of 
$O\left(\alpha_s^2,\Lambda^2_{\mathrm{QCD}}/m_{c,b}^2, \alpha_s \Lambda_{\mathrm{QCD}}/m_{c,b}\right)$ and how 
accurate are the QCDSR results that are used at NLO? 
A guideline for an answer to these questions can be provided by studying the size of NLO corrections in the HQET expansion and by a  
comparison with corresponding available lattice results~\cite{Bigi:2017jbd}.
A definite answer, especially for the pseudoscalar form factor $P_1$, which is needed for the prediction of $R(D^*)$, 
will be given only by future lattice results~\cite{Aviles-Casco:2017nge,Kaneko:2018mcr,Aviles-Casco:2019vin}. 

In all experimental analyses prior to 2017, HQET relations have been employed in terms of a form of the CLN  
parametrization~\cite{Caprini:1997mu} where theoretical uncertainties noted in Ref.~\cite{Caprini:1997mu} were set to zero by fixing 
coefficients to definite numbers. 
Moreover, the slope and curvature of $R_{1,2}(w)$ depend on the same underlying theoretical quantities as $R_{1,2}(1)$, which makes 
the variation of the latter and fixing of the former inconsistent.
In future experimental analyses this has to be taken into account.

Recent preliminary Belle data~\cite{Abdesselam:2017kjf} allowed for a reappraisal of fits to $B\rightarrow D^*l\nu$ by several
groups~\cite{Bigi:2017njr,Bigi:2017jbd,Grinstein:2017nlq,Bernlochner:2017jka,Bernlochner:2017xyx,Jaiswal:2017rve,Harrison:2017fmw}. 
For the first time, Ref.~\cite{Abdesselam:2017kjf} reported deconvoluted $w$ and angular distributions which are independent of the
parametrization.
This allowed to test the possible influence of different parametrizations on the extracted value of $V_{cb}$.
Indeed, based on that data set the central values for $\vert V_{cb}\vert$ varied by up to 6\% between CLN and BGL 
fits~\cite{Bigi:2017njr,Grinstein:2017nlq,Jaiswal:2017rve}.  By floating some additional parameters of the less flexible CLN 
parametrization, the agreement between BGL and CLN could be restored~\cite{Bigi:2017njr,Bernlochner:2017xyx}.
Furthermore, in the literature one could observe  a correlation of smaller central values 
for $V_{cb}$ with stronger HQET+QCDSR input~\cite{Abdesselam:2017kjf,Bigi:2017njr,Bigi:2017jbd,Grinstein:2017nlq,Bernlochner:2017jka,%
Bernlochner:2017xyx,Jaiswal:2017rve,Harrison:2017fmw}.

Recently, on top of the tagged analysis Ref.~\cite{Abdesselam:2017kjf} a new untagged Belle analysis of $B\rightarrow D^*l\nu$ 
appeared~\cite{Abdesselam:2018nnh}. The new, more precise data brought the $|V_{cb}|$ central values of the CLN and BGL fits closer 
together. However, in order to obtain a reliable error, it is necessary to employ the BGL parametrization with a sufficient number of 
coefficients rather than the CLN parametrization. Including the new data,  Ref.~\cite{Gambino:2019sif} obtains
\begin{align}
V_{cb} \cdot 10^3    &= 39.6\left(^{+1.1}_{-1.0}\right)\,, \label{eq:Vcb}  
\end{align}
with a $\chi^2/\mathrm{dof}  = 80.1/72$.
The inclusion of LCSRs or strong unitarity constraints, where input from HQET is used in a conservative way, basically does not change
the fit result~\cite{Gambino:2019sif}. The $V_{cb}$ value in Eq.~(\ref{eq:Vcb}) differs by $1.9\sigma$ from the inclusive result.

\begin{table*}[t]
\centering
\begin{tabular}{ccc}\hline\hline
Ref. & $R(D)$  &  Exp. deviation \\\hline
\cite{Bigi:2016mdz}          & $0.299(3)$                   & $1.4\sigma$ \\ 
\cite{Bernlochner:2017jka}   & $0.299(3)$                   & $1.4\sigma$ \\
\cite{Jaiswal:2017rve}       & $0.302(3)$                   & $1.3\sigma$ \\
\cite{Bordone:2019vic}       & $0.297(3)$                   & $1.4\sigma$ \\\hline\hline 
\end{tabular}
\qquad
\begin{tabular}{|ccc|}\hline\hline
Ref. & $R(D^*)$  &  Exp. deviation \\\hline 
\cite{Bernlochner:2017jka}     & $0.257(3)$     & $2.7\sigma$ \\
\cite{Gambino:2019sif}  & $0.254\left(^{7}_{6}\right)$     & $2.7\sigma$ \\ 
\cite{Jaiswal:2020wer}         & $0.251\left(^{4}_{5}\right)$     & $3.1\sigma$ \\
\cite{Bordone:2019vic}         & $0.250(3)$     & $3.2\sigma$ \\\hline\hline
\end{tabular}
\caption{Recent theory predictions for $R(D^{(*)})$. 
The deviations are calculated from the HFLAV spring 2019 updates 
$R(D)^{\mathrm{exp}} = 0.340(27)(13)$~\cite{Amhis:2019ckw,Lees:2012xj,Lees:2013uzd,Huschle:2015rga,Abdesselam:2019dgh} and
$R(D^*)^{\mathrm{exp}} = 0.295(11)(8)$
\cite{Amhis:2019ckw,Lees:2012xj,Lees:2013uzd,Huschle:2015rga,Sato:2016svk,Aaij:2015yra,Hirose:2016wfn,Hirose:2017dxl,Aaij:2017uff,Aaij:2017deq,Abdesselam:2019dgh},
respectively. 
For older predictions see Refs.~\cite{Fajfer:2012vx,Celis:2012dk,Tanaka:2012nw}.
Table adapted and extended from Ref.~\cite{Schacht:2017vfd}.}
\label{tab:RD-RDstar}
\end{table*}

The shortcomings of the CLN parametrization have been addressed in several recent articles \cite{Bernlochner:2017jka,Jaiswal:2017rve,Jung:2018lfu,Bordone:2019vic,Bordone:2019guc}:
varying the coefficients of the HQE consistently allows for a simultaneous description of the available experimental and lattice data in $B\to D$, while the parametrization
dependence in the extraction of $V_{cb}$ from Ref.~\cite{Abdesselam:2017kjf} remains \cite{Bernlochner:2017jka}. Including additionally contributions at $O(1/m_c^2)$
and higher orders in the $z$ expansion, the extracted values for $V_{cb}$ using the BGL parametrization and the HQE become compatible \cite{Bordone:2019vic}.

For the above reasons, older HFLAV averages, which are based on the CLN parametrization, should not be employed in future analyses, with the 
exception of the total branching ratios, whose parametrization dependence is expected to be negligible. The two most recent experimental analyses of $\bar{B}_{(s)} \rightarrow D_{(s)}^*l^-\bar{\nu}_l$
\cite{Dey:2019bgc,Aaij:2020hsi} 
present results obtained in both CLN and a simplified version of the BGL parametrization.  They did not observe sizeable parametrization dependence, but  found very different values of $V_{cb}$. However, they did not provide data in a format that allows for 
independent reanalyses.

For the lepton flavor nonuniversality observables $R(D^{(*)})$
we list a few recent theoretical predictions in Table~\ref{tab:RD-RDstar}. 
Predictions for further lepton flavor non-universality observables of underlying $b\rightarrow cl\nu$ transitions can be found in Refs.~\cite{Bernlochner:2018bfn,Cohen:2019zev}.
Compared to predictions from before 2016, the predictions in Table~\ref{tab:RD-RDstar} make use of new lattice results and new experimental data.
The results are based on different methodologies and a different treatment of the uncertainties of HQET + QCDSR.
We have a very good consensus for $R(D)$ predictions because in this case the predictions are dominated by 
the recent comprehensive lattice results from Refs.~\cite{Lattice:2015rga,Na:2015kha,Aoki:2016frl}.
QED corrections to $R(D)$ remain a topic which deserves further study~\cite{Becirevic:2009fy,deBoer:2018ipi}. 
In the case of $R(D^*)$, as we do not have yet lattice information on the form factor $P_1$, we can use  the exact endpoint 
relation $P_1(w_{\mathrm{max}}) = A_5(w_{\mathrm{max}})$ and results from HQET and QCDSR.  
Depending on the estimate of the corresponding theory uncertainty one obtains different theoretical errors for the prediction of $R(D^*)$.
As soon as we have lattice results for $P_1$~\cite{Aviles-Casco:2017nge}, the different fits will stabilize and 
we expect a similar consensus as for $R(D)$. 
Despite the most recent experimental results being closer to the SM predictions,  
the $R(D^{(*)})$ anomaly persists and remains a tough challenge for model builders.

\subsection{Semileptonic $B\to D^{**}\ell\bar\nu$ decays \label{ssec::Ddoublestar}}

\def\lqcd{\Lambda_{\rm QCD}}
\newcommand{\dds}{D^{(*)}}
\newcommand{\dss}{D^{**}}
\newcommand{\dsW}{D^{1/2^+}}
\newcommand{\dsN}{D^{3/2^+}}
\newcommand{\dSs}{D^*_0}
\newcommand{\dVs}{D^*_1}
\newcommand{\dV}{D_1}
\newcommand{\dTs}{D^*_2}

Semileptonic $B$ decays to the four lightest excited charm mesons, $D^{**} =
\{D_0^*,\, D_1^*,$ $D_1,\, D_2^*\}$, are important both because they are
complementary signals of possible new physics contributions to $b\to
c\tau\bar\nu$, and because they are substantial backgrounds to the $R(D^{(*)})$
measurements (as well as to some $|V_{cb}|$ and $|V_{ub}|$ measurements).  
Thus, the correct interpretation of future $B\to D^{(*)}\ell\bar\nu$
measurements requires consistent treatment of the $D^{**}$ modes.

The spectroscopy of the $\dss$ states is important, because in addition to the
impact on the kinematics, it also affects the expansion of the form
factors~\cite{Leibovich:1997tu,Leibovich:1997em} in HQET~\cite{Georgi:1990um,Eichten:1989zv}.  
The isospin averaged masses and widths for the six lightest
charm mesons are shown in Table~\ref{tab:charm}.  In the HQS~\cite{Isgur:1989vq,Isgur:1989ed} limit, the spin-parity of the light
degrees of freedom, $s_l^{\pi_l}$, is a conserved quantum number, yielding
doublets of heavy quark symmetry, as the spin $s_l$ is combined with the heavy
quark spin~\cite{Isgur:1991wq}.  The ground state charm mesons containing light
degrees of freedom with spin-parity $s_l^{\pi_l} = \frac12^-$ are the
$\big\{D,\, D^*\big\}$.  The four lightest excited $\dss$ states correspond in
the quark model to combining the heavy quark and light quark spins with $L=1$
orbital angular momentum.  The $s_l^{\pi_l} = \frac12^+$ states are
$\big\{D_0^*,\, D_1^*\big\}$ while the $s_l^{\pi_l} = \frac32^+$ states are 
$\big\{D_1,\, D_2^*\big\}$.  The $s_l^{\pi_l} = \frac32^+$ states are narrow
because their $D^{(*)}\pi$ decays only occur in a $d$-wave or violate heavy
quark symmetry.  In the case of $B_s$ decays, all four $D_s^{**}$ states are
narrow.

\begin{table}[b]
\tabcolsep 6pt
\centerline{\begin{tabular}{ccccc}
\hline\hline
Particle  &    $s_l^{\pi_l}$ &  $J^P$  &  $m$ (MeV)  &  $\Gamma$ (MeV)\\
\hline
$\dSs$ &  $\frac12^+$  &  $0^+$  &  $2349$  &  $236$ \\
$\dVs$ &  $\frac12^+$  &  $1^+$  &  $2427$  &  $384$ \\
\hline
$\dV$ &  $\frac32^+$  &  $1^+$  &  $2421$  &  $31$ \\
$\dTs$ &  $\frac32^+$  &  $2^+$  &  $2461$  &  $47$ \\
\hline\hline
$D^*$ &  $\frac12^-$  &  $1^-$  &  $2009$  &  0. \\
$D$ &  $\frac12^-$  &  $0^-$  &  $1866$  &  0. \\
\hline\hline
\end{tabular}}
\caption{Isospin averaged masses and widths of the six lightest charm mesons,
rounded to 1\,MeV~\cite{PDG2018} (from Ref.~\cite{Bernlochner:2017jxt}).}
\label{tab:charm}
\end{table}

A simplifying assumption used in Refs.~\cite{Leibovich:1997tu,Leibovich:1997em}
to reduce the number of subleading Isgur-Wise functions was to neglect certain
$O(\lqcd/m_{c,b})$ contributions involving the chromomagnetic operator in
the subleading HQET Lagrangian, motivated by the fact that the mass splittings
in both the $s_l^{\pi_l} = \frac12^+$ and $s_l^{\pi_l} = \frac32^+$ doublets
were measured to be much smaller than $m_{D^*} - m_D$.  This is not supported by
the more recent data (see Table~\ref{tab:charm}), so
Ref.~\cite{Bernlochner:2016bci} extended the predictions of
Refs.~\cite{Leibovich:1997tu,Leibovich:1997em} accordingly, including deriving
the HQET expansions of the form factors which do not contribute in the $m_\ell =
0$ limit.  The impact of arbitrary new physics operators was analyzed in 
Ref.~\cite{Bernlochner:2017jxt}, including the $O(\lqcd/m_{c,b})$ and
$(\alpha_s)$ corrections in HQET.  The corresponding results in the
heavy quark limit were obtained in Ref.~\cite{Biancofiore:2013ki}.

The large impact of the $O(\lqcd/m_{c,b})$ contributions to the form
factors can be understood qualitatively by considering how heavy quark symmetry
constrains the structure of the expansions near zero recoil.  It is useful to
think of a simultaneous expansion in powers of $(w-1)$ and $(\lqcd/m_{c,b})$. 
(The kinematic ranges are $0 < w-1 \lesssim 0.2$ for $\tau$ final states, and $0
< w-1 \lesssim 0.3$ for $e$ and $\mu$.) The decay rates to the spin-1 $D^{**}$
states, which are not helicity suppressed near $w=1$, are of the form
\begin{equation}
\frac{{\rm d}\Gamma_{D_1,\, D_1^*}}{{\rm d}w} \sim
  \sqrt{w^2-1}\, \big[ \big( 0_{\rm (HQS)}
  + 0_{\rm (HQS)}\,\varepsilon  + \varepsilon^2 + \ldots \big) 
  + (w-1)\, \big(\varepsilon^0 + \varepsilon + \ldots \big) + \ldots \big] .
\end{equation}
Here $\varepsilon$ is a power-counting parameter of order $\lqcd/m_{c,b}$, and
the 0-s are consequences of heavy quark symmetry.  The $\varepsilon^2$ term in
the first parenthesis is fully determined by the leading order Isgur-Wise
function and hadron mass splittings~\cite{Leibovich:1997tu,Leibovich:1997em,Bernlochner:2016bci,Bernlochner:2017jxt}.  
The same also holds for those new
physics contributions to $B\to D_0^* \ell\bar\nu$, which are not helicity
suppressed. This explains why the $O(\lqcd/m_{c,b})$ corrections to the
form factors are very important, and can make $O(1)$ differences in
physical predictions, without being a sign of a breakdown of the heavy quark
expansion.  The sensitivity of the $D^{**}$ modes to new physics is
complementary and sometimes greater than those of the $D$ and $D^*$
modes~\cite{Biancofiore:2013ki,Bernlochner:2017jxt}.  Thus, using HQET, the
predictions for $B\to D^{**}\tau\bar\nu$ are systematically improvable by better
data on the $e$ and $\mu$ modes, just like they are for $B\to D^{(*)}
\tau\bar\nu$~\cite{Bernlochner:2017jka}, and are being implemented in
HAMMER~\cite{Ligeti:2016npd,Duell:2016maj,Bernlochner:2020tfi}.

\subsection{New physics in $B\to D^{(*)} \tau \nu$}

Independently of the recent discussion on form factor parametrizations and their influence on the extraction of $V_{cb}$ (covered in Sec.~\ref{sec:param}) it is clear from Table~\ref{tab:RD-RDstar} that the SM cannot accomodate the present experimental data on $R(D^{(*)})$. Even after the inclusion of  the most recent Belle measurement~\cite{Abdesselam:2019wbt},  the significance of the anomaly remains $3.1\sigma$. 
This leaves, apart from an underestimation of systematic uncertainties on the experimental side, NP as an exciting potential explanation. The required size of such a contribution comes as a surprise, however: defining $\hat R(X)\equiv R(X)/R(X)_{\rm SM}$, the new average corresponds to $\hat R(D)=1.14\pm0.10$ and $\hat R(D^*)=1.14\pm0.06$; for NP to accommodate these data, a contribution of $5-10\%$ relative to a SM tree-level amplitude is required for NP interfering with the SM, and $O(40\%)$ for NP without interference. 
An effect of this size can be clearly identified with upcoming measurements by LHCb and Belle~II~\cite{Cerri:2018ypt,Kou:2018nap}. It would also immediately imply large effects in other observables.

\begin{figure}[t]	
\includegraphics[height=3.75cm]{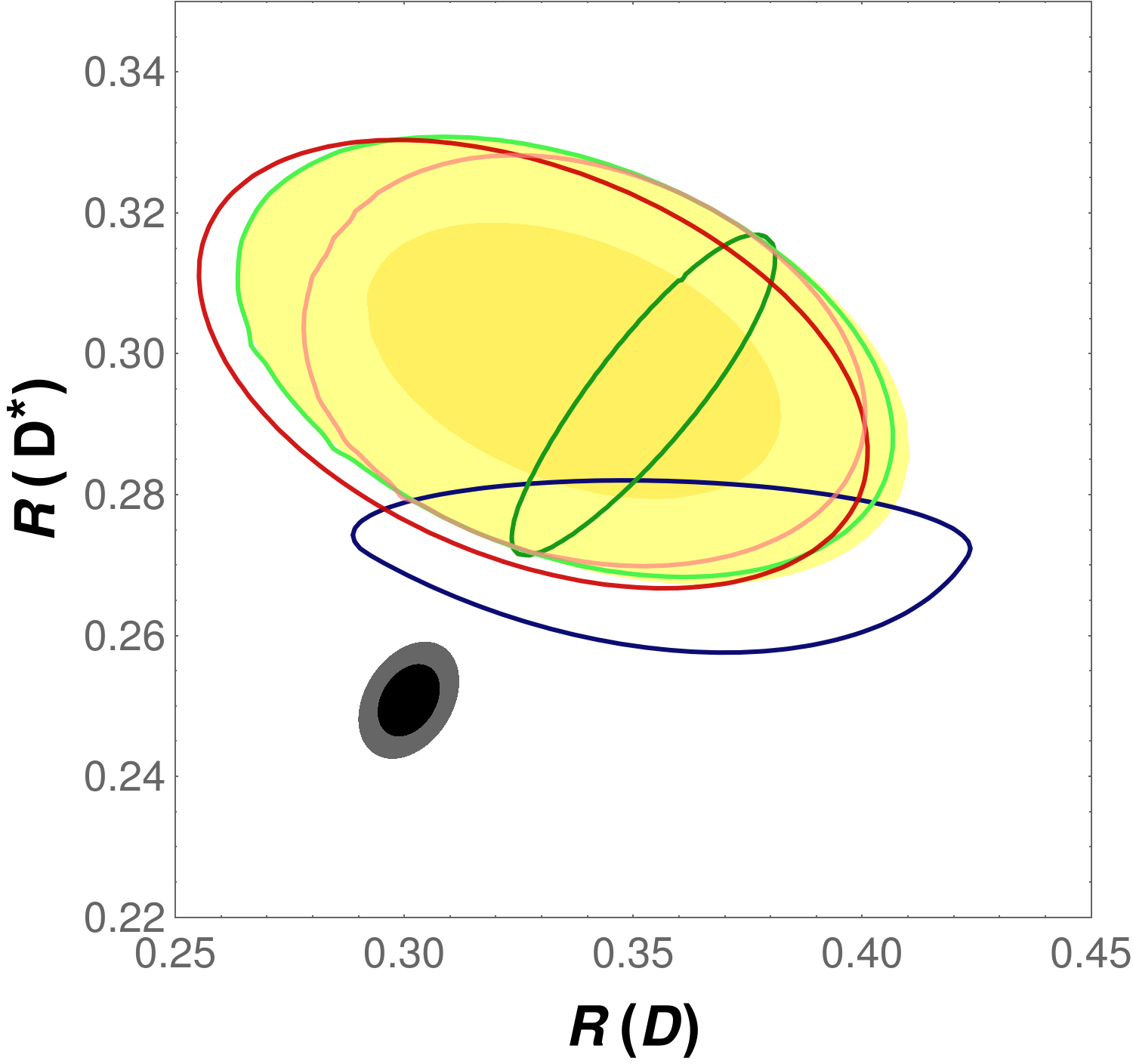}
\includegraphics[height=3.75cm]{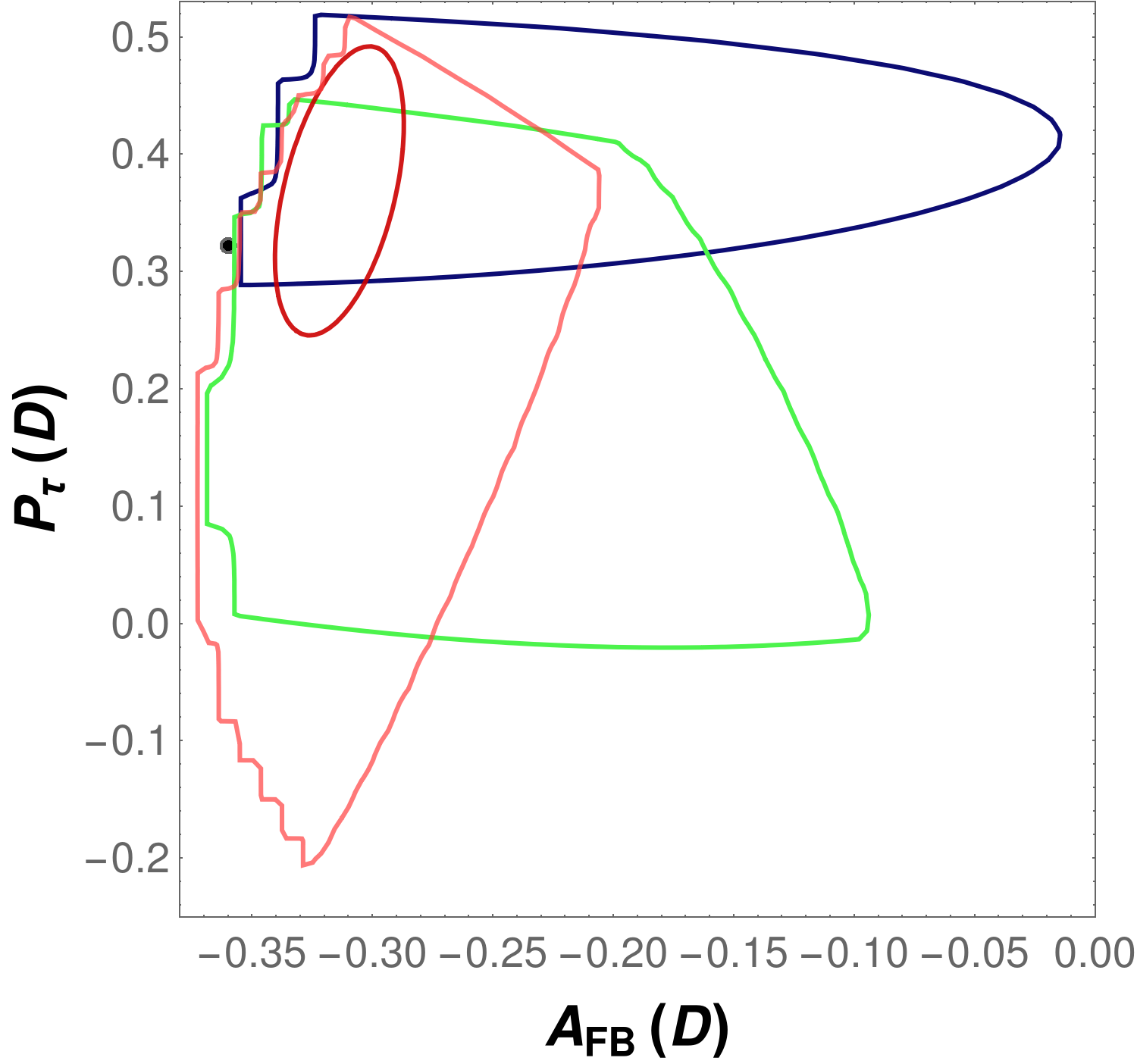}
\includegraphics[height=3.75cm]{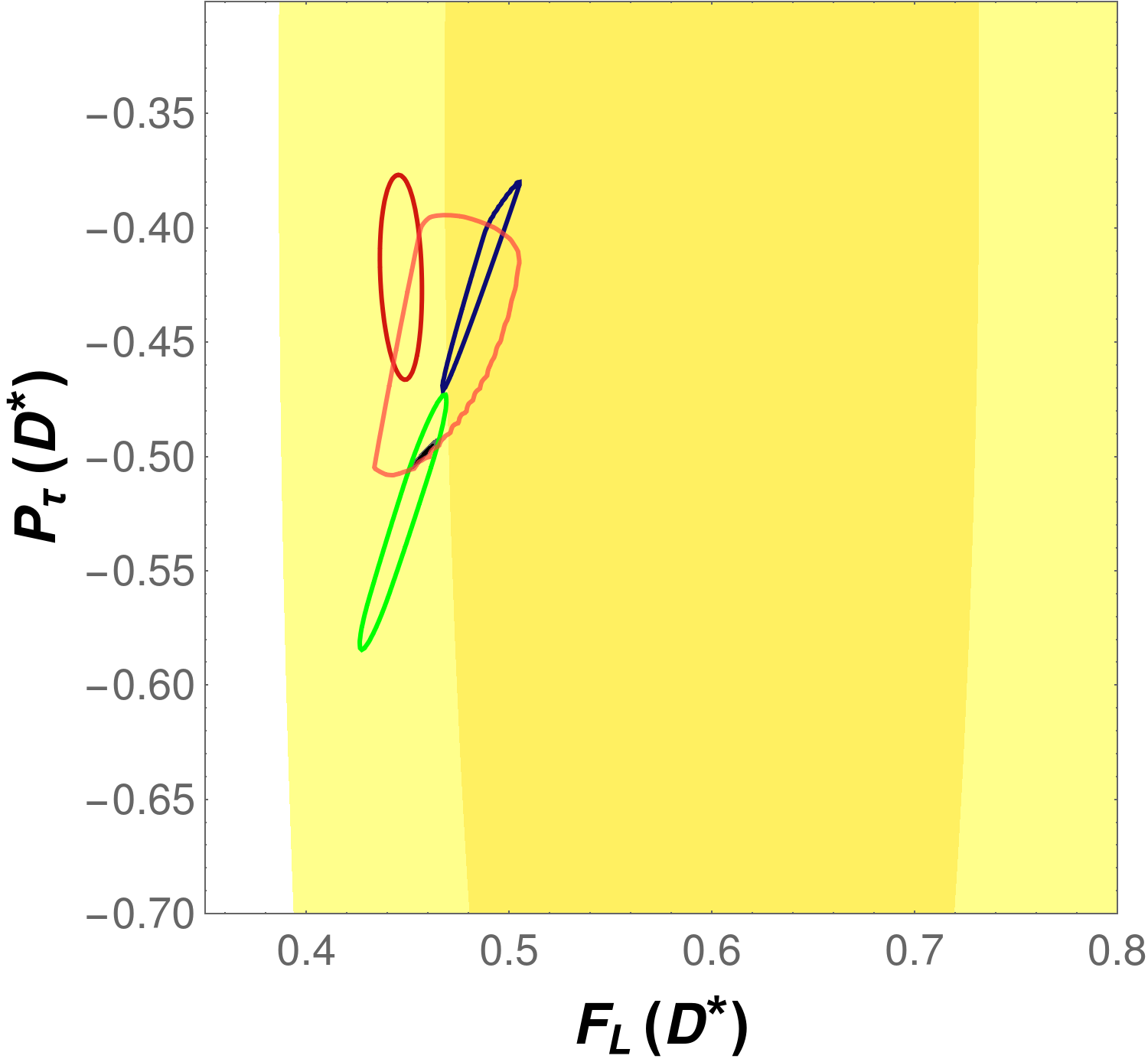}
\caption{\label{fig::FitsNPmodels} State-of-the-art fit results in single-mediator models for selected pairs of observables in $B\to D^{(*)}\tau\nu$ decays (following Ref.~\cite{Murgui:2019czp} for form factor and input treatment). All outer ellipses correspond to $95\%$ confidence level, inner (where present) to $68\%$.
We show the SM prediction in grey, the experimental measurement/average in yellow (where applicable) and scenarios I, II, III IV and V in dark green, green, dark blue, dark red and red, respectively, see text. Contours outside the experimental ellispse imply that the measured central values cannot be accomodated within that scenario. The limit $BR(B_c\to\tau\nu)\leq 30\%$ has been applied throughout, but affects only the fits with scalar coefficients. Dark green contours are missing in the two graphs on the right, because the predictions of scenario I are identical to the SM ones.}
\end{figure}

The potential of $R(D^{(*)})$ as discovery modes does not diminish the importance of additional measurements with $b$-hadrons. Specifically, even with a potential discovery, model discrimination will require measurements beyond these ratios. These additional measurements fall in four categories: 
\begin{itemize}
\item Additional $R(X)$ measurements such as $R(D^{**}), R(\Lambda_c), R(X_c), R(J/\psi)$ and $R(B_s^{(*)})$, are important crosschecks to establish $R(D^{(*)})$ as NP with independent systematics and provide independent NP sensitivity (especially $R(X_c)$ and $R(\Lambda_c)$), as discussed 
in subsections~\ref{ssec::exp} and~\ref{ssec::Ddoublestar}.
Note, however, the existence of an approximate sum rule relating the NP contributions to $R(\Lambda_c)$, $R(D)$, and $R(D^*)$ \cite{Blanke:2018yud}.
\item Integrated angular and polarization asymmetries and polarization fractions are excellent model discriminators. In many models they are completely determined once the measurements of $R(D^{(*)})$ are taken into account. For instance, the recent measurement of the longitudinal polarization fraction of the $D^*$ in $B\to D^*\tau\nu$, $F_L(D^*)$, was able to rule out solutions that remained compatible with the whole set of the remaining $b\to c\tau\nu$ data~\cite{Aebischer:2018iyb,Iguro:2018vqb,Blanke:2018yud,Bardhan:2019ljo,Alok:2019uqc,Murgui:2019czp}. 
The model-discriminating potential of both $R(D^{(*)})$ and selected angular quantities is visualized in Fig.~\ref{fig::FitsNPmodels}, where fit results for pairs of $B\to D^{(*)}\tau\nu$ observables within all phenomenologically viable single-mediator scenarios with left-handed neutrinos to the state-of-the-art data are shown.
\item Differential distributions in $q^2$ and the different angles are extremely powerful in distinguishing between NP models, as can be seen for instance from a recent analysis of data with light leptons in the final state~\cite{Jung:2018lfu}. They require, however, large amounts of data and the insufficient information on the decay kinematics can pose difficulties for the interpretation of the data, as discussed in subsection~\ref{ssec::interpretation}. However, already the rather rough available information on the differential rates $d\Gamma/dq^2(B\to D^{(*)}\tau\nu)$~\cite{Huschle:2015rga,Lees:2012xj} is excluding relevant parts of the parameter space~\cite{Sakaki:2014sea,Freytsis:2015qca,Celis:2016azn,Bhattacharya:2016zcw,Murgui:2019czp}.
\item An analysis of the flavor structure of the observed effect, \emph{e.g.} in $b\to c (e,\mu)\nu$, $b\to u\tau\nu$ and $t\to b\tau\nu$ transitions.
\end{itemize}
In addition to the above observables, the leptonic decay $B_c\to \tau\nu$ plays a special role. Although it is not expected to be measured in the foreseeable future, it provides nevertheless a strong constraint on NP, since the relative influence of scalar NP is enhanced in this mode. A limit can then be obtained even from the total width of the $B_c$ meson~\cite{Li:2016vvp}. Theoretical estimates for the partial width assumed unaffected by NP can be used to strengthen these bounds~\cite{Beneke:1996xe,Alonso:2016oyd,Celis:2016azn}, and also data from LEP~\cite{Akeroyd:2017mhr}. Both approaches rely on additional assumptions, however, see Refs.~\cite{Blanke:2018yud,Bardhan:2019ljo} for recent extensive discussions.

The constraints discussed so far are relevant in any scenario trying to address the existing anomalies. An interesting subclass of such models is that where the existence of a single mediator coupling to only the known SM degrees of freedom is assumed, classified in~\cite{Freytsis:2015qca}, creating only a subset of the possible operators at the $b$ scale. 
Among those, only five scenarios remain that can reasonably well accomodate the data described above, see also 
Refs.~\cite{Tanaka:2012nw,Blanke:2018yud,Freytsis:2015qca,Bhattacharya:2016zcw,Ivanov:2017mrj,Alok:2017qsi,Bifani:2018zmi,%
Blanke:2019qrx,Shi:2019gxi} for comparisons (additional constraints in specific scenarios are commented on below): 
Scenario~I yields only a left-handed vector operator, created by either a heavy color-less vector particle~\cite{Greljo:2015mma,Boucenna:2016wpr,Boucenna:2016qad,Megias:2017ove} (phenomenologically highly disfavored) or a leptoquark, see Refs.~\cite{Li:2016vvp,Fajfer:2012jt,Deshpande:2012rr,Sakaki:2013bfa,Duraisamy:2014sna,Calibbi:2015kma,Fajfer:2015ycq,Barbieri:2015yvd,Alonso:2015sja,Bauer:2015knc,Das:2016vkr,Deshpand:2016cpw,Sahoo:2016pet,Dumont:2016xpj,Becirevic:2016yqi,Barbieri:2016las,DiLuzio:2017vat,Assad:2017iib,Chen:2017hir,Bordone:2017bld,Altmannshofer:2017poe,Calibbi:2017qbu,Becirevic:2018afm,Fornal:2018dqn,Blanke:2018sro,Crivellin:2017zlb} for this and other leptoquark variants. 
Scenario II includes Scenario I, but yields also a right-handed scalar operator, realized for example by a vector leptoquark. 
Scenario III involves both left- and right-handed scalar operators, generated for instance by a charged Higgs~\cite{Crivellin:2012ye,Celis:2012dk,Crivellin:2015hha,Celis:2016azn,Chen:2017eby,Iguro:2017ysu,Chen:2018hqy,Li:2018rax} (with a limited capability to accomodate $R(D^*)$ due to the $B_c$ constraint discussed above). 
Scenarios IV and V involve the left-handed scalar and tensor operator which are generated proportionally to each other ($C_{S_L}=\pm 4C_T$ at the NP scale $\Lambda$), in the latter case with the addition of the left-handed vector operator, again realized in leptoquark models.
It is also possible to analyze the available data in more general contexts. For example, within SMEFT the right-handed vector current is expected to be universal~\cite{Cirigliano:2009wk,Alonso:2014csa,Cata:2015lta}, see~\cite{Murgui:2019czp} for a global analysis in this framework, while this does not hold when the electroweak symmetry breaking is realized non-linearly~\cite{Cata:2015lta}. 
Allowing for additional light degrees of freedom beyond the SM opens the possibility of contributions with right-handed neutrinos, see Refs.~\cite{He:2012zp,Becirevic:2016yqi,He:2017bft,Greljo:2018ogz,Asadi:2018wea,Robinson:2018gza,Azatov:2018kzb,Heeck:2018ntp}.

Once specific models are considered, typically additional constraints apply. Important ones include high-$p_T$ searches, looking for 
collider signatures of the mediators related to the
anomaly~\cite{Faroughy:2016osc,Feruglio:2018fxo,Greljo:2018tzh,Altmannshofer:2017yso}, RGE-induced flavor-non-universal effects in 
$\tau$ decays~\cite{Feruglio:2017rjo}, lepton-flavor violating decays~\cite{Feruglio:2017rjo}, precision universality tests in quarkonia 
decays~\cite{Aloni:2017eny},  charged-lepton magnetic moments~\cite{Feruglio:2018fxo} and electric dipole moments in models with 
non-vanishing imaginary parts~\cite{Dekens:2018bci}.

\subsection{Interpretation of experimental results}
\label{ssec::interpretation}

The reconstructed kinematic distributions used in measurements are sensitive to both the modeling of required non-perturbative inputs 
(e.g., form factors, light-cone meson wave functions), and to assumptions about the underlying fundamental theory (e.g. possible presence 
of operators with chiral structures different from those found in the SM). Current measurements assume the SM operator structure, and 
include the non-perturbative uncertainties as they are known at the time of publication. While this is a valid strategy for testing the 
SM, if in future the presence of a non-SM contribution with a different chiral structure is established then past measurements will 
require reinterpretation. 

In order to present experimental results in such a way to allow a-posteriori analyses to have maximum flexibility in the description of 
non-perturbative inputs and BSM content, the following strategies might be considered. The techniques to allow for reinterpretation of 
results overlap with those used to make differential measurements designed to be sensitive to the chiral structure and non-perturbative 
quantities.

A first possibility, is the publication of unfolded distributions (see, for instance, the $B\to D^* \ell\nu$ spectrum presented in 
Ref.~\cite{Abdesselam:2017kjf}). This method offers the possibility to fit with ease the experimental results to arbitrary 
parametrizations of the form factors~\cite{Bigi:2017njr,Grinstein:2017nlq,Bernlochner:2017jka}; its downside is that it requires 
relatively high statistics and that the unfolded distributions do not contain the whole experimental information.

A second option, which has been employed in the untagged Belle analysis of ref.~\cite{Waheed:2018djm}, is to provide \emph{folded} 
distributions in which detector effects are not removed and no extrapolation is performed, together with experimental efficiencies and
the detector response matrix (which reproduces detector effects to a given accuracy). This allows the use of any parametrization of SM
and BSM effects in comparing with the experimental result. This approach, while requiring slightly more involved a posteriori fitting 
strategies, avoids the statistical problems associated with unfolding and can be extended more easily to higher dimensions.

Finally, the most complete information is contained in the Likelihood function which depends on a set of SM parameters (e.g., for
$B\to \pi\ell\nu$ they could be the coefficients of the $z$-expansion of the form factors and $V_{ub}$) and on the Wilson coefficients
of BSM operators).
This method has not been currently pursued in any $B$ decay measurement, in part because of difficulties related to the extremely large 
amount of information that would need to be presented. Two differing approaches are to publish the full experimental Likelihood in the 
full parameter space of BSM Wilson coefficients and SM non-perturbative coefficients, or to publish the tools for external readers to be 
able to repeat the full experimental fit with the signal model varied. For representing the experimental Likelihood in a high-dimensional 
space, possible approaches include the use of Markov chain sampling, or MVA surface modelling. These are the only strategies which would 
allow the entirety of the experimental information to be available in a posteriori theoretical investigations. It is essential to this 
approach for the experimental measurement to cover the full parameter space in a sufficiently general way, including alternative Likelihoods with 
different parametrizations for nonperturbative effects.

\subsection{HAMMER}

Future new physics searches in $b \to c \, \tau \nu_\tau$ decays are a challenging endeavor: most experimental results make use of kinematic properties of the process to discriminate between the signal of interests and backgrounds. For instance recent measurements from the $B$-factories BaBar and Belle used the lepton momentum spectrum and measurements of LHCb use fits to the four-momentum transfer $q^2$. In new physics scenarios, these distributions change and alter the analysis acceptance, efficiencies, and extracted signal yields. In addition, large samples of simulated decay processes play an integral part in those measurements. In most, one of the leading systematic uncertainties is due to the limited availability of such samples. Thus producing large enough simulation samples for a wide range of new physics points, needed to take into account the aforementioned changes in acceptance, etc. is not a viable path. This is where the tool HAMMER~\cite{Bernlochner:2020tfi,Duell:2016maj} can help: it implements an event-level reweighting, assigning a weight based on the ratio of new physics to simulated matrix element, which allows one to re-use the already generated events. In addition, it is capable of providing histograms for arbitrary new-physics parameter values (including also form factor variations), which can be used, for example, in template fits to kinematic observables. These event weights can completely account for acceptance changes and will enable Belle II and LHCb to directly extract limits on the Wilson coefficients present in $b \to c \, \tau \nu$ transitions.


\section{Heavy-to-light exclusive}
\label{h2l_excl}
In this section we present an overview of $b\to u$ exclusive decays.
We start with a discussion  of the lattice calculations of the $b$ hadron decay form factors to a light pseudoscalar,  vector meson or baryon. We then review the light-cone sum rule calculation of the same form factors and the current experimental situation, as well as the 
prospects at Belle II and LHCb. Finally, we briefly discuss a few related subjects, such as 
the semitauonic heavy to light decays, the decay $b\to \gamma \ell \nu_\ell$, the non-resonant $B\to \pi\pi \ell\nu$ decays, and some subtlety of the $z$-expansion.

\subsection{Form factors for semileptonic $b$-hadron decays into light hadrons from lattice QCD}

\subsubsection{Form factor parametrizations}
\label{sec:HLparametrizations}

\str{
The matrix elements that describe the hadronic part of the semileptonic transitions $B \to P \ell \nu$ or $B \to P \ell \ell$ are 
parametrized in terms of form factors.
Conventionally, one  writes: 
\begin{eqnarray}
\langle P(p_P) |V^{\mu} |B (p) \rangle  & = & \left((p_B+p_P)^\mu - \frac{M_B^2-M_P^2}{q^2}q^\mu\right) \, f_+(q^2)  \nonumber \\ 
&  + & \frac{M_B^2-M_P^2}{q^2}q^\mu \, f_0(q^2),  \\
 \langle P(p_P) |S |B (p) \rangle & = & \frac{M_B^2-M_P^2}{m_b - m_q} f_0(q^2), \\
\langle P(p_P)|T^{\mu\nu}|B (p) \rangle  & = & 2\frac{p_B^\mu p_P^\nu - p_B^\nu p_P^\mu}{M_B+M_P}  f_T(q^2), 
\end{eqnarray}   
where $|B(p_B)\rangle$ denotes a $B^0$, $B^{\pm}$, or $B_s$ meson with four-momentum $p_B$ and mass $M_B$, and $|P(p_P)\rangle$
is a pion or kaon with four-momentum $p_P$ and mass $M_P$, $q^\mu \equiv (p_B-p_P)^\mu$, and the current operators are
$V^\mu = \bar{q} \gamma^\mu b$, $S = \bar{q} b$, and $T^{\mu\nu} =  i\bar{q} \sigma^{\mu\nu} b$.
The form factors $f_+, f_0, f_T$ are the complete set needed to describe the hadronic contributions to charged-current semileptonic 
transitions in and beyond the Standard Model and they also completely describe the factorizable contributions to rare decays. 

For the case of semileptonic $B$-meson decays to vector meson final states, we have:
\begin{eqnarray}
\langle V(p_V, \varepsilon) |V^{\mu}   |B(p_B) \rangle & = & \frac{2i}{M_B+M_V}\epsilon^{\mu\nu\rho\sigma}
    \varepsilon^\ast_\nu p_{V,\rho} p_{B,\sigma} \, V(q^2),      \\
\langle V(p_V,\varepsilon) |A^{\mu}   |B(p_B)  \rangle & = & 2M_V \frac{\varepsilon^\ast\cdot q}{q^2}q^\mu \, A_0(q^2)  \\
        & + &  (M_B+M_V) \left(\varepsilon^{\ast\mu} - \frac{\varepsilon^\ast\cdot q}{q^2}q^\mu\right) A_1(q^2)  \nonumber \\
        & - & \frac{\epsilon^\ast\cdot q}{M_B+M_V}\left[(p_B+p_V)^\mu - \frac{M_B^2 - M_V^2}{q^2}q^\mu\right]A_2(q^2), \nonumber \\
\langle V(p_V,\varepsilon) |T^{\mu\nu}|B(p_B)  \rangle & =  &
\end{eqnarray}
where $V(p_V,\varepsilon)\rangle$ denotes a vector meson ($\rho$, $K^*$, $\phi$) with momentum $p_V$, mass $M_V$, and 
polarization~$\varepsilon$. 
}


The matrix elements that describe the hadronic part of the semileptonic transitions $B\to X\ell\nu$ or $B \to X\ell\ell$ are 
parametrized in terms of the form factors in Eqs.~(\ref{eq:ff-scalar})--(\ref{eq:ff-pseudo-tensor}),
where $X$ now denotes a pion or kaon.
The transitions $B\to X^*\ell\nu$ or $B\to X^*\ell\ell$ are parametrized in terms of the form factors in
Eqs.~(\ref{eq:ff-pseudo})--(\ref{eq:ff-tensor}), where $X^*$ now denotes a $\rho$, $K^*$, or $\phi$~meson.
As discussed in Sec.~\ref{sec:param}, modern theoretical calculations of the form factors
employ $z$-parametrizations to describe their shapes, which can be implemented in a 
model-independent way, being based on analyticity and unitarity constraints. For the case at 
hand, an often used choice for the $z$-parameter defined in Eq.~(\ref{eq:z-def}) is 
$t_0 = (M+m)/(\sqrt{M} - \sqrt{m})^2$, which results in a range $|z| < 0.3$,  centered around 
$z=0$. 
In general, the small range of $z$ coupled with unitarity constraints on the 
coefficients ensure that the polynomial expansions converge quickly. 
As discussed already in Sec.~\ref{sec:param}, for $B$-meson decays to light hadrons with their 
larger $q^2$ range, the BCL parametrization~\cite{Bourrely:2008za} is the standard choice, as 
the resulting forms satisfy the expected asymptotic $q^2$ and near threshold scaling 
behaviors~\cite{Lepage:1980fj,Akhoury:1993uw}:
\begin{eqnarray}
    f_+(q^2) &=& \frac{1}{1-q^2/ M_{B^*(1^-)}^2} \sum\limits_{n=0}^{N_z-1} b_n^+(t_0)
        \left(z^n -(-1)^{n-N_z}\frac{n}{N_z} z^{N_z}\right), 
    \label{eq:BCL_f+} \\
    f_0(q^2) &=& \frac{1}{1-q^2/ M_{B^*(0^+)}^2} \sum\limits_{n= 0}^{N_z} b_n^0(t_0)\, z^n.
    \label{eq:BCL_f0}
\end{eqnarray}

\subsubsection{Lattice QCD results for $B$-meson decay form factors to light pseudoscalars }

Lattice-QCD calculations of the form factors for semileptonic $B_{(s)}$-meson decays  to light hadrons proceed along the same lines as discussed in Sec.~\ref{sec:hth_lat}. In particular, there are a number of different, well-developed strategies for dealing with the heavy $b$-quark in lattice QCD, see Ref.~\cite{Aoki:2019cca} for a review.  
The same two- and three-point functions as for the heavy-to-heavy case are needed here, albeit with the appropriate valence quark propagators, to describe the heavy-to-light decay process. While this affects the statistical errors in the next step, the fits to the spectral representations of the correlation functions to obtain the desired matrix elements on each gauge ensemble and each recoil momentum, the procedure is essentially the same. The resulting ``lattice data'' are then used in combined chiral-continuum fits coupled with a systematic errors analysis to obtain the form factors in the continuum over the range of recoil energies that are included in the simulations. 
Here, a well known challenge is that the recoil energies that are accessible in lattice-QCD calculations  cover only a fraction of the entire kinematic region. A related challenge is that the validity of Chiral Perturbation Theory (used to extrapolate or interpolate to the physical pion mass) is limited to pion energies of $\approx 1 $~GeV. 
The final step is the $z$-expansion fit, from which the form factors are obtained over the entire kinematic range, albeit with larger errors in the region not directly covered in the lattice calculation.  

 Lattice-QCD calculations of the $B\to \pi$ vector current form factors $f_+$ and $f_0$ can be used to determine $|V_{ub}|$ from experimental measurements of the $B\to\pi\ell \nu$ decay rate. 
There are currently two independent, published lattice-QCD computations that employ the modern methods outlined above, including the model-independent  $z$-expansion \cite{Flynn:2015mha,Lattice:2015tia}. The RBC/UKQCD collaboration  \cite{Flynn:2015mha} uses ensembles with $N_f=2+1$ flavors of Domain Wall fermions at two lattice spacings with sea-pion masses in the range $[300,400]$~MeV.  The Fermilab/MILC collaboration  \cite{Lattice:2015tia} uses ensembles with $N_f=2+1$ flavors of asqtad (improved staggered) fermion at four lattice spacings covering the range $a\approx 0.045 -0.12$~fm and a range of sea-pion masses down to $177$~MeV. Earlier work \cite{Bailey:2008wp} used a subset of these ensembles.   
The treatment of the $b$-quark is similar in the two works; Ref.~\cite{Flynn:2015mha} uses a variant of the Fermilab approach, called the relativistic heavy quark (RHQ) action, while Ref.~\cite{Lattice:2015tia} employs the original Fermilab formalism. Both groups also use the mostly nonperturbative renormalization method to compute the renormalization factors. The form factors obtained by the two lattice groups are in good agreement with each other, and can be combined in joint fits together with experimental data for an improved $|V_{ub}|$ determination \cite{Aoki:2019cca}. 
 
Ongoing work by RBC/UKQCD is extending the calculation to include more ensembles  \cite{Flynn:2019jbg}. Ongoing work by the Fermilab/MILC collaboration employs the HISQ $N_f=2+1+1$ ensembles with sea-pion masses at (or near) the physical point, and the Fermilab formalism for the $b$-quark~\cite{Gelzer:2019zwx}.  
The HPQCD collaboration has published a calculation of the scalar form factor for the $B\to\pi$ transition at zero recoil $f_0(q^2_\text{max})$ on a subset of the $N_f=2+1+1$ HISQ ensembles and treating the $b$ -quark in NRQCD~\cite{Colquhoun:2015mfa}, which provides a nice test of the soft-pion theorem, but cannot be used in $|V_{ub}|$ determinations. 
Ongoing work includes a calculation of the $B\to\pi$ form factors over a range of $q^2$ on a subset of the asqtad ensembles using NRQCD $b$-quarks and HISQ light-valence quarks \cite{Bouchard:2013zda}. The JLQCD collaboration has an ongoing project to calculate the $B\to\pi$ form factors on $N_f=2+1$ Domain Wall ensembles using also Domain Wall fermions for the heavy and light valence quarks \cite{Colquhoun:2019tyq}. They focus their calculation on small lattice spacings ($a\approx 0.044 - 0.080$~fm) and include a series of heavy-quark masses to extrapolate to the physical $b$-quark mass. 

The vector current form factors $f_+$ and $f_0$ needed for rare $B\to\pi \ell\ell$ decay are the same as for $B\to \pi \ell \nu$ decay (up to small isospin corrections), but the tensor form factor $f_T$ is also needed to describe the rare process in the SM, while it can contribute to $B\to \pi \ell \nu$ decay only in BSM theories. So far, $f_T$ has been calculated only by the Fermilab/MILC collaboration~\cite{Bailey:2015nbd} using the same ensembles and methods as for the vector current form factors. However, most (if not all) of the ongoing projects described above, now include the complete set of form factors in their analyses, and new results for this form factor will therefore also be forthcoming. 

The $B_s \to K \ell \nu$ process  can be used for an alternate determination of $|V_{ub}|$, and there currently are three independent, published lattice-QCD computations of the vector-current form factors \cite{Bouchard:2014ypa,Flynn:2015mha,Bazavov:2019aom}. In Ref.~\cite{Bouchard:2014ypa} the HPQCD collaboration used NRQCD $b$-quarks and HISQ light-valence quarks to calculate the form factors on a subset of asqtad ensembles. 
The RBC/UKQCD \cite{Flynn:2015mha} work is already described above, since they calculated the $B_s\to K$ and $B\to\pi$ transition form factors together. The Fermilab/MILC collaboration \cite{Bazavov:2019aom} used the same methods and set-up as for their $B\to\pi$ project \cite{Lattice:2015tia} but on a subset of asqtad ensembles. Both Fermilab/MILC \cite{Bazavov:2019aom}  and, in a follow-up paper, HPQCD \cite{Monahan:2018lzv} also computed ratios of $B_s\to K$ and $B_s \to D_s$ observables, which can be used in combination with LHCb measurements to determine $|V_{ub}/V_{cb}|$. 
 
\subsubsection{Challenges of vector mesons }
Lattice calculations of $B_{(s)}$ decay form factors with vector mesons ($\rho$, $K^*$, $\phi$) in the final state are substantially more
challenging, as these vector mesons are unstable resonances for sufficiently light quark masses. The asymptotic final state in the continuum
then contains (at least) two hadrons, and the relation with the finite-volume matrix elements computed on the lattice becomes nontrivial.
The formalism that allows a mapping of finite-volume to infinite-volume $1\to 2$ hadron matrix elements has been
developed~\cite{Lellouch:2000pv,Lin:2001ek,Christ:2005gi,Hansen:2012tf,Briceno:2014uqa,Briceno:2015csa,Agadjanov:2016fbd} and will be discussed in more detail below.
First numerical applications to a form factor with nonzero momentum transfer have been published for the electromagnetic process $\pi \gamma^* \to \pi \pi$, where the $\pi\pi$ final state in a $P$ wave couples
to the $\rho$ resonance~\cite{Briceno:2015dca,Briceno:2016kkp,Alexandrou:2018jbt}.

The lattice QCD calculations of $B_{(s)} \to V$ form factors published to date did not implement this $1 \to 2$ formalism. For the $B \to \rho$ form factors,
there is only an early study by the UKQCD collaboration~\cite{Bowler:2004zb}, performed in the quenched approximation and with heavy up and down quark masses
for which the $\rho$ is stable. For the $B \to K^*$, $B_s \to K^*$, $B_s \to \phi$ form factors, an unquenched lattice QCD calculation is available
\cite{Horgan:2013hoa}. This work used three different ensembles of lattice gauge field configurations with pion masses of approximately
310, 340, and 520~MeV. For the lower two pion masses, the $K^*$ is expected to be unstable, but the analysis was performed as if the $K^*$ were stable.
This entails using only a quark-antiquark interpolating field for the $K^*$, and assuming that the information extracted from exponential fits
to the two-point and three-point correlation functions corresponds to the ``$K^*$'' contribution. The systematic errors introduced by this treatment are difficult to quantify. For unstable $K^*$,
none of the actual discrete finite-volume energy levels directly corresponds to the resonance,
and the actual ground state may be far from the resonance location (for typical lattice volumes, this problem is more severe at nonzero momentum).
However, a quark-antiquark interpolating field couples more strongly to energy levels in the vicinity of the resonance, and ground-state saturation is typically not seen in the correlation
functions before the statistical noise becomes overwhelming. In these cases, exponential fits are still dominated by one or multiple energy levels in the vicinity of the resonance.

In the following, we will denote the vector meson resonance as $V$, and the two pseudoscalar mesons whose scattering shows the resonance as $P_1$ and $P_2$.
The finite-volume energy levels for a given total momentum and irreducible representation of the appropriate symmetry group
are determined by the L\"uscher quantization condition~\cite{Luscher:1990ux} and its generalizations, as reviewed in Ref.~\cite{Briceno:2017max}. In the absence of interactions, they would consist
of $P_1 P_2$ scattering states with energies equal to the sums of the $P_1$ and $P_2$ energies, where the  $P_1$ and $P_2$ momenta take on
the discrete values allowed by the periodic boundary conditions. Through the $P_1 P_2$ interactions, these energy levels are shifted away from their noninteracting
values in a volume-dependent way. In the simplest case (considering only elastic scattering and neglecting the partial-wave mixing induced by the finite volume), each interacting finite-volume energy level
can be mapped to a corresponding value of the infinite-volume $P_1 P_2$ scattering phase shift, or, equivalently, scattering amplitude; more complicated cases with coupled
channels and partial-wave mixing can also be treated. The dependence of the scattering amplitude on the $P_1 P_2$ invariant-mass-squared, $s$, can be described by a Breit-Wigner-type function. By analytically continuing
the scattering amplitude to complex $s$, one finds poles on the second Riemann sheet at $s = (m_V \pm i \Gamma_V/2)^2$, where $\Gamma_V$ is the width of the resonance. This procedure
has been applied successfully to the $\rho$, $K^*$, and other resonances (see Ref.~\cite{Briceno:2017max} for a review).

The $B_{(s)}\to V$ form factors correspond to the residues at the pole at $s = (m_V - i \Gamma_V/2)^2$ in the $B_{(s)}\to P_1 P_2$ form factors, where the $P_1 P_2$
final state is projected to the $\ell=1$ partial wave. These $B_{(s)}\to P_1 P_2$ form factors are functions of $q^2$ and $s$. In the single-channel case, the lattice computation involves the following
steps: (i) Determine the $P_1 P_2$ finite-volume energy spectrum, and the $B_{(s)}\to P_1 P_2$ finite-volume matrix elements both for the ground states and multiple
excited states. (ii) Obtain the infinite-volume $P_1 P_2$ scattering amplitude from the finite-volume
energy spectrum using the L\"uscher method, and fit a suitable parametrisation of the $s$-dependence to the data.
(iii) Map the finite-volume $B_{(s)}\to P_1 P_2$ matrix elements to infinite-volume $B_{(s)}\to P_1 P_2$ matrix elements using the Lellouch-L\"uscher factor, which depends
on the energy-derivative of the scattering phase shift and a known finite-volume function.

The finite-volume
formalism requires the center-of-mass energy $\sqrt{s}$ to be small enough so that no more than two particles can be produced by the scattering through the strong interaction (however, the \emph{total} momentum of the $P_1 P_2$ system
can in principle be arbitrarily large). For example, in the case of the $B \to \pi\pi$ form factors, the formalism requires $\sqrt{s} \lesssim 4\, m_\pi$, which becomes
more restrictive when performing the calculation at lighter quark masses. However, it is likely that the coupling to four pions has negligible effects even at somewhat higher values of $\sqrt{s}$, as needed
to map out the $\rho$ resonance region when using physical quark masses.

\subsubsection{$\Lambda_b \to p$ and $\Lambda_b \to \Lambda^{(*)}$ form factors from lattice QCD }
The $\Lambda_b \to p$ form factors relevant for the decay $\Lambda_b \to p \mu^-\bar{\nu}$ have been computed
in lattice QCD together with the $\Lambda_b \to \Lambda_c$ form factors~\cite{Detmold:2015aaa}; some aspects of
this work were already discussed in Sec.~\ref{sec:LbLcLattice}. The lattice data for $\Lambda_b \to p$ cover the kinematic
range from $q^2\approx 15\:{\rm GeV}^2$ to near $q^2_{\rm max}\approx 22\:{\rm GeV}^2$, and consequently
the predicted $\Lambda_b \to p \mu^-\bar{\nu}_\mu$ differential decay rate is most precise in this range. The integrated
decay rates in the Standard Model were found to be
\begin{equation}
 \frac{1}{|{V_{ub}}|^2}\Gamma (\Lambda_b \to p\: \mu^- \bar{\nu}_\mu)
= (25.7 \pm 2.6_{\,\rm stat} \pm 4.6_{\,\rm syst})\:\:{\rm ps}^{-1}
\end{equation}
and
\begin{equation}
\frac{1}{|{V_{ub}}|^2}\int_{15\:{\rm GeV}^2}^{q^2_{\rm max}}
\frac{\mathrm{d}\Gamma (\Lambda_b \to p\: \mu^- \bar{\nu}_\mu)}{\mathrm{d}q^2} \mathrm{d} q^2
= (12.31 \pm 0.76_{\,\rm stat} \pm 0.77_{\,\rm syst})\:\:{\rm ps}^{-1}.
\end{equation}
The latter has a total uncertainty of 8.8\% (corresponding to a 4.4\% theory uncertainty in a $|V_{ub}|$ determination from this rate),
and the ratio to the partially integrated $\Lambda_b \to \Lambda_c \mu^-\bar{\nu}$ decay rate (\ref{eq:LbLcPartialRate}) has a total uncertainty
of 9.8\%, corresponding to a 4.9\% theory uncertainty in the determination of $|V_{ub}/V_{cb}|$ performed by LHCb~\cite{Aaij:2015bfa}, commensurate with
the experimental uncertainty. The $\Lambda_b\to p$ form factors from Ref.~\cite{Detmold:2015aaa} can also be used to predict the Standard-Model value of the
baryonic $b \to u\ell\bar{\nu}$ lepton-flavor-universality ratio,
\begin{equation}
 \frac{\Gamma (\Lambda_b \to p\: \tau^- \bar{\nu}_\tau)}{\Gamma (\Lambda_b \to p\: \mu^- \bar{\nu}_\mu)} \:=\: 0.689 \:\pm \:0.058_{\,\rm stat} \:\pm\:  0.064_{\,\rm syst}.
\end{equation}
By increasing statistics, removing the partially quenched data sets (cf.~Sec.~\ref{sec:LbLcLattice}), adding one ensemble with physical light-quark masses, and
another ensemble with a third, finer lattice spacing, it will likely be possible to reduce the uncertainties in both the $\Lambda_b \to p$ and $\Lambda_b \to \Lambda_c$ form factors by a factor of 2 in the near future.

The same methods have also been used to compute the $\Lambda_b \to \Lambda$~\cite{Detmold:2016pkz},
$\Lambda_c \to p$~\cite{Meinel:2017ggx}, and $\Lambda_c \to \Lambda$~\cite{Meinel:2016dqj} form factors with lattice QCD.
The latter calculation already includes an ensemble with the physical pion mass, and gave results for the
$\Lambda_c\to\Lambda e^+\nu_e$ and $\Lambda_c\to\Lambda \mu^+\nu_\mu$ branching fractions consistent with, and two times more 
precise than, the measurements performed recently by the \mbox{BESIII} Collaboration~\cite{Ablikim:2015prg,Ablikim:2016vqd}.
This is a valuable test of the lattice methods used to determine the heavy-baryon decay form factors.

A lattice-QCD calculation is also in progress for the $\Lambda_b \to \Lambda^*(1520)$ form factors (in the narrow-width approximation)~\cite{Meinel:2016cxo}, which are relevant for the rare decay $\Lambda_b \to \Lambda^*(\to p\,K) \mu^+\mu^-$.
As with $\Lambda_b \to \Lambda_c^*$, discussed in Sec.~\ref{sec:LbLcLattice}, this initial calculation only reaches $q^2$ in the vicinity of $q^2_{\rm max}$.

\subsection{Light-cone sum rules calculations of heavy-to-light form
factors}

QCD sum rules on the light cone (LCSR) is a non-perturbative method for calculating hadronic quantities~\cite{Balitsky:1986st,Balitsky:1989ry,Chernyak:1990ag}. It has been applied to obtain the form factors for $B$ decays (see the definitions in
Section~\ref{sec:param}). The first LCSR calculations relevant for $V_{ub}$ were performed in 1997 when the next-to-leading order (NLO) twist-2 corrections to $f_+(q^2)$ were calculated~\cite{Khodjamirian:1997ub,Bagan:1997bp}. The leading order (LO) corrections up to twist-4 were calculated in Ref.~\cite{Belyaev:1994zk}. 
Since the LO twist-3 contribution was found to be large, further improvements were made by calculating the smaller NLO corrections~\cite{Ball:2004ye}. A more recent update where the $\overline{\rm MS}$ mass is used in place of the pole mass for $m_b$ can be found in Ref.~\cite{Duplancic:2008zz,Khodjamirian:2011ub} for the $B\to\pi$ case and in Ref.~\cite{Duplancic:2008tk} for the $B_{s}\to K$ case.
Here we will discuss a selection of the more recent LCSR calculations.

For $B\to\pi$, a NNLO ($O(\alpha_s^2\beta_0)$) calculation of $f_+(0)$ was performed, with the result $f_+(0)= (0.262^{+0.020}_{-0.023})$ with uncertainties $\lesssim 9\%$~\cite{Bharucha:2012wy}. This calculation tested the argument that radiative corrections to $f_+f_B$ and $f_B$ should cancel when both calculated in sum rules (the 2-loop contribution to $f_B$ in QCDSR is sizeable). It was found that despite $\sim 9\%$ $O(\alpha_s^2\beta_0)$ change to $f_B$, the effect on $f_+(0)$ was only $\sim 2\%$.

More recently unitarity bounds and extrapolation were used to perform a Bayesian analysis of the form factor $f_+(q^2)$ for $B\to\pi$~\cite{Imsong:2014oqa}. 
Prior distributions were taken for inputs, a likelihood function was constructed
based on fulfilling  the sum rule for $m_B$ to $1\%$, and posterior distributions were obtained using Bayes' theorem. The posterior distributions of the inputs differed only for $s_0$, which was pushed to higher values $s_0=41\pm4$~GeV (mainly due to the choice of $m_b$).
Finally the results were fit to the BCL parametrisation, finding a central value of $f_+(0) = 0.31\pm 0.02$.
Obtaining $f_+(q^2)$ and the first two derivatives at 0 and 10 ${\rm GeV}^2$ has allowed the extrapolation to
higher $q^2$ using improved unitarity bounds.

$V_{ub}$ can also be obtained from the channels $B \to\rho/\omega$, and updated LCSR results were made available in 2015~\cite{Straub:2015ica}. The improvements in these results include: the computation of full twist-4 (+partial twist-5) 2-particle DA contribution to FFs, plus the determination of certain so-far unknown twist-5 DAs in the asymptotic limit;
a discussion of the non-resonant background for vector meson final states;
the determination and usage of updated hadronic matrix elements, specifically the decay constants; fits with full error correlation matrix for the $z$~expansion coefficients, as well as an interpolation to the most recent lattice computation.
The result for $|V_{ub}|$ from $B\to\rho\ell\nu$ has comparable errors to the $B\to\pi$ determination. In general the $B\to V$ results agree with previous exclusive determinations and global fits within errors.

Future prospects for exclusive $V_{ub}$ from LCSR include extending the subset of NNLO corrections calculated both in $q^2$ and to include all NNLO twist 2 and 3 contributions.
It would also be beneficial to perform a Bayesian uncertainty analysis of all $B \to P$,$ D\to P$ LCSRs (along the lines of the aforementioned analysis for $B \to\pi$~\cite{Imsong:2014oqa}).
Finally the measurement of $B_s\to K\ell\nu$ at LHCb/Belle II will allow an important complementary determination of $V_{ub}$ using results from Ref.~\cite{Khodjamirian:2017fxg}.

\subsection{Measuring $|V_{ub}|$ exclusively and the prospects for Belle II}


The most precise exclusive determinations of $|V_{ub}|$ will ultimately come from the most theoretically clean $b \to u \ell^{-} \bar{\nu}_{\ell}$ modes: $\bar{B}^{0} \to \pi^{+} \ell^{-} \bar{\nu_{\ell}}$, $\bar{B}^{0}_{s} \to K^{+} \ell^{-} \bar{\nu_{\ell}}$ and $\Lambda^{0}_{b} \to p \ell^{-} \bar{\nu_{\ell}}$, which involve ground state hadrons in the final state. The main challenge facing measurements of $|V_{ub}|$ from these modes is the large background from $b\to c \ell^{-} \bar{\nu}_{\ell}$ decays, which is  O($|V_{cb}|^{2} / |V_{ub}|^{2}) \approx 100$ more likely to occur. This background is difficult to separate from signal given the need to partially reconstruct the missing signal neutrino.


Several measurements of exclusive $\bar{B}^{0} \to \pi^{+} \ell^{-} \bar{\nu_{\ell}}$ decays were made at the $B$ factories CLEO, BaBar and Belle. These measurements fall in to two categories of tagged and untagged measurements, which exploit the unique $e^{-}e^{+} \rightarrow \Upsilon(4S) \rightarrow B \bar{B}  $ topology and fully hermetic detector design of the $B$ factories. In tagged measurements~\cite{Sibidanov:2013rkk} the non-Signal $B$ meson in the event is first reconstructed in a number of hadronic modes before selecting the signal pion and lepton. Exploiting the known energies and momenta of the interacting $e^{+}e^{-}$ beams allows for neutrino 4-momentum, $p_{\nu}$ to be reconstructed and the signal to be extracted using the missing mass squared of the neutrino, $M^{2} = p^{2}_{\nu}$. In untagged measurements~\cite{Ha:2010rf,delAmoSanchez:2010af} the signal pion and lepton are first selected with a tight selection to reduce background from $b\to c \ell^{-} \bar{\nu}_{\ell} $ decays. The neutrino is then reconstructed by inclusively reconstructing the other $B$ in the event as a sum of remaining tracks and photons. The beam constrained mass, $M_{bc}$, and beam energy difference~\footnote{Here $M_{bc} = \sqrt{E^{*2}_{beam} - P^{*2}_{B}}$ and $\Delta E = E^{*}_{Beam} - E^{*}_{B}$ where $E^{*}_{beam}$ and $E^{*}_{B}$ are beam and $B$ meson energies in the centre of mass frame.}  are used as fit variables to simultaneously extract the signal. While tagged measurements give a high purity and better $q^{2}$ resolution they suffer from a much lower efficiency resulting from the branching fractions and reconstruction efficiencies for tagged modes. 

In both tagged and un-tagged measurements the exclusive $\bar{B}^{0} \to \pi^{+} \ell^{-} \bar{\nu_{\ell}}$ signal is fitted in bins of $q^{2}$ to determine the partial branching fraction in each  bin. These measurements together with LQCD and LCSR predictions can be used as constraints to simultaneously fit the form factors of decays and determine the parameter $|V_{ub}|$. HFLAV performed a fit for $|V_{ub}|$ the $\bar{B}^{0} \to\pi^{+} \ell^{-} \bar{\nu_{\ell}}$ form factor, $f_{+}(q^{2})$, under a BCL parametrisation utilising BaBar and Belle tagged and untaggged datasets and state of the art theory predictions~\cite{Amhis:2016xyh}. This resulted in the most precise determination of $|V_{ub}|$ to date, $|V_{ub}|=3.67\pm 0.09 (exp) \pm 0.12 (theo)$, which has a total uncertainty of~4\%.

Untagged and tagged measurements of $|V_{ub}|$ from $\bar{B}^{0} \to\pi^{+} \ell^{-} \bar{\nu_{\ell}}$ decays at Belle II will significantly improve the precision on $|V_{ub}|$. In order to project the reduction in uncertainty both tagged and untagged analyses were performed on simulated Belle II Monte Carlo. The expected uncertainty on $|V_{ub}|$ was determined for a given luminosity by extracting the partial branching fractions from pseudo-datasets generated from Monto Carlo expectations and fitting these together with LQCD predictions. With $50$~ab$^{-1}$ and future expected improvements in LQCD predictions the projected uncertainties on $|V_{ub}|$ from $\bar{B}^{0} \to \pi^{+} \ell^{-} \bar{\nu_{\ell}}$ decays were 1.7\% (tagged) and 1.3\% (untagged). The dominant systematic for the tagged analysis is the calibration of the tagging efficiency which is assumed irreducible at $1\%$ on $|V_{ub}|$. For the untagged analysis the dominant systematic uncertainty results from the uncertainty on the number of $B\bar{B}$ pairs which is assumed irreducible at $0.5\%$. Several systematics relating to the branch fractions and form factors of $b \to c \ell^{-} \bar{\nu}_{\ell}$ and $b \to u \ell^{-} \bar{\nu}_{\ell}$ decays are also considered irreducible in the untagged analysis given the lower purity than the tagged analysis.  




\subsection{ Measuring  $|V_{ub}|/|V_{cb}|$ at LHCb}
\vspace{-1mm}

All $b$-hadron species are accessible at hadron colliders thus opening to LHCb a wide possibility of $|V_{ub}|$ measurements from exclusive  
$b \to u$ transitions, while inclusive $|V_{ub}|$  measurements do not seem feasible at the moment. 
In proton-proton collision at high energy  $b\bar{b}$ quark pairs are produced mainly from gluon splitting and hadronize independently, 
as a consequence  $b$-hadrons have a wide continuum momentum spectrum and  the reconstruction of semileptonic decays can not profit of 
 the beam-energy constraints used at B-factories. 
 However, thanks to the large boost acquired by the $b$-hadrons, the direction of the momentum can be well determined 
 by the vector connecting the primary vertex of proton-proton interactions and the $b$-hadron decay vertex. 
 By imposing the $b$-hadron mass constraint, the missing neutrino momentum can be calculated with a two-fold ambiguity. 
 A small fraction of unphysical solutions arises from the imperfect reconstruction of vertices positions.
 The best way to choose between the two solutions depends on the specific decay mode under study.
 The choice can be optimized considering additional variables related to the decay kinematics by using linear regression algorithms~\cite{Ciezarek:2016lqu}.
 
The precise determination of an absolute branching fraction requires the precise knowledge of the total $b$-hadron production rate 
and of the experimental detection efficiency, which includes reconstruction, trigger and final states selection. 
To minimize the experimental uncertainty it is preferred to determine ratios of branching fractions, normalizing the $b$-hadron decay mode 
under study to a well-known $b$-hadron decay mode, that has as similar as possible decay topology. 
Choosing a decay of the same $b$-hadron removes the dependence on the production fraction of the specific $b$-hadron. 

The first determination of $|V_{ub}|$ at LHCb was done with baryons, measuring the branching fractions 
for  $\Lambda_b\to p \mu^- {\overline{\nu}}$ and  $\Lambda_b^0 \to \Lambda_c^+ \mu^- {\overline{\nu}}$ decays~\cite{Aaij:2015bfa}.
What is directly determined is the ratio of the CKM matrix elements
\begin{displaymath}
\frac{|V_{ub}|^2} {|V_{cb}|^2} =\frac { \mathcal{B}(\Lambda_b^0 \to p \mu^- {\overline{\nu}}) } { \mathcal{B}(\Lambda_b^0 \to \Lambda_c^+ \mu^- {\overline{\nu}}) }\times R_{FF}
\end{displaymath}
where $R_{FF}$ is the ratio of the relevant form factors, calculated using LQCD. 
The ratio represents a band in the $|V_{ub}|$ versus $|V_{cb}|$ plane and can be converted into a measurement 
of $|V_{ub}|$ using existing measurements of $|V_{cb}|$.
Approximately 10\% of $b$-hadrons produced at LHC are  $\Lambda_b$ and 
a clean signal identification is possible imposing stringent proton identification requirements. 
The large background from $b$-hadron decays with additional charged tracks in the decay products is strongly reduced 
employing isolation criteria by means of multivariate machine-learning algorithms.
The signal yields are determined from a $\chi^2$  fit to the B corrected mass distributions of 
$\Lambda_b^0 \to p \mu^- {\overline{\nu}}$ and  $\Lambda_b^0 \to \Lambda_c^+ \mu^- {\overline{\nu}}$
 candidates. 
The corrected mass is defined as
$m_{corr}=\sqrt{ m_{h\mu}^2+p_{\perp}^2} +p_{\perp}$ 
where $p_{\perp}$ is the momentum of the hadron-$\mu$ pair transverse to the $\Lambda_b^0$ flight direction.

The LQCD form-factors that are used in the calculation of $|V_{ub}|$~\cite{Detmold:2015aaa} are most precise in the 
kinematic region  where $q^2$, the invariant mass squared of the leptonic system, is high.
When the branching fractions of the $b \to u$ ($b \to c$) decays are integrated in the region $q^2>15 (7) \text{GeV}^2$ the              
theory uncertainty on ${|V_{ub}|}/ {|V_{cb}|}$ is 4.9\%. 
This measurements, performed with Run 1 data,  gives ${|V_{ub}|}/ {|V_{cb}|} =0.83\pm0.004$ (stat) $\pm0.004$ (syst), 
consistent with previous exclusive measurements of the two CKM matrix elements.

A new  measurement of this type is currently under study at LHCb.  It uses 
$B_s^0 \to K^+\mu^- {\overline{\nu}}$  decays whose branching fraction is predicted to be of the same order of magnitude of the
$\Lambda_b^0 \to p \mu^- {\overline{\nu}}$ one.

The  signal selection is challenging due to the large background from partially reconstructed decays of all species of $b$-hadrons,
but it can exploit the good efficiency and purity of kaon and muon identification provided by the LHCb detector,
the separation of the $K\mu$ vertex from primary vertex and the already mentioned isolation tools. 
The chosen normalization mode $B_s^0 \to D_s^+\mu^- {\overline{\nu}}$, $D_s^+ \to K^- K^+\pi^+$
benefits of small uncertainty in the $D_s^+$ branching fraction. 
The good identification of this decay mode, despite the large feed-down from $B^0_s$ decays to excited $D_s$ mesons 
with un-reconstructed neutral particles, has been proven to be possible at LHCb with the 
measurement of $B_s^0$  lifetime~\cite{Aaij:2017vqj}.

Form factors for the $B_s^0$ mesons decays to $K$ and $D_s$ have been calculated with LQCD by several groups~\cite{Bouchard:2014ypa,Flynn:2015mha}.
The calculation are performed in the high $q^2$ region and extrapolated to the full region with BGL or BCL z-expansions.
Different calculations agree at high $q^2$,  but there is currently a disagreement in the $q^2=0$ extrapolated value.
For $B_s^0 \to K^+\mu^- {\overline{\nu}}$ in the low $q^2$ region (up to 12~GeV$^2$)  form factors calculated with LCSR 
are also available~\cite{Khodjamirian:2017fxg}.
The uncertainties on the experimental measurement of the $B_s^0 \to K^+\mu^- {\overline{\nu}}$ yield increase at high $q^2$ 
(low kaon momentum) due to the reduced  efficiency and the larger background contamination. 
It is foreseen to perform the measurement in few $q^2$ bins so that the use of different calculations of form factors will be possible.
Larger data samples, accumulated during the LHCb Upgrade period will allow a  differential  measurement in finer $q^2$ bins.

Purely leptonic $B^-\to \mu^- {\bar{\nu}}$  decays are not accessible at LHCb. An alternate way has been tested, searching for the
decay  $B^-\to \mu^- {\bar{\nu}} \mu^+ \mu^-$ where an hard photon is irradiated from the initial state and materializes into two muons.
This decay has the experimental advantage of the presence of additional particles in the final state and of a larger branching fraction, 
due to the removal of the helicity suppression. An upper limit on the branching fraction of 
$1.6 \times 10^{-8}$ has been determined with $4.7~\text{fb}^{-1}$ of integrated luminosity~\cite{Aaij:2018pKa}, making it a possible candidate for a $|V_{ub}|$ measurement 
in the LHCb Upgrade period~\cite{LHCbUpgrade2}.

\subsection{Related issues}

\subsubsection{$R_\pi$}

The experimental signature of $B \to \pi \tau \nu_\tau$ is challenging: low in rate due to CKM suppression, this final state can only be isolated from backgrounds using multivariate analysis techniques. Due to the pseudoscalar nature of the pion in the final state, an increased sensitivity to certain new physics models involving scalar exchange particles is expected and measurements of this branching fraction offer an orthogonal path to probe the anomalies observed in $R(D)$ and $R(D^*)$. The first limit on the branching fraction using leptonic and one-prong $\tau$ decay modes was reported by Ref.~\cite{Hamer:2015jsa}. They reported 
\begin{equation}
    \mathcal{B}(B^0\to  \pi^- \, \tau^+ \, \nu_\tau) < 2.8 \times 10^{-4} \quad \text{at 95\% CL} \, ,
\end{equation}
using a frequentist method. This result can be converted into a value of $R_\pi = \Gamma(B^0 \to \pi^- \, \tau^+ \, \nu_\tau)/\Gamma(B^0 \to \pi^- \, \ell^+ \, \nu_\tau)$ with $\ell = e,\mu$ of
\begin{equation}\label{eq:rpiMeas}
    R_\pi = 1.05 \pm 0.51 \, ,
\end{equation}
which can in turn be compared to the SM prediction of Refs.~\cite{Lattice:2015tia,Bernlochner:2015mya} of
\begin{equation}
    R_\pi = 0.641 \pm 0.016 \, .
\end{equation}
Although the current precision is very limited, this result can already exclude the model parameter space of new physics models e.g. charged Higgs bosons, cf.\ Ref.~\cite{Bernlochner:2015mya}. Albeit a challenging signature, the final state with a charged pion has excellent prospects to be discovered in the large future Belle II data set. A naive extrapolation of Eq.~\ref{eq:rpiMeas} using assuming SM couplings results in evidence with 4~ab${}^{-1}$ and discovery with 11~ab${}^{-1}$ of integrated luminosity. The theoretical precision in $R_\pi$ will further increase with progress in lattice and with combined light lepton and lattice fits (the measured spectra can constrain the low $q^2$ region, which the lattice has difficulties in predicting reliably).

\subsubsection{ Experimental status and prospects of $B\to\ell\nu_\ell\gamma$}

The experimental study of $B\to\ell\nu_\ell\gamma$ with $\ell = e, \mu$ is challenging and requires the clean laboratory of an $e^+ \, e^-$ machinery: in such a setting the known initial state and the full reconstruction of the second $B$-meson produced in the collision provide the necessary constraint to successfully identify this signature. In addition, to not be overwhelmed with background, only photons at high energies ( $\approx 1$~GeV or larger) can be studied this way. The difficulties lie in the low efficiency of the reconstruction performance of the second $B$-meson, which have to happen in low branching fraction hadronic modes, and the still sizeable cross-feed from $B \to \pi^0 \, \ell \bar \nu_\ell$ and $B \to \eta \, \ell \bar \nu_\ell$ decays. These two semileptonic processes produce very similar final states, namely $B\to\ell\nu_\ell\gamma\gamma$, but can be reduced by looking for a unassigned second  high-energetic photon in the collision event under study. To separate $B\to\ell\nu_\ell\gamma$ from such decays successfully a fit to
\begin{eqnarray}
    m_\nu^2 \simeq m_{\rm miss}^2 = \left( p_{B_{\rm sig}} - p_\ell - p_\gamma \right)^2 
\end{eqnarray}
can be carried out. Here $p_\ell$ and $p_\gamma$ denote the reconstructed four-vectors of the visible final states of $B\to\ell\nu_\ell\gamma$. The four-vector of the decaying signal $B$-meson, $p_{B_{\rm sig}}$, can be reconstructed using the information from the reconstructed tag-side $B$-meson. Correctly reconstructed signal decays peak at $m_\nu^2 \approx 0$~GeV${}^2$, whereas the dominant semileptonic decays are shifted to higher values due to the absence of the additional photon in the four-vector sum. The sensitivity can be further increased by explicitly reconstructing the semileptonic backgrounds and combine this information into a global analysis. This was the strategy pursuit by Ref.~\cite{Gelb:2018end}, which constrained the $\pi^0$ semileptonic background this way. The current experimental limit with a lower photon energy cut of $1$~GeV is
\begin{equation}\label{eq:dBFlnug}
 \Delta \mathcal{B}(B \to \ell \nu_\ell \gamma ) < 3.0 \times 10^{-6} \quad \text{at 95\% CL} \, .
\end{equation}
The above limit was determined using a flat Bayesian prior. 

\begin{figure}[t]
	\begin{center}
			\includegraphics[width=0.7\textwidth]{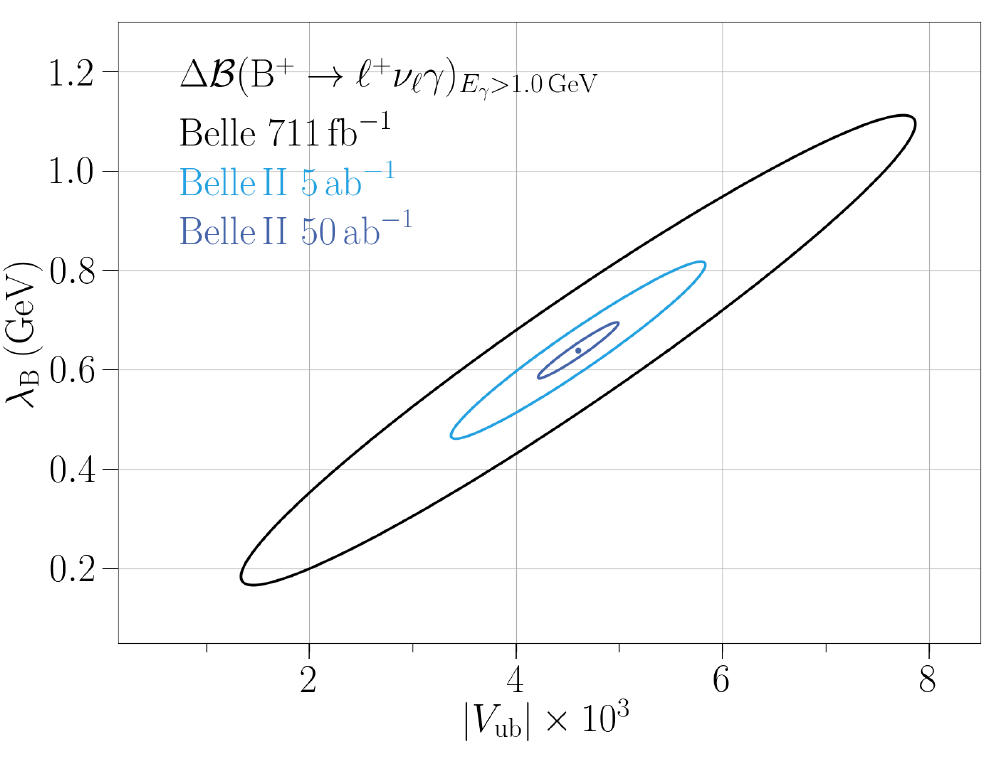}
	\end{center}
	\caption{Projection of the extraction of $\lambda_B$ and $\left| V_{ub} \right|$ for the expected Belle II data sets. The ellipses correspond to the expected uncertainty. The figure is from Ref.\cite{Gelb:49050}.}
	\label{fig:lnugamma_lb_vub_belle2_prospects}
\end{figure}

The discovery prospects for this decay at Belle II are excellent: the improved tracking capabilities, better calorimeter electronics, and the continuous development of modern tagging algorithms such as Ref.~\cite{Keck:2018lcd} will help improving the sensitivity. Extrapolating from the central value and uncertainty of the currently most precise limit of Eq.~\ref{eq:dBFlnug} of $\Delta \mathcal{B}(B \to \ell \nu_\ell \gamma ) = \left(1.4 \pm 1.1 \right) \times 10^{-6}$, evidence should be possible with 5~ab${}^{-1}$ and a discovery is possible with 50~ab${}^{-1}$~\cite{Gelb:49050}. In principle, after discovery the value of $\left| V_{ub} \right|$ could be extracted from this decay as well, along with the first inverse momentum of the light-cone distribution amplitude, $\lambda_B$. An extrapolation from the current sensitivity is shown in Figure~\ref{fig:lnugamma_lb_vub_belle2_prospects}, based on the numbers from Ref.~\cite{Gelb:49050}. The sensitivity for $\left| V_{ub} \right|$ will not be competitive with other methods (leptonic and semileptonic), but the achievable precision on $\lambda_B$ will help measurements and interpretations, which rely on our understanding of the light-cone distribution amplitude properties. 

\subsubsection{ Theoretical progress for $B\to \gamma \ell\nu_\ell$}

The photoleptonic decay $B\to \gamma \ell\nu_\ell$  
determined by two independent form factors 
is the simplest probe of the $B$-meson light-cone distribution amplitudes (LCDAs), 
which represent one of the most important inputs in the theory of semileptonic and 
nonleptonic $B$-decays based on QCD factorization and LCSRs. The  
calculation of the form factors in HQET  and at large photon recoil in the leading power
is well developed and can be found in Ref.~\cite{Beneke:2011nf}. The 
$1/m_b$ and $1/E_\gamma$  power suppresed effects,  expressed in a form
of the soft overlap part of the form factors, were quantified using 
a technique~\cite{Braun:2012kp} based on dispersion relations and quark-hadron duality
(see also Ref.~\cite{Wang:2016qii}). The most advanced calculation of the $B\to \gamma \ell\nu_\ell$ form factors,
including power suppressed terms, was done recently~\cite{Beneke:2018wjp} resulting in the prediction of the decay branching 
fraction at $E_\gamma >1.0$~GeV as a function of the key unknown 
theoretical quantity: the inverse moment $\lambda_B$  of the B-meson LCDA.
An alternative approach~\cite{Wang:2018wfj} calculates the power-suppressed corrections due to photon emission at long distances
in terms of the photon LCDAs in the LCSR framework. 
The proof of concept for a lattice QCD calculation of radiative leptonic decays was recently done in~\cite{Kane:2019jtj},
see also~\cite{deDivitiis:2019uzm}.

\subsubsection{$B\to \pi\pi \ell\nu_\ell$ decay beyond $\rho$} 
Calculations of $B\to \rho$ form factors both in lattice QCD and from LCSRs usually
adopt a narrow $\rho$ approximation and by default ignore the influence of nonresonant
effects (radially excited $\rho$'s) in the mass interval around  $\rho$.
The role of these effects has to be assessed at a quantitative level.
In Refs.~\cite{Hambrock:2015aor,Cheng:2017sfk} the first
attempt to calculate more general $B\to\pi\pi$ form factors from LCSRs, using
two-pion LCDAs at low mass of dipion system and at large recoil, was undertaken.
The currently limited knowledge of these nonperturbative inputs 
calls for their further development and also for alternative methods.
In Ref.~\cite{Cheng:2017smj} a different version  of LCSRs with $B$ meson LCDAs was obtained
which predicts the convolutions of the 
$\bar{B}^0\to \pi^+\pi^0$ form factors in $P$ wave with the timelike pion 
form factor. In the narrow $\rho$-meson limit
these sum rules reproduce analytically the known LCSRs for $B\to \rho$ form factors. Using data for the pion vector form factor from $\tau$ decay, the  finite-width effects and the contribution of excited $\rho$-resonances
to the $B\to\pi\pi$ form factors were found to amount up to $\sim 20\%$ 
in the small dipion  mass region where they can be interpreted as a nonresonant ($P$-wave) background to the $B\to\rho$ transition. For a more general analysis
of $B\to \pi\pi\ell\nu_\ell$ decays see e.g.\ Refs.~\cite{Faller:2013dwa,Kang:2013jaa}.

\subsubsection{Remarks on the $z$ expansion} 
\label{sec:z-remarks}
The use of the so-called $z$ expansion for form factors has become a standard practice for semileptonic decays, see
Refs.~\cite{Boyd:1997qw,Hill:2006ub} for a pedagogical discussion.
In the workshop several issues concerning it were discussed, in particular its application to baryon form factors.  

Form factors which parametrize matrix elements of the form $\langle L|J|H\rangle$ have known analytic structure.
In particular, they are analytic in the complex $t=q^2$ plane outside a cut  on the real axis.
The cut starts at  some positive $t_\text{cut}$ equals to the invariant mass squared of the lightest state the current $J$ can produce.  The domain of analyticity can be mapped onto the unit circle via  $z=\left(\sqrt{t_\text{cut}-t}-\sqrt{t_\text{cut}-t_0}\right)/\left(\sqrt{t_\text{cut}-t}+\sqrt{t_\text{cut}-t_0}\right)$, where $t_0$ is a free parameter denoting the point that is mapped to $z=0$. The form factor can be expanded as a Taylor series in $z$ which is a model-independent parametrization. For heavy-to-light form factors the maximum value of $z$ is related to the distance between $(m_H-m_L)^2$ and $t_\text{cut}$. As a result, increasing $t_\text{cut}$ decreases the maximum value of $z$  leading to a faster convergence of the series. 

Naively one might assume that the lightest state is the two-particle state $\bar H L$. This would imply that $t_\text{cut}=(m_H+m_L)^2$, but this is not the case in general. For example, for the proton electric and magnetic  form factors $(H=L=p)$ the cut starts at the two-pion threshold and not at the $p\bar p$ threshold. As another example, for one of the $B\to \pi$ form factors ($f_+)$ the cut starts at $m^2_{B^*}$.  Since this is a simple pole, it can be easily ``removed" by considering $(t-m_{B^*}) f_+$ as a Taylor series in $z$. For $(t-m_{B^*}) f_+$  the cut starts at $(m_B+m_\pi)^2$. If one uses a higher value of $t_\text{cut}$ than the physical one, one faces the danger of trying to expand the form factor in a region where it is not analytic. One of the immediate results of the workshop was the identification of such a problem in the literature. For baryon form factors, e.g. $\Lambda_B\to p$, analyses have used the wrong value of $t_\text{cut}=(m_{\Lambda_B}+m_p)^2$, see Ref.~\cite{Khodjamirian:2011jp} and arXiv.org version 2 of Ref.~\cite{Detmold:2015aaa}. In fact, $t_\text{cut}$ for the baryon form factors is the same as for the meson form factors of analogous decays. 

Another issue discussed in the workshop is the use (or lack of use) of  bounds on the coefficients of the $z$ expansion.  Although the form factor is expressed as an infinite series, in practice the series is truncated after a few terms.  One would like to ensure that the value of a physical parameter such as $|V_{ub}|$ is independent of the number of parameters used, by bounding  the coefficients. For example,  one can use a unitarity bound~\cite{Bourrely:1980gp} or a bound from the heavy quark expansion~\cite{Becher:2005bg}. It seems that currently there is no consistent use of bounds in the extraction of $|V_{ub}|$. As the analysis \cite{Hill:2010yb} shows, this can be a problem as the data improve and the number of necessary parameters increases. This can be especially problematic  if one needs to use the $z$-expansion for extrapolation.  The community needs to be aware of this issue and at least test that results do not change if bounds are applied to the coefficients.  

The unitarity bounds for meson decays such as $B\to \pi$ rely on the fact that for $(t-m_{B^*}) f_+$  the cut starts at the  $(m_B+m_\pi)^2$. For baryon decays  such as  $\Lambda_B\to p$, unitarity can only constrain the region above $(m_{\Lambda_B}+m_p)^2$. The region between $(m_B+m_\pi)^2$ and  $(m_{\Lambda_B}+m_p)^2$ is left unconstrained. Following the analysis of Ref.~\cite{Hill:2010yb}, one might worry that the contribution of the latter region is the dominant one. While considering together mesons and baryons contributions to the dispersive bounds might overcome the problem \cite{Cohen:2019zev},  further study is warranted.


\section{Quark masses and leptonic decays}
\label{dc_qm}

\subsection{Quark masses}
\label{sec:qm}

\newcommand{\fpiPDG}{\ensuremath{f_{\pi, \rm PDG}}}
\newcommand{\MeV}{\text{MeV}}

In the Standard Model (and many extensions), quark masses and the CKM matrix all stem from Higgs-Yukawa couplings between the quark
fields and the Higgs doublet.
It is therefore natural to consider the bottom-quark mass, $m_b$, in this report.
As discussed in Sec.~\ref{h2h_incl}, $m_b$ can be extracted from the inclusive semileptonic $B$ decay distributions, along
with~$|V_{cb}|$.
In the theory of inclusive decays, the charm-quark mass, $m_c$, is also needed to control an infrared sensitivity; see
Sec.~\ref{h2h_incl}.

\begin{figure}[b]
\centering
    \includegraphics[width=0.48\textwidth]{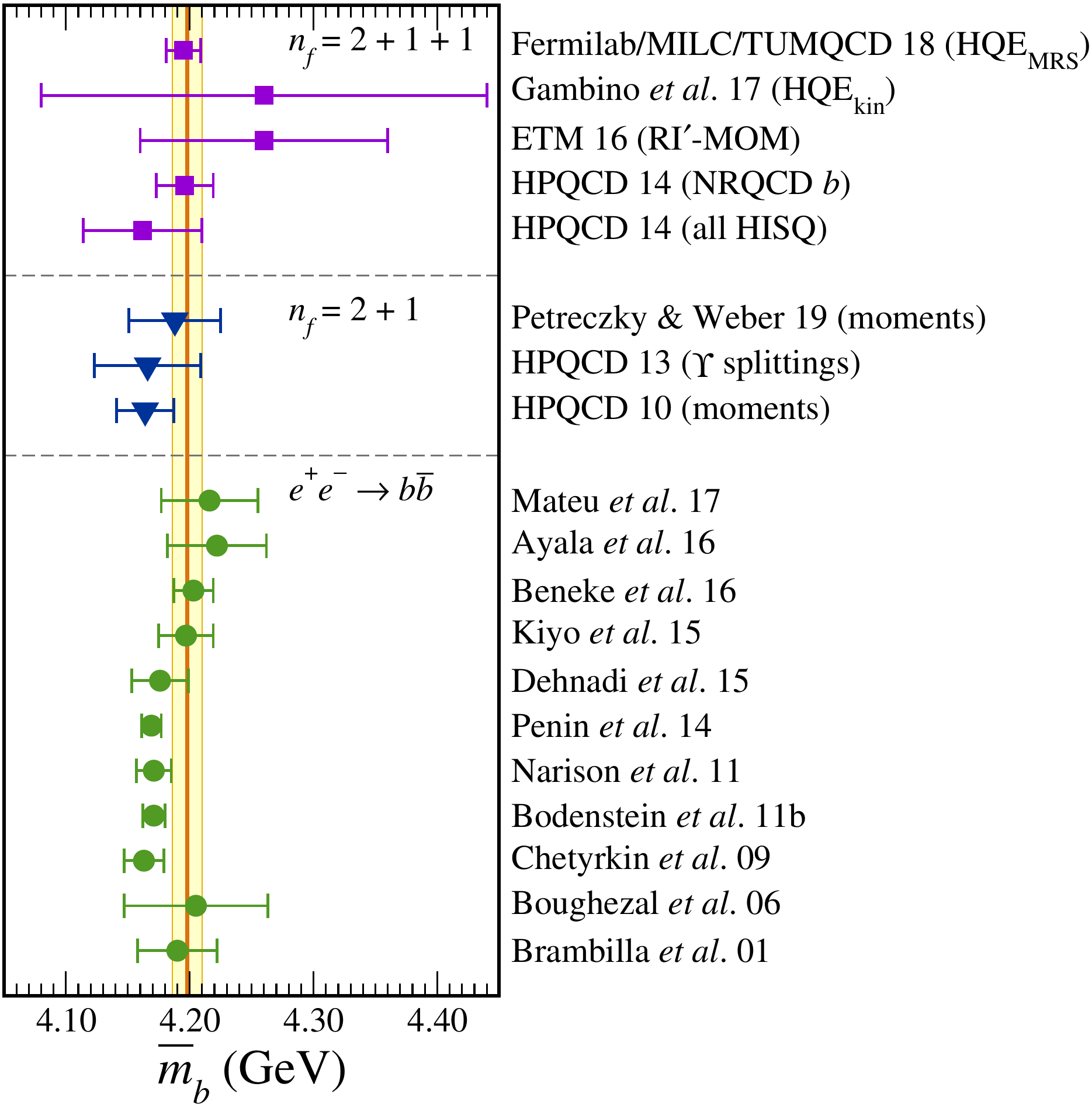} \hfill
    \includegraphics[width=0.48\textwidth]{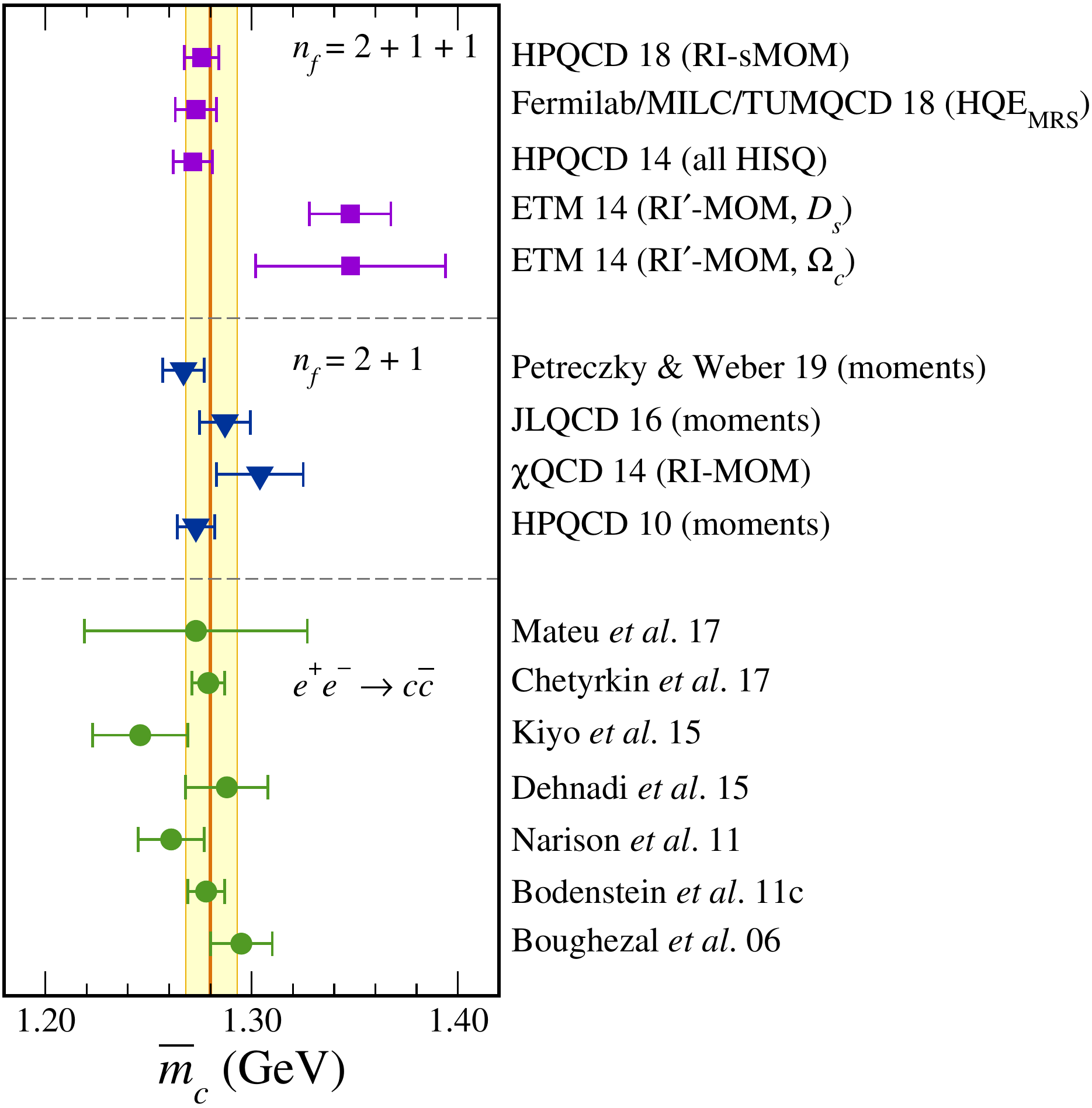}
    \caption{Comparison of results for the bottom-quark mass $\bar{m}_b=m_{b,\MSbar}(m_{b,\MSbar})$ (left) and the charm-quark mass 
    $\bar{m}_c=m_{c,\MSbar}(m_{c,\MSbar})$ (right).
    Squares denote lattice-QCD calculations with $2+1+1$ flavors of sea quark~\cite{Lytle:2018evc,Bazavov:2018omf,Gambino:2017vkx,%
    Bussone:2016iua,Colquhoun:2014ica,Chakraborty:2014aca,Alexandrou:2014sha,Carrasco:2014cwa}; 
    triangles denote lattice-QCD calculations with $2+1$ flavors of sea quark~\cite{Petreczky:2019ozv,Nakayama:2016atf,%
    Yang:2014sea,Lee:2013mla,McNeile:2010ji};
    circles denote results extracted from $e^+e^-$ collisions near $Q\bar{Q}$ threshold~\cite{Mateu:2017hlz,Chetyrkin:2017lif,%
    Ayala:2016sdn,Beneke:2016oox,Kiyo:2015ufa,Dehnadi:2015fra,Penin:2014zaa,Narison:2011xe,Bodenstein:2011fv,Bodenstein:2011ma,%
    Chetyrkin:2009fv,Boughezal:2006px,Brambilla:2001qk}.
    The vertical band shows the FLAG~2019 average for $2+1+1$ sea flavors.
    Note that $2+1$-flavor calculations are in rough (good) agreement for bottom (charm).}
    \label{fig:qm}
\end{figure}

Figure~\ref{fig:qm} compares results from lattice QCD with realistic sea content of $n_f=2+1+1$ or $2+1$ sea quarks with the 
FLAG~2019~\cite{Aoki:2019cca} average for the $2+1+1$ sea.
The average for $\bar{m}_b$ is dominated by the very precise result from the Fermilab Lattice, MILC, and TUMQCD Collaborations,
while that for $\bar{m}_c$ is dominated by the corresponding Fermilab/MILC/TUMQCD result together with two separate results from the
HPQCD Collaboration.
The FLAG~2019~\cite{Aoki:2019cca} averages (for $2+1+1$ sea flavors) are
\begin{eqnarray}
    \bar{m}_b = m_{b,\MSbar}(m_{b,\MSbar}) &=& 4.198(12)~\text{GeV} , \\
    \bar{m}_c = m_{c,\MSbar}(m_{c,\MSbar}) &=& 1.280(13)~\text{GeV} ,
\end{eqnarray}
based on Refs.~\cite{Bazavov:2018omf,Gambino:2017vkx,Bussone:2016iua,Colquhoun:2014ica,Chakraborty:2014aca} 
and~\cite{Lytle:2018evc,Bazavov:2018omf,Chakraborty:2014aca,Alexandrou:2014sha,Carrasco:2014cwa}, respectively.
Another recent review~\cite{Komijani:2020kst} finds averages with somewhat smaller uncertainties
\begin{eqnarray}
    \bar{m}_b = m_{b,\MSbar}(m_{b,\MSbar}) &=& 4.188(10)~\text{GeV} , \\
    \bar{m}_c = m_{c,\MSbar}(m_{c,\MSbar}) &=& 1.2735(35)~\text{GeV} ,
\end{eqnarray}
based on the same original sources.
In the case of $\bar{m}_c$, two results~\cite{Alexandrou:2014sha,Carrasco:2014cwa} agree poorly with the others, increasing
$\chi^2/\text{dof}$ of the average by a factor of around~5.
FLAG~2019~\cite{Aoki:2019cca} stretches the error bar by $\sqrt{\chi^2/\text{dof}}$, while Ref.~\cite{Komijani:2020kst} discards 
them: the resulting error bar is smaller because of the compatibility of the inputs as well as the lack of stretching.
Other differences in averaging methodology are quantitatively unimportant.
In either case, the quoted averages are much more precise than those in the~PDG.

A last remark is that the most precise results~\cite{Lytle:2018evc,Bazavov:2018omf,Chakraborty:2014aca} all use the very high
statistics MILC HISQ ensembles with staggered fermions for the sea quarks~\cite{Bazavov:2012xda,Bazavov:2017lyh}.
In the future, other groups~\cite{Baron:2010bv,Baron:2010th,Aoki:2010dy,Boyle:2017jwu} will have to collect similar statistics to
enable a complete cross check.


Four distinct methods are used in the results shown in Fig.~\ref{fig:qm}:
1)~converting the bare lattice mass to the \MSbar\ scheme,
2)~fitting to a formula for the heavy-light hadron mass in the heavy-quark expansion~\cite{Kronfeld:2000ck,Kronfeld:2000gk}, and
3)~computing moments of quarkonium correlation functions~\cite{Bochkarev:1995ai,Allison:2008xk}.%
\footnote{Lattice methods with no results in Fig.~\ref{fig:qm} are not discussed here.} The first two require an intermediate
renormalization scheme that can be defined for any ultraviolet regulator: quark masses defined this way can be computed with lattice
gauge theory or dimensional regularization.
For example, HPQCD~13 ($\Upsilon$ decays)~\cite{Lee:2013mla} uses two-loop lattice perturbation theory to convert the bare NRQCD
mass to the pole mass~\cite{Hart:2004bd,Hart:2009nr}, and dimensional regularization to convert the pole mass into the \MSbar\ mass.

Instead of the pole mass, one can use a regularization-independent momentum-subtracted mass~\cite{Martinelli:1994ty}.
Like the \MSbar\ scheme these RI-MOM schemes are mass-independ\-ent renormalization schemes, but they depend on the gauge.
In lattice gauge theory, Landau gauge is easily obtained on each gauge-field configuration via a minimization
procedure~\cite{Davies:1987vs}.
The mass renormalization factor, $Z_m$, can be computed from the three point function for the scalar or pseudoscalar density,
because $Z_m^{-1}=Z_S=Z_P$ (up to technical details for Wilson fermions).
For example, the matrix element $\langle p'|P|p\rangle$, between gauge-fixed quark states, can be used to define $Z_P$ using the
same formulas for lattice gauge theory as for continuum gauge theory (with dimensional regularization)~\cite{Martinelli:1994ty}.
The schemes labeled RI-MOM and RI$'$-MOM use $p'=p$ and slightly different definitions of the quark-field normalization~$Z_2$;
for a review see Ref.~\cite{Aoki:2010yq}.
The momentum transfer $q\equiv p'-p=0$ here, namely it is ``exceptional'' in the sense of Weinberg's theorem~\cite{Weinberg:1959nj}.
On the other hand, the RI-sMOM scheme~\cite{Sturm:2009kb} chooses $p'$ and $p$ such that $p^2=q^2=p^{\prime2}\equiv\mu^2$.
Without the exceptional momentum, the extraction of $Z_P$ is more robust.
It would be interesting to see whether RI-sMOM on the ETM 2+1+1 ensembles yields $\bar{m}_b$ favoring the RI$'$-MOM results or the
RI-sMOM results on MILC's ensembles.

The HQE method starts with the HQE formula for a heavy-light hadron mass~\cite{Falk:1992wt,Bigi:1994ga},
\begin{equation}
    M = m + \bar{\Lambda} + \frac{\mu_\pi^2}{2m} - d_J\frac{\mu_G^2(m)}{2m} + \cdots,
    \label{eq:hqe}
\end{equation}
where $M$ is the hadron mass, which is computed in lattice QCD as a function of the quark mass, $m$, and $d_J$ depends on the spin
of the hadron.
The quantities can be identified with the energy of gluons and light quarks, $\bar{\Lambda}$, the Fermi motion of the heavy quark,
$\mu_\pi^2$, and the hyperfine splitting, $\mu_G^2$.
($\mu_G^2$ depends logarithmically on~$m$.) Although this idea is not new~\cite{Kronfeld:2000ck,Kronfeld:2000gk}, to be precise one
has to confront the definition of~$m$.
Although the pole mass is natural in the context of the HQE, it is not suitable in practice, because of its infrared sensitivity.
The $\MSbar$ mass, on the other hand, breaks the power counting: $m_\text{pole}-m_{\MSbar}\propto \alpha_sm_\text{pole}$.
Instead, one chooses mass definitions that, in some sense, lie in between these two choices.
Gambino \emph{et al.}~\cite{Gambino:2017vkx} choose the kinetic mass~\cite{Bigi:1996si}, while
Fermilab/MILC/TUMQCD~\cite{Bazavov:2018omf} choose the minimal renormalon subtracted (MRS) mass~\cite{Brambilla:2017hcq}.
After extracting $m_\text{kin}$ or $m_\text{MRS}$ from fitting Eq.~(\ref{eq:hqe}), the result can be converted to the \MSbar\ scheme
with three- and four-loop perturbation theory, respectively.
In addition to the different matching, the error bar from Fermilab/MILC/TUMQCD is so small because it is based on the largest data
set of all calculations in Fig.~\ref{fig:qm}.
See Sec.~\ref{sec:HQE-LQCD} for further discussion and results for $\bar\Lambda$, $\mu_\pi^2$, $\mu_G^2$, and higher-dimension
corrections to the HQE.

One can avoid an intermediate scheme by computing a short-distance quantity in lattice QCD, taking the continuum limit, and 
analyzing the result with \MSbar\ perturbation theory.
For example, on can compute moments of quarkonium correlation functions~\cite{Bochkarev:1995ai,Allison:2008xk},
\begin{eqnarray}
    G_\Gamma^{(n)}  &=& \sum_t t^n G_\Gamma(t), \\
    G_\Gamma(t) &=& c_\Gamma \sum_{\bm{x}} \langle \bar{Q}\Gamma Q(\bm{x},t)\,\bar{Q}\Gamma Q(\bm{0},0) \rangle,
\end{eqnarray}
for some Dirac matrix $\Gamma$.
In lattice gauge theory, the pseudoscalar density needs no renormalization if $\Gamma=\gamma^5$ and $c_{\gamma^5}=m_Q^2$.
The moments $G_\Gamma^{(n)}$ are physical observables with a good continuum limit, which is proportional to $m_Q$ to the appropriate
power, multiplied by a dimensionless function of $\alpha_s(m_Q)$.
Thus, these moments also yield determinations of the strong coupling as well as quark masses.
In Fig.~\ref{fig:qm}, results obtained in this way are labeled ``moments''.

The same moments $G_\Gamma^{(n)}$ can be obtained from the cross section for $e^+e^-$ annihilation into ${Q\bar{Q}}$ hadrons via a
suitably subtracted dispersion relation.
In this case, $\Gamma=\gamma^\mu$ for the electromagnetic current, and $c_{\gamma^\mu}=1$ because the electromagnetic current is
conserved.
Thus, the same perturbative calculations (only changing $\Gamma$) can be used to extract the bottom- and charm-quark masses and
$\alpha_s$ from experimental measurements.
The dispersion relation, related sum rules, and the perturbative series for the moments are the basis of the result labeled
$e^+e^-\to{}b\bar{b}$ and $e^+e^-\to{}c\bar{c}$ in Fig.~\ref{fig:qm}.
The order $\alpha_s^p$, $p=1$, $2$, $3$, became available in 1993~\cite{Broadhurst:1993mw}, 1997~\cite{Chetyrkin:1997mb}, and
2006~\cite{Chetyrkin:2006xg,Boughezal:2006px}, respectively.

\subsection{Leptonic decays}
\label{sec:dc}

Instead of semileptonic decays, CKM matrix elements can also be determined from purely leptonic decays.
For example, a goal of Belle~II is to improve the determination of $V_{ub}$ from $B^+\to\tau^+\nu$,
as well as $V_{cd}$ from $D^+\to\ell^+\nu$ and $V_{cs}$ from $D_s^+\to\ell^+\nu$, and a goal of LHCb is to observe $B_c\to\tau\nu$.
The rates for leptonic decays suffer a helicity suppression, making tauonic and muonic decays preferred experimentally.
Leptonic decays are mediated by the axial-vector part of the electroweak current, as well as possible pseudoscalar currents, so 
they complement semileptonic decays in this way.

The hadronic quantity describing the decay is known as the \emph{decay constant}, defined by
\begin{equation}
    \langle0|\bar{b}\gamma^\mu\gamma^5u|B^+(p)\rangle = ip^\mu f_{B^+},
    \label{eq:AfB}
\end{equation}
where $p^\mu$ is the four-momentum of the $B$ meson and $f_{B^+}$ is the decay constant.
For other mesons, the axial currents and notation change in obvious ways.
From the partial conservation of the flavor-nonsinglet axial current, the pseudoscalar density can also be used to compute the 
decay constant:
\begin{equation}
    (m_b+m_u)\langle0|\bar{b}\gamma^5u|B^+(p)\rangle = M_{B^+}^2 f_{B^+},
    \label{eq:mPfB}
\end{equation}
where $m_b$ and $m_u$ are bare quark masses.

Equations~(\ref{eq:AfB}) and (\ref{eq:mPfB}) are the basis of lattice-QCD calculations.
In general, the axial current used is not a Noether current, so it is not absolutely normalized.
Fermion formulations with good chiral symmetry (staggered, overlap, domain wall) provide an absolutely normalized pseudoscalar 
density.
Until recently, however, lattice spacings have not been small enough to use these approaches for the $b$~quark.
Methods developed especially for heavy quarks have therefore been used, and they do not provide any absolutely normalized 
$\bar{b}\Gamma u$ bilinears.

\begin{figure}
    \centering
    \includegraphics[width=0.8\textwidth]{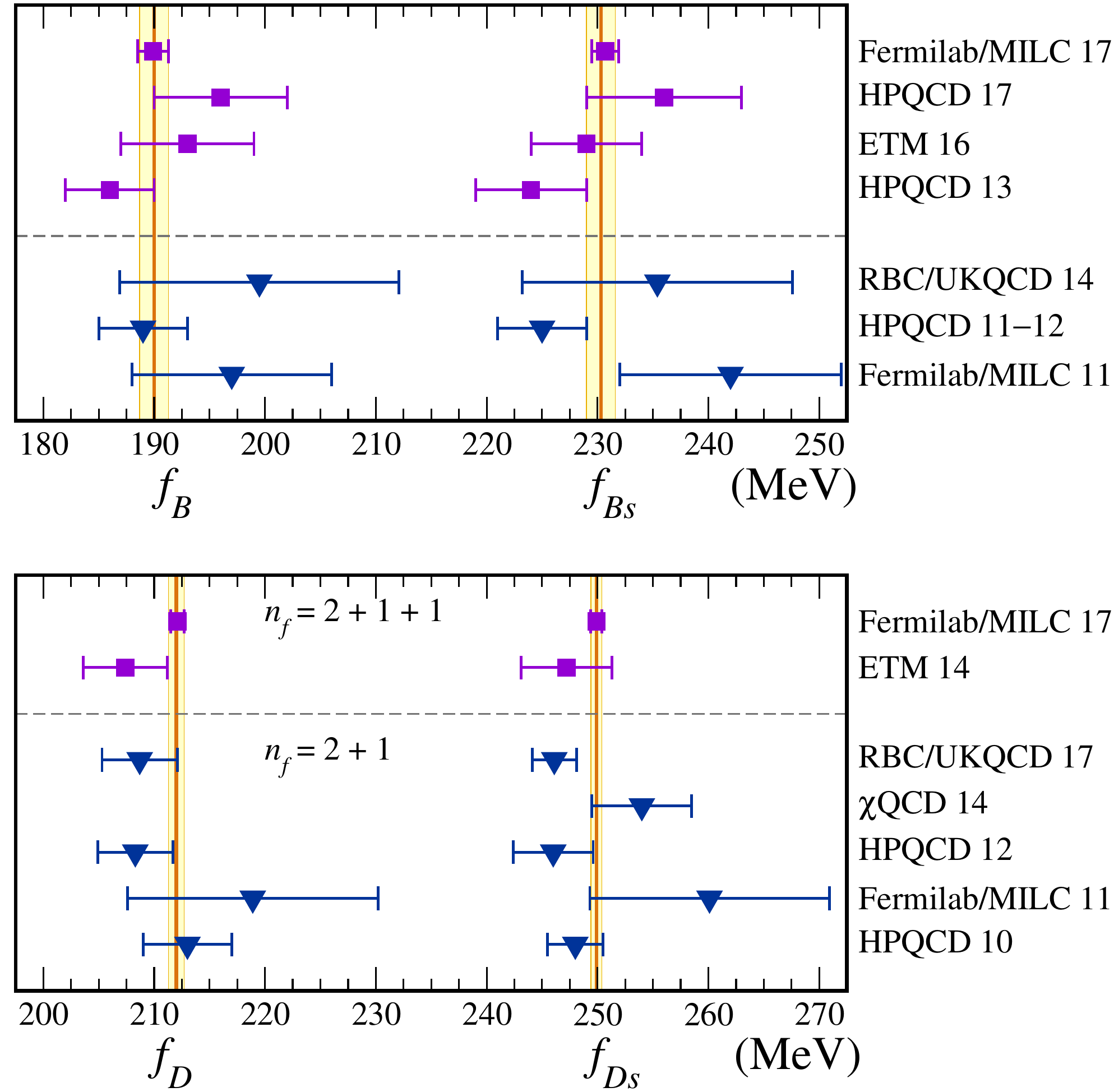}
    \caption{Comparison of results for the $B_{(s)}$-meson (top) and the $D_{(s)}$-meson (bottom) decay constants.
    Squares denote lattice-QCD calculations with $2+1+1$ flavors of sea quark~\cite{Bazavov:2017lyh,Hughes:2017spc,Bussone:2016iua,%
    Carrasco:2014poa,Dowdall:2013tga};
    triangles denote lattice-QCD calculations with $2+1$ flavors of sea quark~\cite{Boyle:2017jwu,Christ:2014uea,Yang:2014sea,%
    Na:2012iu,Na:2012kp,McNeile:2011ng,Bazavov:2011aa,Davies:2010ip}.
    The vertical bands show the FLAG~2019 average for $2+1+1$ sea flavors~\cite{Aoki:2019cca}.}
    \label{fig:dc}
\end{figure}

Figure~\ref{fig:dc} compares results from lattice QCD with realistic sea content of $n_f=2+1+1$ or $2+1$ sea quarks with the 
FLAG~2019~\cite{Aoki:2019cca} average for the $2+1+1$ sea.
Because the Fermilab/MILC results dominate the FLAG average, we simply quote them~\cite{Bazavov:2017lyh}:
\begin{eqnarray}
    f_{B^+}  &=& 189.4  (0.8)_\text{stat}  (1.1)_\text{syst}  (0.3)_{\fpiPDG} [0.1]_\text{EM scheme}~\MeV , \label{eq:fB+}\\ 
    f_{B^0}  &=& 190.5  (0.8)_\text{stat}  (1.0)_\text{syst}  (0.3)_{\fpiPDG} [0.1]_\text{EM scheme}~\MeV , \label{eq:fB0}\\ 
    f_{B_s}  &=& 230.7  (0.8)_\text{stat}  (1.0)_\text{syst}  (0.2)_{\fpiPDG} [0.2]_\text{EM scheme}~\MeV . \label{eq:fBs}\\
    f_{D^0}  &=& 211.6  (0.3)_\text{stat}  (0.5)_\text{syst}  (0.2)_{\fpiPDG} [0.2]_\text{EM scheme}~\MeV , \label{eq:fD0}\\ 
    f_{D^+}  &=& 212.7  (0.3)_\text{stat}  (0.4)_\text{syst}  (0.2)_{\fpiPDG} [0.2]_\text{EM scheme}~\MeV , \label{eq:fD+}\\ 
    f_{D_s}  &=& 249.9  (0.3)_\text{stat}  (0.2)_\text{syst}  (0.2)_{\fpiPDG} [0.2]_\text{EM scheme}~\MeV , \label{eq:fDs}
\end{eqnarray}
where the systematic uncertainties stem from different choices in choosing fit ranges for the correlation functions and checking 
the continuum extrapolation by adding a coarser lattice; the third ``\fpiPDG'' error comes from converting from lattice units to 
MeV with the pion decay constant of the PDG~\cite{PDG2018}; the last uncertainty stems from ambiguities in estimating 
electromagnetic effects in the context of a QCD calculation omitting QED.
The results are arguably precise enough for the foreseeable future.

The results in Eqs.~(\ref{eq:fB+})--(\ref{eq:fDs}) again use the very high statistics MILC HISQ ensembles with staggered fermions
for the sea quarks.
Here the lattice spacing is, for some ensembles, small enough to reach the $b$ quark, so the calculation uses the HISQ action for
all $b$ and light quarks alike.
Thus, an absolutely normalized pseudoscalar density is available, so the uncertainty is essentially statistical, as propagated
through a fit to the continuum limit with physical quark mass.
Again, other groups will have to collect similar statistics in the future to enable a complete cross check.

To go beyond the precision quoted here, analyses of leptonic decays will have to include QED radiative corrections to the measured
rates.
The issues and an elegant solution for light mesons (pion and kaon) can be found in
Refs.~\cite{Carrasco:2015xwa,Giusti:2017dwk,DiCarlo:2019thl}.
Radiative corrections for heavy-light mesons will be more difficult to incorporate, because of the hierarchy of soft scales
$\Lambda_\text{QCD}$, $\Lambda_\text{QCD}^2/m_Q$, $\Lambda_\text{QCD}^3/m_Q^2$, etc.


\section{Heavy-to-heavy inclusive}
\label{h2h_incl}


\subsection{Heavy Quark Expansion for $b\to c$}


\subsubsection{Review of the Current Status}
The heavy quark expansion (HQE) for the inclusive semileptonic $b\to c$ transitions 
starts form a correlation function for the $b\to c$ current 
\begin{eqnarray}
&& d \Gamma  \propto   \sum_X (2 \pi)^4 \delta^4 (P_B - P_X -q)
\langle B(v)  | \bar{b} \gamma_\mu (1-\gamma_5) c | X \rangle \, \langle X | \bar{c} \gamma_\nu (1-\gamma_5) b | B(v) \rangle   
\nonumber \\
&&  =
\int d^4 x  \, e^{iq\cdot x} \, \langle B(v) |  \bar{b}(x)  \gamma_\mu (1-\gamma_5) c (x)  \bar{c} \gamma_\nu (1-\gamma_5) b    | B(v) \rangle
\nonumber \\
&&=  2 \mbox{ Im} 
\int d^4 x  \, e^{iq\cdot x} \,  \langle B(v) |T \{ \bar{b}(x)  \gamma_\mu (1-\gamma_5) c (x)  \bar{c} \gamma_\nu (1-\gamma_5) b \} | B(v) \rangle 
 \\
	    &&=  2 \mbox{ Im} 
\int d^4 x \,  e^{-i (m_b v - q) \cdot x}
\langle B(v) |T \{ \bar{b}_v (x)  \gamma_\mu (1-\gamma_5) c (x)  \bar{c} \gamma_\nu (1-\gamma_5) b_v  \} | B(v) \rangle \nonumber
\end{eqnarray}
with 
$$
b(x) = e^{-im_b v\cdot x} b_v (x) \, . 
$$
The time ordered product in the last line can be expanded in an operator product expansion which for large $m_b$ and $m_c$ yields
an expansion in terms of local hadronic matrix elements which parametrize the hadronic input. Within this approach, the 
differential rate can be expressed as a series in $1/m$  
\begin{eqnarray}
d \Gamma &=& d \Gamma_0 + \left(\frac{\Lambda_{\rm QCD}}{m_b}\right)^2  d \Gamma_2 
+ \left(\frac{\Lambda_{\rm QCD}}{m_b}\right)^3  d \Gamma_3 + 
\left(\frac{\Lambda_{\rm QCD}}{m_b}\right)^4  d \Gamma_4 
\nonumber \\ \nonumber 
&& + d \Gamma_5 \left( a_0  \left(\frac{\Lambda_{\rm QCD}}{m_b}\right)^5 
+ a_2 \left(\frac{\Lambda_{\rm QCD}}{m_b}\right)^3 \left(\frac{\Lambda_{\rm QCD}}{m_c}\right)^2 \right) \\
&& + ... + d \Gamma_7  \left(\frac{\Lambda_{\rm QCD}}{m_b}\right)^3  \left(\frac{\Lambda_{\rm QCD}}{m_c}\right)^4
\label{HQE1} 
\end{eqnarray}
The coefficients $d\Gamma_i$ are given by 
\begin{equation}
d \Gamma_i = \sum_k  C_i^{(k)}   \langle B(v) | O_i^{(k)}  | B(v) \rangle 
\end{equation} 
where the $O_i^{(k)}$ are operators of mass-dimension $i+3$ and the sum over $k$ runs over all elements of the operator basis, 
$C_i^{(k)}$ are coefficients that can be calculated in QCD perturbation theory as a series in $\alpha_s (m_b)$. 
Note that starting at order $1/m_b^3$ the $b \to c$ HQE exhibits an infrared sensitivity to the charm quark mass; for the total
rate, $\Gamma_3$ contains a $\log(m_c^2)$ while $ \Gamma_5$ contains inverse powers of $m_c^2$  which are explicitly shown in 
Eq.~(\ref{HQE1}). 

The leading term $d \Gamma_0$ is the partonic result which turns out to be independent of any unknown hadronic matrix element.   
This term is fully known (triple differential rate) at tree level, at order $\alpha_s$~\cite{Trott:2004xc,Aquila:2005hq} and
order $\alpha_s^2$~\cite{Aquila:2005hq,Pak:2008qt,Biswas:2009rb,Melnikov:2008qs,Gambino:2011cq}. 

Due to heavy quark symmetry, there is no term $d\Gamma_1$ and the leading power corrections appear at order $1/m^2$.
These are given in 
terms of two non-perturbative matrix elements 
\begin{align}
    2 M_B \mu_\pi^2  &= - \langle B (v)|\bar{b}_v  (iD)^2  b_v | B (v) \rangle
    \label{MuPi} \\
    2 M_B \mu_G^2    &= -i\langle B (v)|\bar{b}_v \sigma_{\mu \nu} (iD^\mu )( iD^\nu) b_v| B (v) \rangle
    \label{MuG}                                    
\end{align} 
The coefficients of these two matrix elements are known to order
$\alpha_s$~\cite{Becher:2007tk,Alberti:2012dn,Alberti:2013kxa,Mannel:2014xza,Mannel:2015jka}. 
At order $1/m_b^3$ there are again only two matrix elements which are given by 
\begin{align}
    2 M_H \rho_D^3     &= - \langle B (v) |\bar{b}_v (iD_\mu)  (ivD) (iD^\mu)  b_v | H (v) \rangle \\
    2 M_H \rho_{LS}^3  &= -i\langle B (v) |\bar{b}_v\sigma_{\mu \nu} (iD^\mu ) (ivD) ( iD^\nu) b_v| B (v) \rangle
\end{align}
For these matrix elements only the tree level coefficients are known. 
Furthermore, if the matrix elements are defined as above\footnote{More commonly used definitions differ by $O(1/m_b)$ terms.},
the coefficient of $\rho_{LS}^3 $ vanishes for the total rate, 
which is related to reparametrization invariance of the HQE~\cite{Mannel:2018mqv}. 

The HQE predictions of the inclusive semileptonic rates depend on $m_b$ and $m_c$, and the size of the perturbative QCD corrections
depends on the choice of the quark-mass scheme. The quark masses are discussed in detail in a different section of this paper, and we refer the 
reader to this section.

\subsubsection{Higher power corrections} 
At order $1/m_b^4$ and higher the number of independent nonperturbative parameters starts to proliferate. In addition, due to the 
dependence on powers of $1/m_c$ the power counting needs to be re-defined: since we have parametrically
$m_c^2 \sim \Lambda_{\rm QCD} m_b$ one has to count the term  $d \Gamma_5 a_2$  as a part of $d \Gamma_4$, see (\ref{HQE1}). 
Thus the full complexity of the dim-8 operators already enters an analysis of the $1/m_b^4$ contribution.  

We shall not list the independent matrix elements appearing at order $1/m_b^4$ and $1/m_b^5$, rather we refer the reader to the list given in Refs.~\cite{Mannel:2010wj,Heinonen:2014dxa}.
However, the proper counting of the number of independent operators has been settled only recently~\cite{Kobach:2017xkw},
using the method of Hilbert series.
It turns out that at tree level there are 9 dimension 7 operators~\cite{Mannel:2010wj} while QCD 
corrections will increase this number to 11~\cite{Kobach:2017xkw}. 

The reason is very simple. At order $1/m_b^4$ we have operators with four covariant derivatives, which can be written as 
$ \langle \bm{E}^2 \rangle $ (chromoelectric field squared) and  
$ \langle \bm{B}^2 \rangle $  (chromomagnetic field squared)  
where $\bm{E}$  and $\bm{B}$ are both color-octets. Thus the combination appearing at tree level is 
\begin{equation}
\bm{E}^2 =  \bm{E}^a \cdot \bm{E}^b \,\, T^a T^b  \quad \mbox{and likewise for} \, \bm{B}^2 \, . 
\end{equation}  
However, the symmetric product of $T^a$ and $T^b$ contains a singlet and an octet component 
\begin{equation}
\frac{1}{2} ( T^a T^b + T^b T^a ) = \delta^{ab} + d^{abc} T^c  \, . 
\end{equation} 
The two terms on the right-hand side acquire different coefficients once QCD corrections are taken into account, and thus become 
independent operators. Although this observation~\cite{Kobach:2017xkw} is correct, it has no impact unless QCD corrections are 
considered at order $1/m_b^4$.
The same argument explains the different counting at order $1/m_b^5$ where we have 18 parameters at tree 
level~\cite{Mannel:2010wj}, while the general case involves 25 matrix elements~\cite{Kobach:2017xkw}. 

Clearly the number of independent parameters appearing at order $1/m_b^{4,5}$ is too large to extract them from experiment,
even if data will become very precise in the future. To this end, one has to rely on some  additional theoretical input,
which should better   be model dependent.
A systematic approach has been proposed in Ref.~\cite{Mannel:2010wj} and refined in Ref.~\cite{Heinonen:2014dxa}:
it is based on the 
``lowest-lying state saturation Ansatz'' (LLSA) and corresponds to a naive factorization of the matrix elements.
The LSSA allows us to write 
all matrix elements appearing in $1/m_b^4$ and $1/m_b^5$  in terms of four parameters, which are $\mu_\pi^2$ and $\mu_G^2$ 
(see Eqs.~(\ref{MuPi}) and~(\ref{MuG})) and $\epsilon_{1/2}$ and  $\epsilon_{3/2}$, where $\epsilon_j$ are the excitation energies of the lowest 
orbitally excited spin symmetry doublets with $j$ the spin of the light degrees of freedom. Note that in this setup also $\rho_D$ and 
$\rho_{LS}$ can be computed which may serve as a check, since these parameters  can also be extracted from experiment. 

The LLSA has been used to study the impact of the $1/m_b^{4,5}$ terms on the extraction of $V_{cb} $ in 
Ref.~\cite{Gambino:2016jkc}. 
It turns out that, even if a generous margin is allowed for the uncertainties, the shift in the extracted $V_{cb}$ remains well 
below~1\%, and with the default choices of Ref.~\cite{Gambino:2016jkc} a shift of $-0.25\%$ is found. 
 
Recently the impact of the reparametrization invariance on the HQE has been re-investiga\-ted.
In Ref.~\cite{Mannel:2018mqv,Fael:2018vsp} it has been shown that the number of independent parameters in higher orders can be
reduced by reparametrization invariance, for the total rate and the $q^2$ moments. While the number of HQE parameters up to order
$1/m_b^2$ is still two, there is only one parameter at $1/m_b^3$, since the spin-orbit term can be absorbed into $\mu_G^2$.
At order $1/m_b^4$ there will be only four parameters, which opens up the possibility of constraining the higher dimensional matrix
elements directly with experimental data, at least if Belle~II will be able to measure several moments of the $q^2$ distribution.


\subsubsection{Heavy Quark Expansion for $B \to X_c \tau \bar{\nu}$}
The recent data on the exclusive decays $B \to D^{(*)} \tau \bar{\nu}$ indicate that the branching ratios of these channels 
lie above the prediction of the SM. This issue is discussed in detail in sec.~\ref{sec:Vcb-RD-RDs}, but we may also consider the 
inclusive decay $B \to X_c \tau \bar{\nu}$ for which the HQE provides us with a precise prediction. 

While a new measurement of $B \to X_c \tau \bar{\nu}$ 
has to wait until Belle II has collected a sufficient data sample, we may compare with a measurement
performed at LEP resulting in~\cite{PDG2018}
$$
{\rm Br}(b\mbox{-admix} \to X \tau \bar{\nu})  = (2.41 \pm 0.23)\%
$$  
where $b\mbox{-admix}$ refers to the $b$-hadron admixture  produced in a $Z$ decay. Since to leading order the 
inclusive semitauonic branching fraction of all $b$-hadrons are the same, we may take this as an estimate of $B \to X_c \tau \bar{\nu}$.  

This has to be compared with the measured sum of $B \to D  \tau \bar{\nu}$ and $B \to D^{*} \tau \bar{\nu}$
$$
{\rm Br}(B \to [D + D^*] \tau \bar{\nu})  = (2.68  \pm 0.16)\%,
$$
indicating that the two ground states tend to oversaturate the inclusive decay. 

The decay $B \to X_c \tau \bar{\nu}$ has been studied in the HQE~\cite{Ligeti:2014kia} up to $1/m_b^2$ and $\alpha_s^2$ in the $1S$ scheme, resulting in  
$$
{\rm Br}(B^- \to X_c \tau \bar{\nu}) = (2.42 \pm 0.05)\% .
$$
More recently,  sizable effects of order $1/m_b^3$ have been found~\cite{Mannel:2017jfk}, which using the kinetic scheme, but without $O(\alpha_s^2)$ contributions, found
$$
{\rm Br}(B^- \to X_c \tau \bar{\nu}) = (2.26 \pm 0.05)\%  .
$$
The additional inclusion of  $O(\alpha_s^2)$ effects in the kinetic scheme appears to lead to a very similar 
value~\cite{Bhattacharya:2018kig}.  
These HQE calculations are compatible with the LEP measurement. 

\begin{figure}
\centering
\includegraphics[scale=0.6]{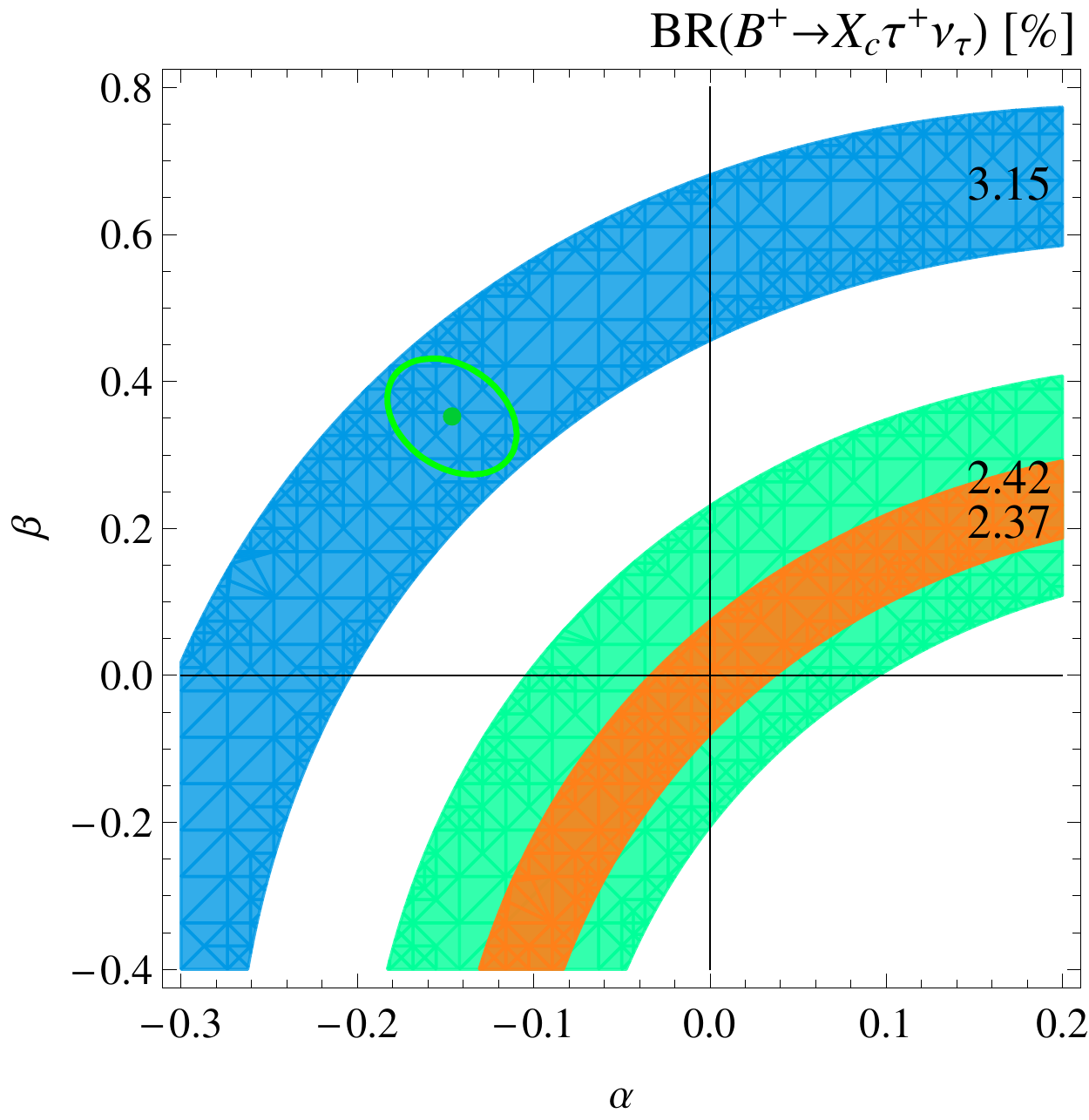}
\caption{Fit of the data to the parameters $\alpha$ and $\beta$. The green ellipse represents the fit result to the 
exclusive channels, the green band represents the LEP measurement, the red band the SM result obtained form the HQE.}
\label{fig:albe}  
\end{figure} 

However, the LEP measurement is not very precise and thus leaves room for new physics contributions. In the context 
of $R(D^{(*)})$ many new physics scenarios have been discussed, and we will not repeat any of this here.
Instead we use a very simple ansatz to explore qualitatively the effect of new physics.
To this end, we add an additional interaction of the form
\begin{equation} \label{HNP}
\mathcal{H}_{\rm NP}  = \frac{G_F V_{cb}}{\sqrt 2} 
\left( \alpha  \, O_{V+A} + \beta \, O_{S-P} \right)
\end{equation} 
with 
\begin{eqnarray}
O_{V+A}  & = &  \left( \bar c \gamma_\mu (1 + \gamma_5) b \right) 
\left(\bar \tau \gamma^\mu (1 - \gamma_5) \nu \right), 
\label{eq:operators} \\
O_{S-P} & = &  \left( \bar c (1 - \gamma_5) b \right) 
\left(\bar \tau (1 - \gamma_5) \nu \right) 
\nonumber
\end{eqnarray} 
We may fit the two parameters $\alpha$ and $\beta$ to the data on $B \to D^{(*)} \tau \bar{\nu}$ and find 
$\alpha = -0.15 \pm 0.04$ and $\beta = 0.35 \pm 0.08$~\cite{Mannel:2017jfk}. This may be inserted back into the calculation of the
total rate for  $B \to X_c \tau \bar{\nu}$ for which we find 
\begin{equation}
 {\rm Br}(B^- \to X_c \tau \bar{\nu}) = (3.15 \pm 0.19)\%
\end{equation} 
indicating a significant shift of the inclusive rate. This result is graphically presented in fig~\ref{fig:albe} and indicates that generically 
the exclusive and inclusive data are in tension, unless the new physics is such that it almost cancels in the inclusive rate.

\subsection{Inclusive processes in lattice QCD}

Until recently, the application of lattice QCD has been limited to the 
calculation of form factors of exclusive processes such as
$B\to D^{(*)}\ell\nu$ or $B\to\pi\ell\nu$,
for which initial and final states contain a single hadron.
A first proposal to evaluate the structure
functions relevant to the inclusive decays $B\to X_{u,c}\ell\nu$ in lattice QCD
was put forward in \cite{Hashimoto:2017wqo}.
As mentioned above, 
the differential decay rate for the inclusive decay 
$B(p_B)\to X_c(p_X)\ell(p_\ell)\nu(p_\nu)$
may be written in terms of the structure functions of $W_{\mu\nu}(p_B,q)$, 
which contains the sum over all possible final states:
\begin{equation}
  W_{\mu\nu}(p_B,q) = \sum_X (2\pi)^3\delta^4(p_B-q-p_X)
  \frac{1}{2M_B}
  \langle B(p_B)|J_\mu^\dagger|X(p_X)\rangle
  \langle X(p_X)|J_\nu|B(p_B)\rangle,\nonumber
\end{equation}
where $J_\mu$ stands for the $b\to c$ weak current and 
$q^\mu=(p_\ell+p_\nu)^\mu$ is the momentum transfer.
The optical theorem relates this to the forward scattering matrix
element $T_{\mu\nu}(p_B,q)$,
\begin{equation}
  T_{\mu\nu}(p_B,q) = i\int d^4\! x\, e^{-iqx}
  \frac{1}{2M_B}
  \langle B(p_B)|T\{J_\mu^\dagger (x) J_\nu(0)\}|B(p_B)\rangle,
\end{equation}
as $-(1/\pi)\mathrm{Im} T_{\mu\nu}=W_{\mu\nu}$, see for instance
\cite{Manohar:1993qn,Blok:1993va}.

One can calculate these forward matrix elements on the lattice as long as the
momenta $p_B$ and $q$ are in the region where no singularity develops.
It means that the lattice calculation is possible in an unphysical
kinematical region where no real decay is allowed.
This kinematical region corresponds to the situation where the energy
given to the final charm system $p_X^0$ is too small to create 
real states such as the $D$ and $D^*$ mesons or the $D\pi$ continuum
states. 
The connection to the physical region can be established by 
using Cauchy's integral on the complex plane of $p_X^0$.
An alternative method is to reconstruct the spectral density
(of the states $X$ appearing in the sum)
directly from the lattice correlation function 
\cite{Hansen:2017mnd}.

An exploratory lattice calculation has been performed at relatively
light $b$ quark masses
\cite{Hashimoto:2017wqo}.
The numerical results suggest that the matrix element is nearly
saturated by the ground state $D^{(*)}$ meson contribution
at the zero-recoil limit.

Since the non-perturbative lattice calculation may be obtained at the
kinematical point away from the resonance region, it may also be used
to validate the heavy quark expansion (HQE) method.
So far, the HQE calculation is available in the unphysical region only at the
tree-level, $O(\alpha_s^0)$.
The one-loop and two-loop corrections have been calculated for the
differential decay rate.
They have to be transformed to the unphysical kinematical point
by applying the Cauchy integral. Such work is in progress.

As already mentioned, the lattice calculation can be made only in the 
unphysical kinematical region and its comparison with the experimentally 
observed $B$ decay distribution is not straightforward. One should first 
perform an integral of the experimental data with an appropriate weight 
to reproduce Cauchy's integral in the complex plane of $p_X^0$, which 
requires the experimental data obtained as a function of two kinematical 
variables $q^2$ and $p_B\cdot q$. It still doesn't cover the whole 
complex plane, and one need to supplement by a perturbative QCD 
calculation for the region of $p_X^0>p_B^0$. The perturbative expansion 
in this unphysical region should be well-behaved, but the details should 
be investigated further.

More recently, a different approach that in principle allows  to calculate the total decay rate has been 
proposed \cite{Gambino:2020crt}.
 In the new method, the integral corresponding to the phase space of 
the $B\to X_c\ell\nu$
 is directly performed rather than the Cauchy's integral. As a result, 
 information about the
 unphysical kinematical region is no longer necessary. A first comparison with the HQE 
with a  small $m_b\sim 2.7$GeV shows good agreement with the lattice calculation, despite large uncertainties.
This method may open an opportunity to 
 compute the inclusive decay rate fully non-perturbatively using 
lattice QCD, and  
 can also be applied to calculate various moments of the $B\to X_c\ell\
\nu$ decays, as well as
the  more challenging $B\to X_u\ell\nu$ decays.

\subsection{HQE matrix elements from lattice QCD}
\label{sec:HQE-LQCD}

The same hadronic parameters appearing in the OPE analysis of inclusive semileptonic $B$-meson decays appear also in the HQE of
the pseudoscalar (PS) and vector (V) heavy-light meson masses.  
Therefore, one can try to determine them from a lattice calculation of the latter at different values of the heavy quark mass. 
After the pioneering work of Ref.~\cite{Kronfeld:2000gk}, new unquenched results have been presented 
recently~\cite{Bazavov:2018omf,Gambino:2017vkx}.
These papers are mentioned in Sec.~\ref{sec:qm} for their results on quark masses.

In Ref.~\cite{Gambino:2017vkx}  a precise lattice computation of PS and V heavy-light meson masses has been performed for 
heavy-quark masses ranging from the physical charm mass up to $\simeq 4$ times the physical $b$-quark mass, adopting the gauge
configurations generated by the European Twisted Mass Collaboration (ETMC) with $N_f = 2+1+1$ dynamical quarks at three values
of the lattice spacing ($a \simeq 0.062, 0.082, 0.089$~fm) with pion masses in the range $M_\pi \simeq 210$--450~MeV. 
The heavy-quark mass is simulated directly on the lattice up to $\simeq 3$ times the physical charm mass. 
The interpolation to the physical $b$-quark mass is obtained with the ETMC \emph{ratio 
method}~\cite{Blossier:2009hg,Bussone:2016iua}, based on ratios of the spin-averaged meson masses computed at nearby heavy-quark 
masses, and the kinetic scheme is adopted.
The extrapolation to the physical pion mass and to the continuum limit yields $m_b^{\rm kin}(1~\mbox{GeV}) = 4.61 (20)$~GeV, 
corresponding to $\overline{m}_b(\overline{m}_b) = 4.26 (18)$~GeV in the $\overline{\rm MS}$ scheme, in agreement with other
$m_b$ determinations; see Sec.~\ref{sec:qm}.
The ratio method is applied above the physical $b$-quark mass to provide heavy-light meson masses towards the static point. 
The lattice data are analyzed in terms of the HQE and the matrix elements of dimension-4 and dimension-5 operators are determined 
with good precision, namely:
 \begin{align}
    \label{eq:dim4_final}
    \overline{\Lambda} &= 0.552 ~ (26) ~\text{GeV} , \\
    \label{eq:dim5_1_final}
    \mu_\pi^2 &= 0.321 ~ (32)~\text{GeV}^2  , \\
    \label{eq:dim5_2_final}
    \mu_G^2(m_b) &= 0.253 ~ (25)~\text{GeV}^2 .
\end{align}
The size of two combinations of the matrix elements of dimension-6 operators is also determined: 
\begin{align}
     \label{eq:dim6_1_final}
     \rho_D^3 - \rho_{\pi \pi}^3 - \rho_S^3 &= 0.153 ~ (34) ~\text{GeV}^3  ~ , \\
     \label{eq:dim6_2_final}
     \rho_{\pi G}^3 + \rho_A^3 - \rho_{LS}^3 &= -0.158 ~ (84) ~\text{GeV}^3 ~ ,
\end{align}
with the full covariance matrix provided in Ref.~\cite{Gambino:2017vkx}.
Although all the above results refer to the asymptotic limit, namely to infinitely heavy quarks, and differ from the matrix 
elements extracted in the inclusive fits described above by higher power corrections, they are found to be mutually consistent.
In the future lattice results could  be used as additional constraints in the semileptonic fits. Another interesting future 
application concerns the heavy-quark sum rules for the form factor entering the semileptonic decay $B \to D^* \ell \nu$ at 
zero-recoil; here the non-local correlators $\rho_{A, S, \pi \pi, \pi G}$ play an important role; see Ref.~\cite{Gambino:2012rd}.

The analysis by the Fermilab, MILC and TUMQCD Collaborations~\cite{Bazavov:2018omf}, based on~\cite{Brambilla:2017hcq}, employs 
only PS mesons and the minimal renormalon subtracted (MRS) heavy quark mass. The results are obtained using MILC ensembles  with 
five values of lattice spacing ranging from approximately 0.12~fm to 0.03~fm, enabling good control over the continuum 
extrapolation, and both physical and unphysical values of the two light and the strange sea-quark masses.
This leads to 
\bea
\overline{\Lambda}_{\rm MRS}= 0.555~(31)~\text{GeV} 
\eea
while power corrections are controlled by the difference $\mu_\pi^2-\mu_G^2(m_H)$.
Assuming  $\mu_G^2(m_b)=0.35(7)\mbox{GeV}^2$ as a prior, the authors find $\mu_\pi^2=0.05(21)\mbox{GeV}^2$.
Notice that the definition of $\mu_\pi^2$ used here still has a renormalon ambiguity of order $\Lambda_{\rm QCD}^2$.

\subsection{Experimental status}


\subsubsection{Measurements of inclusive observables in $B\to X_c\ell\nu$}

Several experiments have measured the partial branching fraction of the
inclusive decay~$B\to X_c\ell\nu$ ($\ell=e,\mu$) as a function of the lower
threshold on the lepton momentum ($E_\mathrm{cut}$), or other inclusive
observables in this decay such as the moments of the lepton energy and of the
$X_c$~mass distribution. Available measurements are listed in
Table~\ref{tab:mom_exp}, where it should be noted that the most recent
experimental result is from the year 2010.
\begin{table}
  \caption{List of available measurements of inclusive moments in
    $B\to X_c\ell\nu$. We also specify the types of the lepton energy
    $E_\ell$ and hadronic mass $M(X_c)$ spectrum moments which have been
    determined in the respective publications. The zeroth order moment of the
    lepton energy spectrum ($n=0$) refers to a measurement of the partial
    branching fraction.} \label{tab:mom_exp}
\begin{center}
\begin{tabular}{lll}
  \hline
  Experiment &
  Lepton spectrum moments $\langle E^n_\ell\rangle$ &
  Hadron spectrum moments $\langle M^{2n}_X\rangle$\\
  \hline \hline
  BaBar &
  $n=0,1,2,3$~\cite{Aubert:2009qda,Aubert:2004td} &
  $n=1,2,3$~\cite{Aubert:2009qda}\\
  Belle &
  $n=0,1,2,3$~\cite{Urquijo:2006wd} &
  $n=1,2$~\cite{Schwanda:2006nf}\\
  CDF & &
  $n=1,2$~\cite{Acosta:2005qh}\\
  CLEO & &
  $n=1,2$~\cite{Csorna:2004kp}\\
  DELPHI &
  $n=1,2,3$~\cite{Abdallah:2005cx} &
  $n=1,2$~\cite{Abdallah:2005cx}\\
  \hline
\end{tabular}
\end{center}
\end{table}

The Belle collaboration has measured spectra of the lepton energy~$E_\ell$ and
the hadronic mass $M(X_c)$ in $B\to X_c\ell\nu$ using 152~million
$\Upsilon(4S)\to B\bar B$ events~\cite{Urquijo:2006wd,Schwanda:2006nf}. These
analyses proceed as follows: first, the decay of one $B$~meson in the event is
fully reconstructed in a hadronic mode ($B_\mathrm{tag}$). Next, the
semileptonic decay of the second $B$~meson in the event ($B_\mathrm{sig}$) is
identified by searching for a charged lepton amongst the remaining particles
in the event. In Ref.~\cite{Urquijo:2006wd}, the electron momentum spectrum in the
$B$~meson rest frame is measured down to 0.4~GeV. In ~\cite{Schwanda:2006nf},
all remaining particles in the event, excluding the charged lepton
(electron or muon), are combined to reconstruct the hadronic 
$X$~system. The $M(X_c)$ spectrum is measured for different lepton energy
thresholds in the $B$~meson rest frame. The observed spectra are distorted by
resolution and acceptance effects and cannot be used directly to obtain the
moments. In the Belle analyses, acceptance and finite resolution
effects are corrected by unfolding the observed spectra using the
Singular Value Decomposition (SVD) algorithm~\cite{Hocker:1995kb}. Belle
measures the energy moments $\langle E^k_\ell\rangle$ for $k=0,1,2,3,4$ and
minimum lepton energies ranging from 0.4 to 2.0~GeV. Moments of the hadronic
mass~$\langle M^k_X\rangle$ are measured for $k=2,4$ and minimum lepton
energies from 0.7 to 1.9~GeV.
\begin{figure}
  \centering
  \includegraphics[width=0.45\columnwidth]{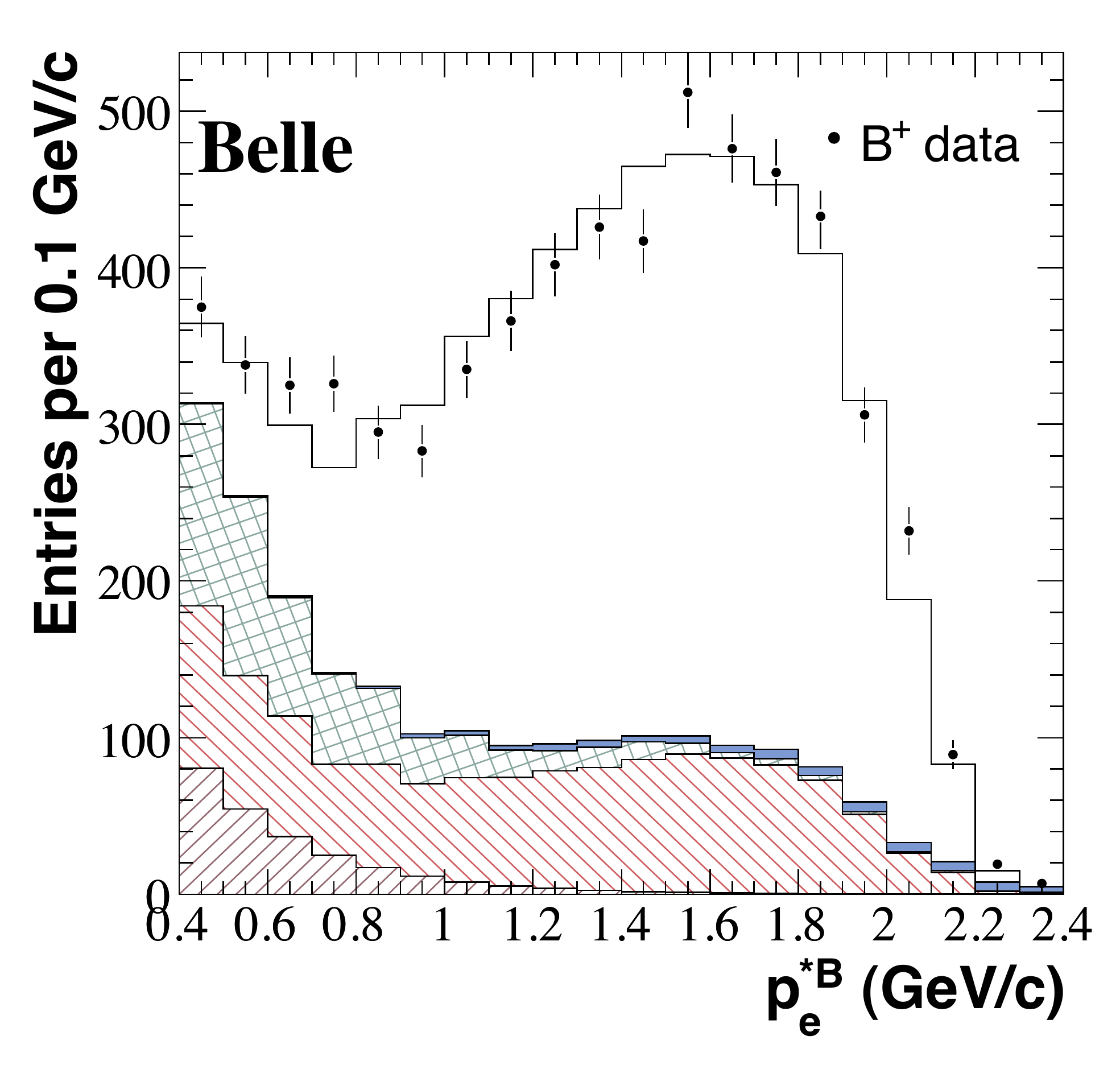}
  \includegraphics[width=0.43\columnwidth]{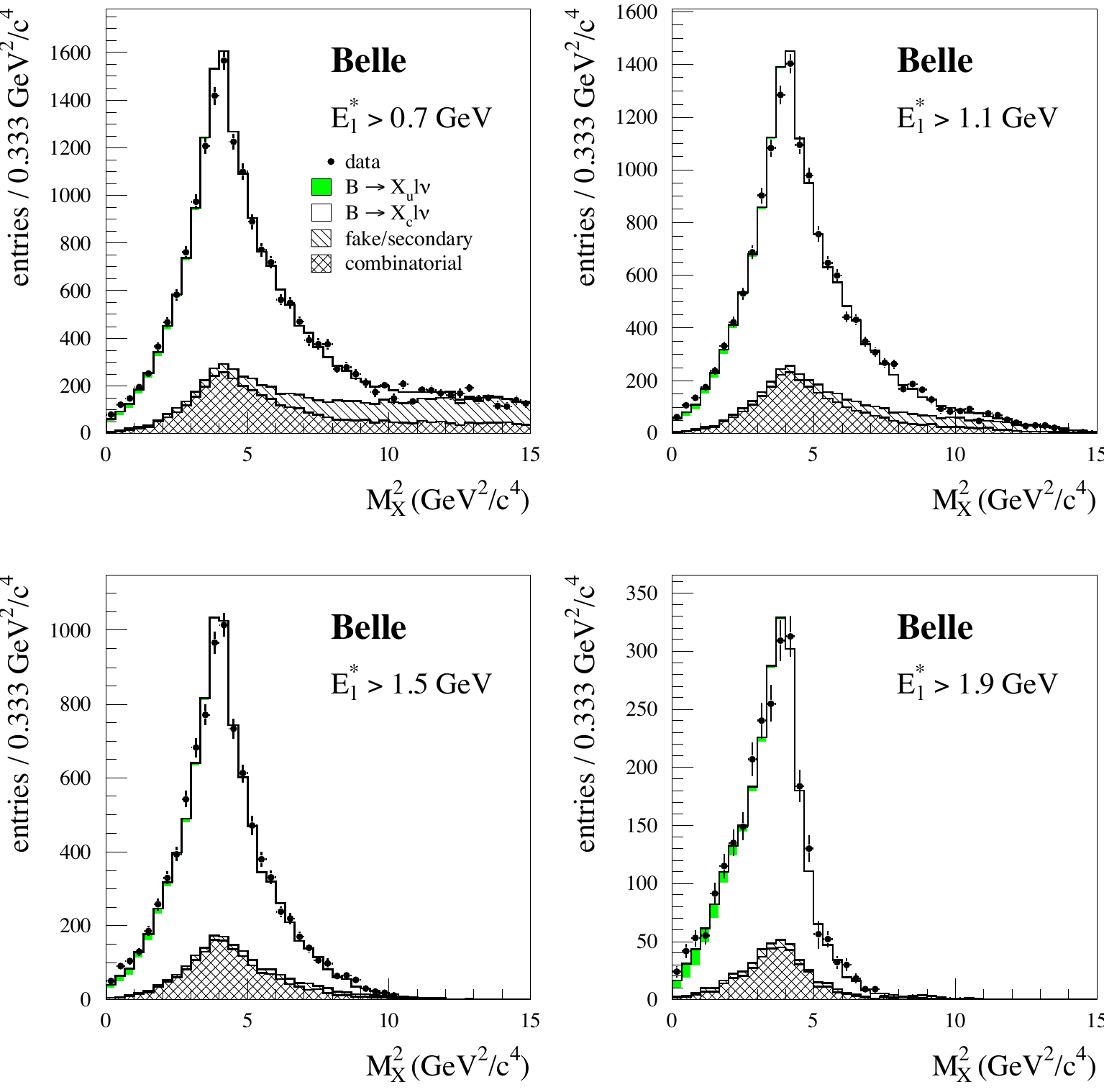}
  \caption{Belle measurements of the electron energy (left) and hadronic mass
    (right) spectra~\cite{Urquijo:2006wd,Schwanda:2006nf}.}
\end{figure}

BaBar has measured the lepton energy and hadronic mass moments in
$B\to X_c\ell\nu$~\cite{Aubert:2004td,Aubert:2009qda}. Furthermore, first
measurements of combined hadronic mass and energy moments of the form
$\langle n^k_X\rangle$ with $k=2,4,6$ are presented. They are
defined as $n^2_X=M^2_X-2\widetilde\Lambda E_X+\widetilde\Lambda^2$,
where $M_X$ and $E_X$ are the mass and the energy of the $X$~system and the
constant $\widetilde\Lambda$ is taken to be 0.65~GeV. The most recent analysis
is the one of hadronic mass $M(X_c)$ moments, which are determined using a
data sample of 232 million $\Upsilon(4S)\to B\bar B$
events~\cite{Aubert:2009qda}. The experimental method is similar to
the Belle analysis discussed previously, \emph{i.e.}, one $B$~meson is
fully reconstructed in a hadronic mode and a charged lepton with
momentum above 0.8~GeV in the $B$~meson frame identifies
the semileptonic decays of the second $B$. The remaining particles
in the event are combined to reconstruct the hadronic system $X$.   
The resolution in $M(X_c)$ is improved by a kinematic fit to the whole event,
taking into account 4-momentum conservation and constraining the missing 
mass to zero. To derive the true moments from the reconstructed ones, BaBar
applies a set of linear corrections. These corrections depend on 
the charged particle multiplicity of the $X$~system, the normalized missing
mass, $E_\mathrm{miss}-p_\mathrm{miss}$, and the lepton momentum. In this way,
BaBar measures the moments of the hadronic mass spectrum up to
$\langle M^6_X\rangle$ for minimum lepton energies ranging from 0.8 to 1.9~GeV.

\subsubsection{Determination of $|V_{cb}|$ from inclusive decays}

The Heavy flavor Averaging Group (HFLAV) has used the measurements
discussed in the previous section to determine $|V_{cb}|$ from a fit to HQEs of
inclusive observables~\cite{Amhis:2016xyh}. Using expressions in the so-called
kinetic
scheme~\cite{Benson:2003kp,Gambino:2004qm,Gambino:2011cq,Alberti:2013kxa,Alberti:2014yda}
and a precise determination of the $c$-quark mass,
$m_c^{\overline{\rm MS}}(3~{\rm GeV})=0.986\pm 0.013$~GeV~\cite{Chetyrkin:2009fv},
as external input, HFLAV obtains 
\begin{eqnarray}
  |V_{cb}| & = & (42.19\pm 0.78)\times 10^{-3}~, \\
  m_b^{\rm kin} & = & 4.554\pm 0.018~{\rm GeV}~, \\
  \mu^2_\pi & = & 0.464\pm 0.076~{\rm GeV^2}~.
\end{eqnarray}
The $\chi^2$ of the fit is 15.6 for $43$ degrees of freedom. Using
expressions in the so-called $1S$ scheme~\cite{Bauer:2004ve,Bauer:2002sh} the same set of
measurements results in
\begin{eqnarray}
  |V_{cb}| & = & (41.98\pm 0.45)\times 10^{-3}~, \\
  m_b^{1S} & = & 4.691\pm 0.037~{\rm GeV}~, \\
  \lambda_1 & = & -0.362\pm 0.067~{\rm GeV^2}~,
\end{eqnarray}
with a $\chi^2$ of the fit of 23.0 for $59$ degrees of freedom. This analysis
uses measurements of the photon energy moments in $B\to
X_s\gamma$~\cite{Aubert:2005cua,Aubert:2006gg,Limosani:2009qg,Chen:2001fja}
to constrain the $b$-quark mass and does not include higher order corrections of $O(\alpha_s^2)$ and $O(\alpha_s/m_b^2)$.

As mentioned above, the semileptonic moments have been analysed also including higher order power corrections 
estimated using the LSSA~\cite{Gambino:2016jkc}.
In this case a kinetic scheme fit to the experimental data that additionally includes 
a constraint  $m_b^{kin} = 4.550(42)$GeV from PDG (after scheme conversion) leads to a slightly more precise value,
\begin{eqnarray}
  |V_{cb}| & = & (42.00\pm 0.64)\times 10^{-3}~. 
\end{eqnarray}


\section{Heavy-to-light inclusive}
\label{h2l_incl}

\subsection{Introduction and theoretical background}

Inclusive semileptonic heavy to light decays can in principle be analyzed similarly to $B\to X_c \ell\nu$ by using a local OPE.
In practice,  due to the large charm background, experimental cuts are generally imposed and reduce the ``inclusivity" of the
theoretical prediction.
In particular, the local OPE does not converge well when the invariant mass of the hadronic system is $M_X\lesssim M_D$.  
In such a case the decay spectra are described using a 
``non-local" OPE~\cite{Neubert:1993ch,Neubert:1993um,Bigi:1993ex}, where perturbative coefficients are convoluted with 
non-perturbative ``Shape Functions" (SFs), the $B$ meson analogs of parton distribution functions.
In this SF region, the perturbative coefficients themselves can be factorized  into ``hard" and ``jet" pieces, where the former has
a typical scale of $m_b$ and the latter has a typical scale of $\sqrt{m_b\Lambda_{\mbox{\scriptsize QCD}}}$.
In the infinite mass limit $m_b\to\infty$ there is a single non-perturbative SF.
Power corrections start at $1/m_b$ and include multiple ``subleading" 
SFs~\cite{Lee:2004ja,Bosch:2004cb,Beneke:2004in,Bauer:2001mh,Leibovich:2002ys}.

One can classify the terms based on their suppression by $1/m_b$ and $\alpha_s$.
The perturbative components of the leading power term are known at $O(\alpha_s^2)$~\cite{Gambino:2006wk,Bonciani:2008wf,%
Asatrian:2008uk,Beneke:2008ei,Bell:2008ws,Brucherseifer:2013cu}.
The $1/m_b$ power corrections include terms convoluted with the leading power SF whose perturbative parts are known at
$O(\alpha_s)$~\cite{Paz:2009ut} and terms convoluted with subleading SFs whose perturbative parts are known at
$O(\alpha_s^0)$~\cite{Lee:2004ja,Bosch:2004cb,Beneke:2004in}.
At this order one can still use subleading functions of one light-cone variable. The inclusion of $O(\alpha_s)$ contributions
of subleading SFs requires functions of multiple light-cone momenta in analogy to higher twist effects in Deep Inelastic Scattering~\cite{Ellis:1982cd}. 
Schematically, in the SF region we have the  factorization formula
\begin{equation}\label{factorization}
    d\Gamma\sim H\cdot J\otimes S+\frac{1}{m_b}\sum_{i}\,h\cdot J_0\otimes s_i
        +\frac{1}{m_b}\sum_{k}\,h\cdot j_k\otimes S+\,O\left(\frac{1}{m_b^2}\right)\,,
\end{equation}
where $H$ is the leading power hard function, $J$ is the leading power jet function, both known at $O(\alpha_s^2)$,
$J_0$ is the $O(\alpha_s^0)$ part of $J$, $h=1+O(\alpha_s)$, $s_i$ are given in
Refs.~\cite{Lee:2004ja,Bosch:2004cb,Beneke:2004in}, and $j_k$ in Ref.~\cite{Paz:2009ut}.
The symbol $\otimes$ denotes an integral over the light-cone momentum.   

The moments of the leading and subleading SFs are related to the HQE parameters measured in the inclusive semileptonic decays to charm.
The relations are known for the leading SF up to at least the fifth moment~\cite{Gunawardana:2017zix}, although the current large 
uncertainty of higher HQE parameters~\cite{Mannel:2010wj,Gambino:2016jkc} might limit the use of higher moments relations.
The formalism in Ref.~\cite{Gunawardana:2017zix} allows to construct such relations for the subleading SFs too, but at present 
only the first three moments are known \cite{Bauer:2001mh,Bauer:2002yu}.
A detailed knowledge of the SFs is necessary only in a portion of the phase space where $p_+=E_X-p_X\sim \Lambda_{\rm QCD}$; 
elsewhere only the first few moments of the SFs are relevant and one recovers the local OPE description.

The present $|V_{ub}|$ determination by HFLAV~\cite{Amhis:2016xyh} is based on various approaches which are all rooted in 
(\ref{factorization}) and differ in the inclusion and treatment of perturbative and nonperturbative contributions,
see Ref.~\cite{Antonelli:2009ws} for a detailed discussion.

The approach known as BLNP (Bosch-Lange-Neubert-Paz)~\cite{Lange:2005yw} aimed at  a precision extraction of $|V_{ub}|$ from  $B\to X_u \ell\nu$ and
$B\to X_s\gamma$, based on the knowledge in 2005.  It used the first two terms in (\ref{factorization}), in particular the
$O(\alpha_s)$ expression for $H\cdot J\otimes S$ and the $O(\alpha^0_s)$ expression for the
$h\cdot J_0\otimes s_i$ terms. Kinematical corrections  that scale as $\alpha_s/m_b$ and $\alpha_s/m^2_b$~\cite{DeFazio:1999ptt},
as well as $1/m_b^2$ corrections~\cite{Manohar:1993qn,Blok:1993va}, for which factorization formulas were not known, were also 
included by convolution with the leading power shape function. Using Renormalisation Group methods $H$ is evolved from ``hard" to the ``jet" scale
to resum Sudakov double logs.  As for the non-perturbative inputs, the leading order SF
was to be taken  from $B\to X_s\gamma$ and subleading SFs $s_i$ to be modeled using $\sim700$ models. In practice, the current 
treatment of $S$  by experiments is to use an exponential or Gaussian model constrained by the first two moments of $S$ obtained 
from the global fit of HQE parameters in the kinetic scheme~\cite{Amhis:2016xyh}.

Since Ref.~\cite{Lange:2005yw} appeared, there have been many theoretical advances.
Two-loop calculations of  $H$~\cite{Bonciani:2008wf,Asatrian:2008uk,Beneke:2008ei,Bell:2008ws} and $J$~\cite{Becher:2006qw} as well
as one-loop calculation of $j_k$~\cite{Paz:2009ut} became available.
The free quark  differential decay rate were calculated at $O(\alpha_s^2\beta_0)$~\cite{Luke:1994du,Bauer:2001rc,Hoang:2005pj,Gambino:2006wk} and at complete
$O(\alpha_s^2)$~\cite{Brucherseifer:2013cu}.
Running effects from the ``hard" to the ``jet"  at $O(\alpha_s^2)$ were studied~\cite{Greub:2009sv}.
It was found there that the factorization of the perturbative coefficient into jet and hard functions is not strictly necessary. 
More recently,  three loop calculations of $J$ \cite{Bruser:2018rad} and the \emph{partonic} $S$ \cite{Bruser:2019yjk} were performed. Implementing these within the BLNP framework would probably require also the calculation of $H$ at three-loops, which is not available yet.
There were also theoretical advances in the description of non-perturbative effects in $B\to X_s\gamma$~%
\cite{Lee:2006wn,Benzke:2010js,Benzke:2010tq}.
In particular, new subleading shape functions unique to $B\to X_s\gamma$ were identified~\cite{Benzke:2010js}, making it more 
difficult to use data from radiative $B$ decays as input for the extraction of $|V_{ub}|$.
These new features are not yet implemented in the BLNP approach.
An alternative implementation of the same conceptual framework has been presented in Ref.~\cite{Ligeti:2008ac}, together with a 
systematic procedure to account for the uncertainties in the modelling of the leading SF, to be discussed below.

The GGOU (Gambino-Giordano-Ossola-Uraltsev) approach~\cite{Gambino:2007rp} avoids the expansion in $1/m_b$ and the introduction of subleading SFs.
The perturbative coefficients are computed at fixed order to  $O(\alpha_s^2\beta_0)$ in the kinetic scheme.
The effect of RGE evolution in the SF region and all subleading SFs are absorbed into three $q^2$-dependent SF $F_i(k, q^2)$,
whose first moments are fixed by present semileptonic fits. The  uncertainty due to the functional form is estimated comparing
$\sim100$ models.

The emergence of the SF can also be seen  in perturbation theory: soft-gluon resummation together with an infrared prescription gives
rise to a $b$ quark SF.
In the DGE (Dressed-Gluon Exponentiation) approach~\cite{Andersen:2005mj,Gardi:2008bb} this is achieved by an internal resummation
of running coupling corrections in the Sudakov exponent, thus providing a perturbative model for the leading SF.
A somewhat similar line of action is followed in Ref.~\cite{Aglietti:2007ik} where the infrared prescription is provided by the 
so-called analytic QCD coupling.

The so-called  Weak Annihilation (WA) contributions are a  source of theoretical uncertainty common to all approaches.
In the local OPE they emerge at $O(1/m_b^3)$ but are enhanced by a large Wilson coefficient~\cite{Bigi:1993bh} and may give rise to 
a difference between $B^+$ and $B^0$ decays.
As they are expected to be much more important in charm decays, the latter constrain them most effectively at present.
In particular, the $D^0$, $D^+$ and $D_s$ total semileptonic rates and the electron spectra measured by the  CLEO 
Collaboration~\cite{Asner:2009pu} have been employed~\cite{Bigi:2009ym,Ligeti:2010vd,Gambino:2010jz}.
From the absence of  clear indications for WA effects in semileptonic charm decays, one can conclude that the  WA correction to the 
total rate of $B\to X_u \ell\nu$ must be smaller than about 2\%~\cite{Gambino:2010jz}.
However, WA is localized in the high $q^2$ region and therefore the related uncertainty on $|V_{ub}|$ depends on the kinematical
cuts, and this is taken into account in the current HFLAV averages.
Because the high $q^2$ tail is particularly sensitive to higher power corrections (and not to the SFs), see for instance
Refs.~\cite{Bauer:2000xf,Bauer:2001rc,Gambino:2007rp}, one might eventually expect the cleanest determinations of $|V_{ub}|$ to come from the
low~$q^2$ region only.
An upper cut on $q^2$ might therefore be beneficial \cite{Lange:2005yw,Gambino:2007rp}. 

A few recent experimental analyses \cite{Urquijo:2009tp,Lees:2011fv} have relaxed the kinematic cuts, making use of experimental 
information to subtract the background.
As a result, most of the $B\to X_u \ell\nu$ phase space is taken into account and the sensitivity to the SFs is substantially 
reduced, while a description based on the local OPE sets in.
In these cases the quoted theoretical uncertainties are smaller, but one should keep in mind that these analyses still depend
on the SFs treatment and modelling for the determination of the reconstruction efficiencies, whose uncertainty contribute to the 
final experimental systematic error.
As will be discussed later on, a realistic signal simulation requires the implementation of so-called hybrid models that transform
the inclusive predictions of the approaches mentioned above into individual final hadronic states.
The uncertainties related to such hybrid models remain a major issue for the inclusive determination of $|V_{ub}|$.


\subsection{Status of the experimental results} 


The most difficult task of the inclusive measurements is the discrimination between the $B\to X_u\ell\nu$ signal and the much more abundant decays involving Cabibbo-favoured $B\to X_c\ell\nu$ decays. 
The signal events are studied in restricted regions of the phase space to improve the signal-to-background ratio.
Compared to $B\to X_c\ell\nu$ events, the signal tends to have higher lepton momenta $p_\ell$, lower invariant mass of the $X_u$ state $M_X$, 
higher $q^2$, and smaller values of the light-cone momentum $P_+=E_X-|\bm{p}_X|$, where $E_X$ and $\bm{p}_X$ are energy and momentum of 
the hadronic system $X_u$ in the $B$ meson rest frame. As explained above, these restrictions introduce difficulties in the
calculation of the expected partial branching fraction, enhancing perturbative and nonperturbative QCD corrections which lead to large 
theoretical uncertainties in the measurement of~$|V_{ub}|$. 

The measurement of the partial branching fraction $\Delta\cal{B}$ can be obtained with \emph{tagged} or \emph{untagged} analyses. 

\subsubsection{Tagged Analyses}

In tagged analyses, the $\Upsilon(4S)\to B{\overline B}$ events are identified by reconstructing one of the $B$ mesons, $B_{reco}$, via fully hadronic decays. The signal decay of the second $B$ meson ($B_{signal}$) is identified just by the presence of an electron or a muon.
The tracks and neutral objects not associated with the $B_{reco}$ can be uniquely assigned to the signal side, so that the inclusive $X_u$ state can be clearly reconstructed. The neutrino four-momentum $p_\nu$ can be estimated from the missing momentum $p_{miss}=p_{e^+e^-}-p_{B_{reco}}-p_{X_u}-p_\ell$, where $p_{e^+e^-}$ is the initial state four-momentum. From this, all the kinematic variables of the signal state can be easily computed. 

Because the momentum of the signal $B$ meson is determined from of the $B_{reco}$, the signal decay products can be computed directly in the $B$-meson rest frame, resulting in an improved resolution of the accessible observables. Moreover, the constrained kinematics allow for a better separation of the signal from the background. 

The downside of the tagged analysis is the low signal efficiency (about 0.3-0.5\%) which implies that
for kinematic variables like the lepton momentum $p_\ell$, the untagged analyses at the B-factories can give competitive or better 
results. 
Undetected and poorly reconstructed tracks or photons lead to irreducible background from the dominant $B\to X_c$ decays even in regions 
of the phase space potentially free of such background, and this can affect the final resolution on the signal kinematics. 

The hadronic $B$-tagging approach was used for the first time by BaBar to extract the $B\to X_u\ell\nu$ signal in the phase space region $M_X<1.55$~GeV, with the further requirement that $p_\ell>1$~GeV \cite{Aubert:2003zw}. Using the same sample, BaBar removed the constraint on $M_X$ and obtained the $B\to Xu\ell\nu$ partial branching ratio requiring only $P_\ell>1$~GeV, which covers about $90\%$ of the signal phase space \cite{Aubert:2006qi}. This challenging analysis was affected by the large statistical uncertainties and limited by the knowledge on the $B\to X_c$ background components and  the signal composition available when it was published.

Exploiting the full datasets collected, both Belle ~\cite{Urquijo:2009tp}  and BaBar ~\cite{Lees:2011fv} published measurements of $B\to X_u\ell\nu$ partial branching fraction, performing a fit in $M_X$ and $q^2$, and requiring only $p_\ell > 1$~GeV. BaBar determined also the partial branching fractions in other several restricted regions of the phase space.
 

\subsubsection{Untagged Analyses}

The untagged measurements allow to collect large samples but are affected by considerable backgrounds.
The untagged measurements have access only to a few kinematic variables, namely the lepton momentum $p_\ell$, and the $q^2$ spectra,
\begin{itemize}
\item lepton spectrum: this can be studied inclusively without requirements on the rest of the event.
In this case the momentum spectrum can only be given in the $\Upsilon(4S)$ rest frame. 
\item $q^2$ distribution: this requires the reconstruction of the neutrino 4-momentum, which exploits the high hermeticity of the
$B$~factories' detectors. 
The neutrino 4-momentum is given by the event missing 4-momentum, $p_{miss}=p_{e^+e^-}-p_{vis}$, where $p_{e^+e^-}$ is the initial state 
4-momentum, and $p_{vis}$ is the total visible 4-momentum determined by all the charged tracks from the collision point, identified pairs 
of charged tracks from $K_s$, $\Lambda$ and $\gamma\to e^+e^-$, and energy deposits in the electromagnetic calorimeter. 
\end{itemize}

The lepton momentum spectrum is affected by large backgrounds  
from $B\to X_c\ell\nu_\ell$ via the $D\ell\nu$, $D^{*}\ell\nu$, $D^{**}\ell\nu$ (where by $D^{**}$ is a mixture of charm excited state and non resonant $D^{(*)}-n\pi$ transitions), $D_s K\ell\nu X$ and also secondary leptons from $D$ mesons decays, and a background from $e^+e^-\to q\overline{q}$ events, where the main contribution comes from $c\overline{c}$, which is assessed from control data samples recorded below the $\Upsilon(4S)$ resonance. 
Because of the large background, usually the signal is extracted only for regions with high momentum lepton, typically $p_\ell>1.9-2.1$ \mbox{GeV}. Old analyses of the lepton endpoints are from CLEO \cite{Bornheim:2002du}, Belle \cite{Limosani:2005pi} and BaBar \cite{Aubert:2005mg}.

Recently, BaBar published  a study \cite{TheBABAR:2016lja} of the lepton spectrum using the full data set, and exploiting all the knowledge about the rate and the form factors of the various $B\to X_c\ell\nu$ exclusive decays which are the major source of backgrounds. 
The signal is extracted from a fit to the electron momentum spectrum, which is described as the sum of predicted signal (model dependent shape) and various specific backgrounds yields with shapes fixed by MC. The fit covers lepton momentum in the $\Upsilon(4S)$ rest frame from 0.8 to 2.7 \mbox{GeV}, in 50 \mbox{MeV} bins, except for the data in the interval 2.1 to 2.7 \mbox{GeV} which are combined in a single bin to avoid effects from differences in the shape of the theoretically predicted signal spectrum. In a given momentum interval, the excess of events above the sum of the fitted background contributions is taken as the number of signal events. 

An important difference of this analysis with respect to the other ones is that different theoretical models are considered in the extraction of the partial branching fractions. Instead, all other measurements determine the partial branching fraction by using a single model, and its partial rate is then converted in a measurement of $|V_{ub}|$ by taking the corresponding partial rate predicted by the theory calculations.

The extracted inclusive signal branching fractions and the values of $|V_{ub}|$ agree well for GGOU and 
DGE, although they are about 13\% smaller than the average of the other measurements. This difference can be attributed to the shape of the predicted signal spectrum and/or the shapes of some of the large background contributions above 2 \mbox{GeV} where the signal fraction is largest. On the other hand, the value of $|V_{ub}|$ based on BLNP agrees well with other measurements. 

\begin{figure}
\centerline{\includegraphics[scale=0.60]{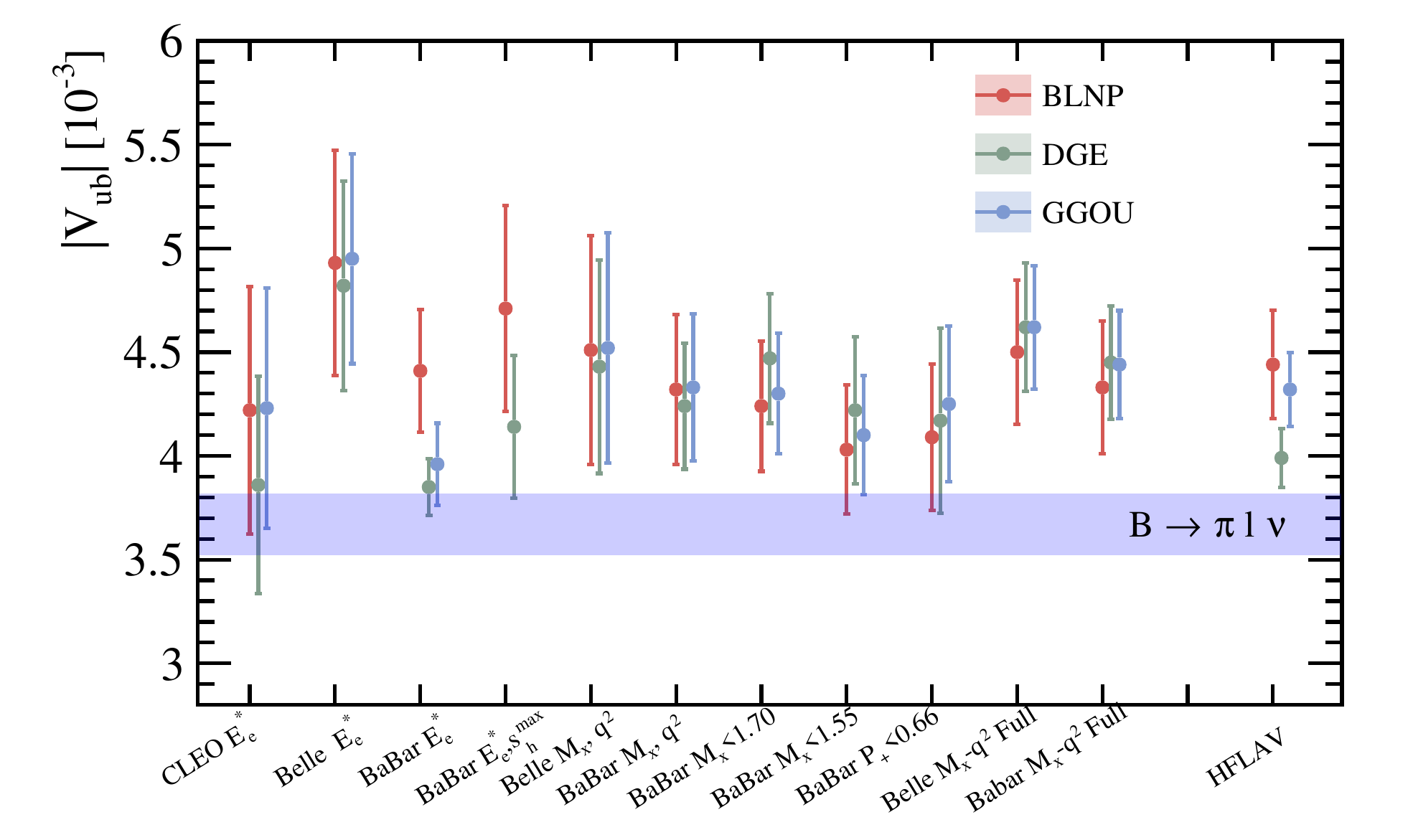} }
\caption{Measurements of inclusive $|V_{ub}|$ and their averages
based on BLNP, DGE and GGOU calculations. The HFLAV average of $|V_{ub}|$ results from  $B\to\pi\ell\nu_\ell$ decay is also reported for comparison.}
\label{fig:vub_incl_results}  
\end{figure}

A subset of all the measurements of the inclusive $|V_{ub}|$ are reported in Fig.\ref{fig:vub_incl_results} for the various frameworks considered, see \cite{Amhis:2019ckw} for more details. 

\subsubsection{Lessons learned from the past }

The measurements based on tagged samples have considerably larger statistical uncertainties. The sample size allows for only a few bins in the 2D fit, but there are regions of the phase space (e.g. low $M_X$) where the background fractions are modest. The current sensitivity to the details of the shapes of the signal and background distributions is however limited.

For untagged measurements only the high end of the spectrum is sensitive to the signal and also to the background near their kinematic endpoints.
Both approaches have their pros and cons, given the size of the currently available data.
The latest BaBar measurement of the lepton spectrum, shows a high dependence of the result from the signal model. The same effect, even if not directly evident, was observed also in tagged measurements from the sensitivity of the signal yield extraction on the shape function parameters in the analyses that cover larger portion of the phase space.  

\begin{figure}
\centerline{\includegraphics[scale=0.3]{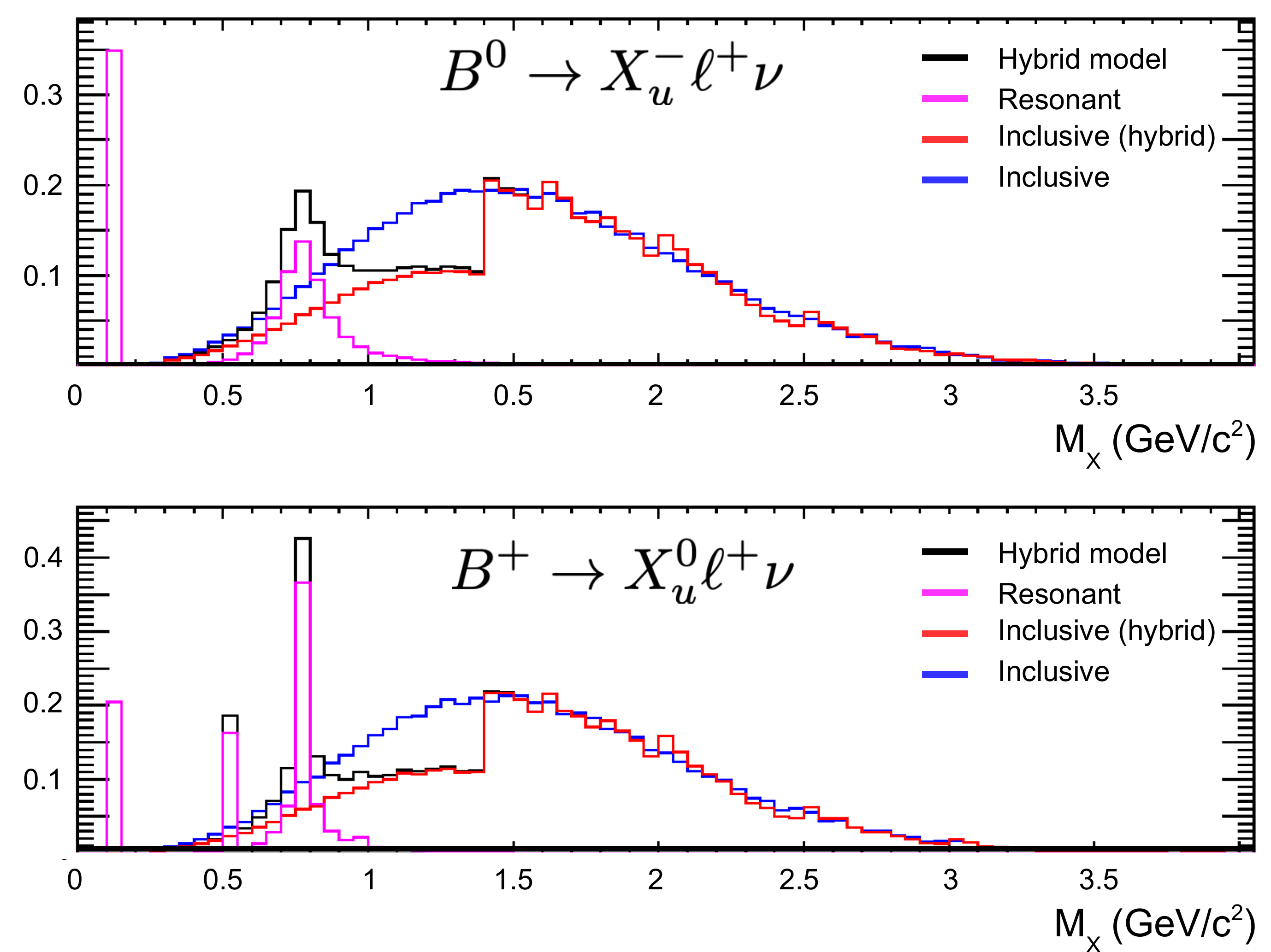} }
\caption{Model of the hadronic invariant mass $M_X$ for the signal $B\to X_u\ell\nu$ events, separately for $B^0$ (top) and $B^+$ (bottom). }
\label{fig:vub_incl_hybrid}  
\end{figure}

Semileptonic $B\to X_u\ell\nu$ decays are simulated as a combination of resonant decays with $X_u=\pi,\eta,\eta',\rho,\omega$, and decays to nonresonant  hadronic final states $X_u$. 
The latter is simulated with a continuous invariant mass spectrum following the  theory predictions by De Fazio and Neubert \cite{DeFazio:1999ptt}, which depend on the SF parameters and $m_b$. 
The nonresonant and the resonat part are combined such that the sum of their branching fractions is equal to the measured one for the inclusive $B\to X_u\ell\nu$. 
The events generated with this model, are reweighted to obtain predictions for different SF parameters and different branching fraction of the resonant states. This model is usually called "hybrid model". Belle in \cite{Urquijo:2009tp}, corrects the hybrid model to match the moments of the $M_X$ and $q^2$ distributions predicted by the the GGOU model. A picture of the model of the invariant mass $M_X$ shape used to describe the $B\to X_u\ell\nu$ is reported in Fig.\ref{fig:vub_incl_hybrid}.

Another effect not considered so far, is the impact of the fragmentation of the generated $u$ quark into final state hadrons, which is performed using JETSET. The modeling of the final state multiplicity could affect both the signal efficiency and the signal templates used to separate signal from background.


The measurement of the partial branching fraction separately for neutral and changed $B$ mesons has been used 
to constrain the WA contribution. Both tagged approach, in various regions of the phase space \cite{Lees:2011fv}, and untagged approach, in the high lepton region \cite{Aubert:2007tw}, have been used, but these have given weak upper limits mainly because of the large statistical uncertainties.
More stringent upper limit on WA has been obtained by CLEO which used a model dependent approach studying the high $q^2$ region in $B\to X_u\ell\nu$ decays \cite{Rosner:2006zz}. Both these bounds are milder than those estimated from $D$ and $D_s$ semileptonic decays in Refs.\cite{Ligeti:2010vd,Gambino:2010jz} which were mentioned above.

In the tagged measurements the suppression of the $b\to c$ background is performed by vetoing events where a $K^+$
or a $K_s^0$ is detected in the hadronic $X$ system. This causes a loss in the signal contribution where a $s{\bar s}$ pair is produced  (usually called $s {\bar s}$-popping). The fraction of these events is about $12\%$ of the non-resonant component and it is
fixed in the fragmentation parameters of JETSET/PYTHIA. The uncertainty on this fraction is assumed to be about $30\%$, so for analyses that aim to cover larger regions of the phase space, with higher statistics this could be an irreducible source of systematic uncertainty. This is another point that should be improved in future analyses at Belle~II.

\subsection{Fitting distributions: SIMBA and NNVub}
\label{subsect:simba}

As we discussed above, SFs modelling is an important source of theoretical 
uncertainty in the study of $B\to X_u \ell \nu$ and particularly in the extraction 
of $|V_{ub}|$ from these decays. While the first few moments of the SFs must 
satisfy OPE constraints, direct experimental information on the SFs is somewhat limited. 
Indeed, the measured photon spectrum in $B\to X_s\gamma$ is sensitive to a different set of subleading SFs. 
However differential distributions in $B\to X_u \ell \nu$ such as the lepton energy and the invariant mass distributions depend directly on all the SFs and can therefore be used to constrain them. Conversely, they can be used to validate SFs models and approaches
where the SFs are calculated, such as DGE. The high luminosity expected  makes the measurement of differential distributions possible at Belle~II.

The extraction of $|V_{ub}|$ performed by HFLAV in the BLNP and GGOU frameworks assumes a set of two-parameter functional forms, and it is unclear to what extent the chosen set is representative of the available functional space, and whether the estimated uncertainty really reflects the limited knowledge of the SFs.
This point was first emphasized in Ref.~\cite{Ligeti:2008ac}, where a different strategy was proposed, based on the expansion of the leading SF in a basis of orthogonal functions, whose coefficients are fitted to the $B\to X_s \gamma$ spectrum, and on the modeling of the subleading SFs.  The SIMBA project 
\cite{Bernlochner:2013gla} aims at performing a global fit to $B\to X_s \gamma$ and $B\to X_u \ell\nu$ spectra, to simultaneously determine $|V_{ub}|$, $m_b$,
the leading SF, as well as the Wilson coefficient of radiative $b$ decays. Additional external constraints, such from $B\to X_c\ell\nu$, can also be employed.

Another strategy, called NNVub and explored in  \cite{Gambino:2016fdy} for the GGOU approach, employs artificial neural networks as unbiased interpolants for the SFs, in a way similar to what the NNPDF Collaboration do in fitting for Parton Distribution Functions \cite{Ball:2008by}. This method allows for unbiased estimates of the SFs functional form uncertainty, and for a straightforward implementation of new experimental data, including $B\to X_s \gamma$ and $B\to X_u \ell\nu$ spectra and other inputs on quark masses and OPE matrix elements. 
Both SIMBA and NNVub appear well posed to analyse the Belle~II data in a model independent and efficient way. 

\subsection{Prospect for the future: Belle~II outlook}

The measurements of fully differential spectra on the kinematic variables, e.g. $q^{2}$, $M^{2}_{X}$, $p^{\pm}_{X}$, $E_{l}$, and separate measurements for charged and neutral B-meson decays are required to allow for an improved extraction of  $|V_{ub}|$ in the long term. Therefore, the future measurements should provide these unfolded spectra  independent of theoretical assumptions.

\begin{figure}
\centerline{\includegraphics[scale=0.25]{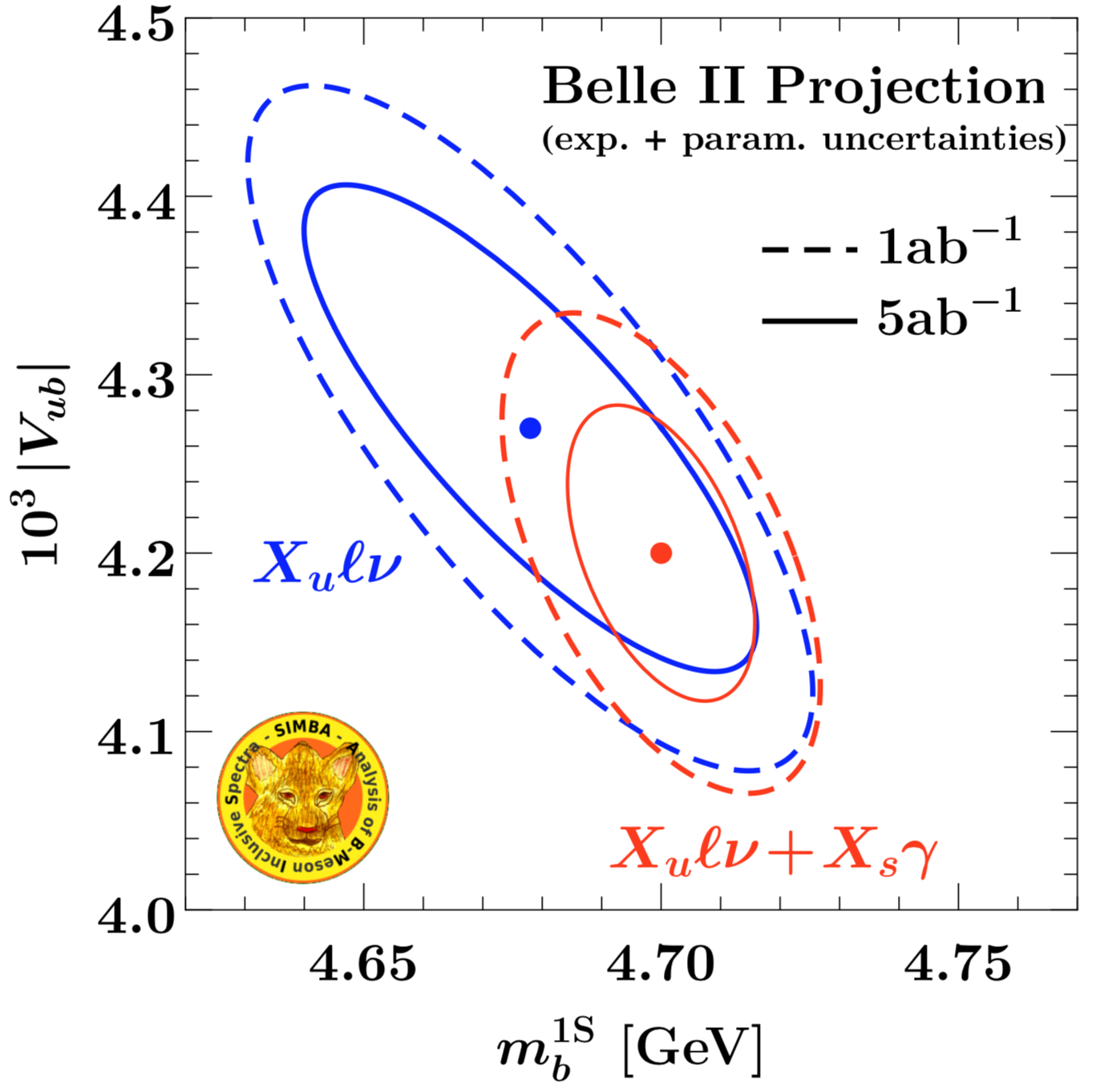} }
\caption{Belle~II projection for a global fit in the SIMBA approach of $|V_{ub}|$ with 1~ab$^{-1}$ and 5~ab$^{-1}$. Theory uncertainties are not included in the fit and are expected to be of similar size.}
\label{fig:simba_belle2_vub}  
\end{figure}

Combining both $B\to X_{u} \ell \nu$ and $B\to X_{s}\gamma$ as well as constraints on the SF moments  from $B\to X_{c} \ell \nu$ in a global fit can simultaneously provide the inclusive $|V_{ub}|$ and  the leading SF functional form with its uncertainties as they follow from the uncertainties in the included experimental measurements. Fig.~\ref{fig:simba_belle2_vub} shows the projections for a global fit in the SIMBA framework with two projected single-differential spectra of $M_{X}$ and $E_{\ell}$ for $B\to X_{u} \ell \nu$ and a $E_{\gamma}$ spectrum for $B\to X_{s}\gamma$ from 1~ab$^{-1}$ and 5~ab$^{-1}$ Belle~II data set~\cite{Kou:2018nap}.

The new tagging algorithm developed for Belle~II can perform better than the old neural network method used in the previous Belle publications with about 3 times higher efficiency~\cite{Keck:2018lcd}. With a larger data set, the systematic uncertainties counted for reconstruction efficiencies, fake leptons and continuum background knowledge are expected to improve for this measurement. The projections for inclusive $|V_{ub}|$ are summarized in Table~\ref{tab:belle_vub_error}.

\begin{table} 
\tabcolsep 4pt
\centering
\caption{Expected percentage uncertainties in inclusive $|V_{ub}|$ measurements with the Belle full data sample, 5~ab$^{-1}$ and 50~ab$^{-1}$ Belle~II data~\cite{Kou:2018nap}. 
}
\label{tab:belle_vub_error}
    \begin{tabular}{cccccc}
    \hline\hline
    Int. Luminosity           & Statistical & \begin{tabular}[c]{@{}l@{}}Systematic\\ (reducible, irreducible)\end{tabular} & Total Exp. & Theory  & Total   \\
    \hline
    605 fb$^{-1}$ (old B tag) & 4.5         & (3.7, 1.6)                                                                    & 6.0        & 2.5-4.5 & 6.5-7.5 \\
    5 ab$^{-1}$               & 1.1         & (1.3, 1.6)                                                                    & 2.3        & 2.5-4.5 & 3.4-5.1 \\
    50 ab$^{-1}$              & 0.4         & (0.4, 1.6)                                                                    & 1.7        & 2.5-4.5 & 3.0-4.8 \\
    \hline\hline
    \end{tabular}
\end{table}

\section{Outlook}
\label{summ}
We have summarized our main results in Sec.~1. In this final Section, we would like to
look at the prospects of our field over the next five years.
What can we expect for semileptonic $b$ decays at the two main experiments? What kind of progress can we  reasonably 
anticipate in lattice QCD and continuum calculations?

Belle II has started data taking with a complete detector in March 2019 and recorded about 10/fb in its first year of operation. By introducing the crab waist scheme at the collision point, SuperKEKB achieved the world's highest instantaneous luminosity of $2.4\times 10^{34}/$cm$^2$/s in June 2020 with acceptable background conditions for Belle II to take data. In total 64/fb of $\Upsilon(4S)$~data were recorded in the spring 2020 run, bringing the total to 74/fb. Data taking will resume in October 2020 with the goal to reach a total integrated luminosity of more than 100/fb before the end of the year break. Belle II plans to accumulate a data set equivalent to the Belle luminosity of about 1/ab by the end of 2021. In 2022 the experiment will enter a long shutdown to install the second pixel detector layer and replace the silicon photomultipliers in the barrel particle identification device. Data taking will resume in 2023 and by 2025 Belle II expects to have recorded a data sample exceeding 10/ab.

Given these luminosity prospects, competitive Belle II results for semileptonic $B$~decays can be expected in the years to follow. In addition, a three times more efficient hadronic tag and better low momentum tracking of the slow pion from the $D^*$~decay will further benefit semileptonic analyses in particular. This will allow to take a fresh look at the CKM matrix element magnitudes $|V_{cb}|$ and $|V_{ub}|$ and to improve measurements which are still statistically limited, such as $R(D)$ and $R(D^*)$.


The LHCb experiment has shown great capabilities with the results on $R(D^*)$, $|V_{ub}|/|V_{cb}|$ with $\Lambda_b$ decays, and $|V_{cb}|$ with $B_s$ decays. These measurements are based on the data collected in 2011 and 2012 (Run 1), corresponding at $3/$fb of integrated luminosity. 
The data collected in 2015-2018 (Run 2) at $pp$ collision energy of $\sqrt{s}=13~$ TeV, correspond to about $6/$fb of integrated luminosity. There are various ongoing analyses on the full dataset. 
Most of the measurements are limited by systematic uncertainties,
among which the largest ones are generally due to 
external inputs from other experiments and to the limited available samples of Monte Carlo simulations. Nevertheless the large dataset available is going to be fully exploited. 

The LHCb experiment is at present undergoing a major upgrade of the detector.
The construction and commissioning should end in 2021, when LHC will resume the activity.
The upgrade will allow to collect data at higher instantaneous luminosity, so about five $pp$ collisions per bunch crossing are foreseen,
to be compared with about one-two $pp$ collisions in Run1 and Run2.
To handle the higher occupancy expected in the detector, besides the improvements in the various subdetectors, a full software L0 trigger 
will be employed.
The software L0 trigger will add flexibility to the data taking, allowing to reduce the thresholds for muon and hadron trigger decisions,
enlarging in this way the physics capabilities.
The analyses of semileptonic decays with taus and electrons will benefit from the lower trigger thresholds in terms of signal
efficiencies.
With this upgraded detector, LHCb is planning to integrate a luminosity of $23/$fb by 2024, and
to collect a total sample of $50/$fb by 2028-2029, after LHC will have switched to higher luminosity. 

By now, lattice QCD is the tool of choice for the form factors describing semileptonic decays of $b$-hadrons.
At present, the most urgent need is the $q^2$ (or, equivalently, $w$) dependence of the form factors of $B\to D^*l\nu$, both to see how 
the form-factor slopes affect the $|V_{cb}|$ determination and to solidify the SM prediction of~$R(D^*)$.
A few such calculations are underway.
Given the success of LHCb with $\Lambda_b$ semileptonic decays, updates of the baryon form factors are desirable, and we encourage other 
lattice-QCD practitioners to turn their attention to these decays.
Another topic for future research are rigorous calculations with a $\rho$ or $\phi$ vector meson in the final state.

The leptonic decay constants are now at the subpercent level of uncertainty, and the prospects for extending these methods to semileptonic
form factors are underway.
In general, near-term lattice-QCD calculations of this precision will be based on the MILC collaboration's HISQ ensembles, which, among
all lattice data sets, span the largest range of lattice spacing at physical light-quark masses and with high statistics.
We consider it important that other ensemble sets be extended to a similar range, to enable further (sub)percent-level calculations with 
different systematics from the fermion discretization.

The inclusive determination of $|V_{cb}|$ will benefit from the calculation of new higher order effects, such as the $O(\alpha_s^3)$ contributions to the total width,  and from a reassessment of QED effects. However, the next frontier is represented by the integration with lattice QCD calculations
to improve the determination of HQE matrix elements, and eventually by the calculation of the inclusive rates directly on the lattice. 
For what concerns inclusive charmless decays, the general theoretical framework appears solid but needs to be updated in the light of
recent higher order calculations and should be extensively validated by experimental data which will become available at Belle~II.
In particular, the measurement of the lepton energy and hadronic invariant mass distributions will provide important information on the
Shape Functions, while the $q^2$ distribution will allow us to constrain and possibly avoid the effect of  Weak Annihilation.
The wealth of data expected at Belle~II, a close cooperation between theorists and experimentalists, and hopefully new lattice data should help resolve various open issues, so that we might 
eventually expect the uncertainty on inclusive $|V_{ub}|$ to become lower than 3\%. 

\begin{acknowledgement}
This work was  supported by the Mainz Institute for Theoretical Physics (MITP) of the Cluster of Excellence PRISMA+ (Project ID 39083149) which hosted the workshop. We acknowledge the friendly effectiveness of its staff and thank the scientific  coordinator  Tobias Hurth and  director Matthias Neubert for their encouragement. We are grateful to  G.~Caria, B.~Dey, A.~Greljo, B.~Grinstein, N.~Gubernari, G.~Herdoiza, Y.~Kwon, H.~Meyer, R.~Laha, W.~Lee,    P.~Owen, S.~Stefkova, P.~Urquijo, who also participated in the workshop.
F.~Bernlochner and L.~Cao  were supported by the DFG Emmy-Noether Grant No.\ BE~6075/1-1.
 C.~Davies was supported by the UK Science and Technology Facilities Council. 
A.~El-Khadra was supported by the U.S.\ Department of Energy, Office of Science, Office of High Energy Physics under Award Number DE-SC0015655 and by the Fermilab Distinguished Scholars Program. 
P.~Gambino and M.~Jung were supported by the Italian Ministry of Research (MIUR) under grant PRIN 20172LNEEZ.
S.~Hashimoto was supported  by JSPS KAK- ENHI Grant Number JP26247043 and by the Post-K and Fugaku supercomputer project through the Joint Institute for Computational Fundamental Science (JICFuS).
The work of A.~Khodjamirian and T.~Mannel was supported by the DFG (German Research
Foundation) under  grant 396021762 - TRR 257 "Particle Physics Phenomenology after the Higgs Discovery".
Z.~Ligeti was supported in part by the Office of High Energy Physics of the U.S.\ Department of Energy under contract DE-AC02-05CH11231. S.~Meinel was supported by the U.S.~Department of Energy, Office of Science, Office of High Energy Physics under Award Number DE-SC0009913.
G.~Paz was supported by the U.S. Department of Energy grant DE-SC0007983 and by a Career Development Chair award from Wayne State University.
S.~Schacht was supported by a DFG For\-schungs\-stipen\-dium under contract No.\ SCHA 2125/1-1. A.~Vaquero was supported by the U.S. National Science Foundation under grants PHY14-14614 and PHY17-19626.
This manuscript has been authored by Fermi Research Alliance, LLC under Contract No.~DE-AC02-07CH11359 with the U.~S.\ Department of 
Energy, Office of Science, Office of High Energy Physics.
\end{acknowledgement}

\bibliographystyle{spphys}
\bibliography{mitp,mitp_exp,ask}

\end{document}